%% file: main.tex
\documentclass[prd,nofootinbib,preprintnumbers]{revtex4-2}
\usepackage{epsfig,float}
\usepackage{amsfonts}
\usepackage{amsmath}
\usepackage[colorlinks=true,linkcolor=red,citecolor=blue]{hyperref}
\usepackage{wrapfig}
\usepackage{enumitem}
\usepackage{nicefrac}
\usepackage[parfill]{parskip}
\usepackage{orcidlink}
\usepackage{booktabs}
\usepackage{subcaption}

\usepackage[normalem]{ulem}

\newcommand{\re}{\mathrm{Re}}
\newcommand{\im}{\mathrm{Im}}
\newcommand{\dr}{\mathrm{DR}}
\newcommand{\dd}{\mathrm{d}}
\newcommand{\MSb}{\overline{\text{MS}}}
\newcommand{\LambdaQCD}{\Lambda_{\textrm{QCD}}}

\newcommand{\zmax}{z_{\textrm{max}}}

\usepackage{array}
\newcommand{\PreserveBackslash}[1]{\let\temp=\\#1\let\\=\temp}
\newcolumntype{C}[1]{>{\PreserveBackslash\centering}p{#1}}
\newcolumntype{R}[1]{>{\PreserveBackslash\raggedleft}p{#1}}
\newcolumntype{L}[1]{>{\PreserveBackslash\raggedright}p{#1}}

\begin{document}

\title{Nucleon tomography and total angular momentum of valence quarks \\ from synergy between lattice QCD and elastic scattering data}

\author{
Krzysztof Cichy$^{1}$\orcidlink{0000-0002-5705-3256},
Martha Constantinou$^{2}$\orcidlink{0000-0002-6988-1745},
Paweł Sznajder$^{3}$\orcidlink{0000-0002-2684-803X},
Jakub Wagner$^{3}$\orcidlink{0000-0001-8335-7096}
}

\affiliation{
$^1$ Faculty of Physics and Astronomy, Adam Mickiewicz University, ul.\ Uniwersytetu Poznańskiego 2, 61-614 Poznań, Poland \\
$^2$ Department of Physics, Temple University, Philadelphia, PA 19122 - 1801, USA \\
$^3$ National Centre for Nuclear Research, NCBJ, 02-093 Warsaw, Poland 
}

\begin{abstract}
We present an exploratory work on integrating lattice-QCD results with experimental data for elastic scattering. Within the framework of generalized parton distributions (GPDs), this approach allows for the extraction of detailed information about nucleon tomography and the total angular momentum carried by valence quarks. To accomplish this reliably, we introduce a new quantity, the ``double ratio'', which significantly reduces the systematic uncertainties inherent in lattice QCD computations. Moreover, we introduce a ``shadow'' term in the extraction procedure, which is sensitive only to lattice-QCD results. This term allows us to investigate the model dependence of the extraction, providing a more flexible description of the nucleon that goes beyond the previously considered bell-shaped distribution.

\end{abstract}

\date{\today}

\maketitle

\input{sec_intro}
\input{sec_basics}

\input{sec_lattice}
\input{sec_comparison}
\input{sec_extraction}
\input{sec_results}
\input{sec_summary}

\begin{acknowledgments}
The authors would like to thank N.~Nurminen for his contribution to the preparatory work and H.~Dutrieux for providing his insights. 

K.~C.\ is supported by the National Science Centre (Poland) grant OPUS No.\ 2021/43/B/ST2/00497.
M.~C. acknowledges financial support from the U.S. Department of Energy, Office of Nuclear Physics, under Grant No.\ DE-SC0025218. The work benefits from collaboration within the Quark-Gluon Tomography (QGT) Topical Collaboration funded by the U.S. Department of Energy, Office of Science, Office of Nuclear Physics with Award DE-SC0023646.

The generation of lattice data for this work was supported in part by the PLGrid Infrastructure (Prometheus supercomputer at AGH Cyfronet in Cracow), the Poznan Supercomputing and Networking Center (Eagle/Altair supercomputer), the Interdisciplinary Centre for Mathematical and Computational Modelling of the Warsaw University (Okeanos supercomputer), the Academic Computer Centre in Gda\'nsk (Tryton supercomputer), facilities of the USQCD Collaboration, funded by the Office of Science of the U.S. Department of Energy and resources of the National Energy Research Scientific Computing Center, a DOE Office of Science User Facility supported by the Office of Science of the U.S. Department of Energy under Contract No. DE-AC02-05CH11231 using NERSC award NP-ERCAP0022961.
\end{acknowledgments}

\bibliography{bibliography}

\end{document}

%% file: sec_intro.tex
\section{Introduction}
\label{sec:intro}

Generalized parton distributions (GPDs)~\cite{Muller:1994ses,Ji:1996ek,Ji:1996nm,Radyushkin:1996ru,Radyushkin:1997ki} provide a well-established framework within the factorization theorems of perturbative quantum chromodynamics and offer a wealth of unique information about the partonic content of the nucleon. In particular, GPDs are needed for the so-called nucleon tomography~\cite{Burkardt:2000za,Burkardt:2002hr,Burkardt:2004bv}, where the density of partons carrying a specific fraction of the nucleon's momentum is mapped in a plane perpendicular to the direction of the nucleon's motion. Additionally, within the GPDs framework, one can access components of the nucleon's energy-momentum tensor and, from there, evaluate the total angular momentum carried by specific partons~\cite{Ji:1996ek,Ji:1996nm} as well as the ``mechanical'' forces induced in a partonic medium~\cite{Goeke:2007fp,Polyakov:2018zvc}. GPDs may be inferred from experimental data in exclusive processes. An alternative source of information on GPDs is provided from first principle lattice-QCD calculations, with $x$-dependent determinations having become possible only very recently due to the development of novel techniques, such as the quasi- and pseudo-distributions~\cite{Ji:2013dva,Radyushkin:2017cyf}.

Experiments at DESY, CERN, and JLab have already delivered data for exclusive reactions sensitive to GPDs, such as deeply virtual Compton scattering (DVCS), deeply virtual meson production (DVMP), and timelike Compton scattering (TCS), the latter of which has only recently been measured~\cite{CLAS:2021lky}. Additionally, the GPDs topic is a cornerstone of future experimental programmes, particularly those at JLab~\cite{Accardi:2023chb}, EIC~\cite{AbdulKhalek:2021gbh} and EIcC~\cite{Anderle:2021wcy}. Despite this progress, it is clear that extracting GPDs from experimental data will be very challenging, and even data obtained in the foreseeable future may not result in precise constraints on these objects. The reason is the limited information that can be accessed from DVCS, DVMP, and TCS, as evidenced by the presence of so-called ``shadow'' GPDs~\cite{Bertone:2021yyz,Moffat:2023svr}, which decouple from observables for these processes. 

One potential solution to this problem is the measurement of processes with enhanced sensitivity \cite{Qiu_2023}, such as double DVCS (DDVCS)~\cite{Deja:2023ahc}, exclusive production of di-photons~\cite{Grocholski:2021man,Grocholski:2022rqj} or electro- and meso-production of the photon-meson pairs \cite{Duplancic:2018bum, Duplancic:2022ffo, Duplancic:2023kwe, Boussarie:2016qop, Qiu:2024mny}, which either provide complementary information on GPDs or allow for mapping them in the full kinematic domain. However, the feasibility of measuring these processes remains uncertain, primarily due to their small cross-sections. 

Another potential solution is to complement information with lattice-QCD results.
Quantities describing the structure of hadrons have been computed on the lattice for several years.
However, the focus was on moments of partonic distributions, which can be accessed via local matrix elements. 
In practice, this approach is constrained to only the lower moments, since the higher ones suffer from poor signal-to-noise ratios and unavoidable power-divergent mixings with lower-dimensional operators.
In turn, direct access to light-front correlations defining the $x$-dependent distributions is prohibited on a Euclidean lattice due to the Wick rotation to imaginary time.
As mentioned above, novel techniques were introduced to allow for indirect access, where an appropriate lattice observable can be factorized perturbatively into a light-front distribution.
Starting with seminal papers of Ji \cite{Ji:2013dva,Ji:2014gla} that introduced quasi-distributions, several other approaches were proposed or revived \cite{Liu:1993cv,Detmold:2005gg,Braun:2007wv,Chambers:2017dov,Radyushkin:2017cyf,Ma:2017pxb}, see Refs.~\cite{Cichy:2018mum,Radyushkin:2019mye,Ji:2020ect,Constantinou:2020pek,Cichy:2021lih,Cichy:2021ewm} for reviews.
While the initial center of attention were parton distribution functions (PDFs), the introduced indirect approaches allow for a simple generalization to include a momentum transfer between the initial and the final state, thus giving access to GPDs. 
Obviously, GPDs are more difficult to explore due to their dependence on additional variables. 
However, much work has already been performed, see, e.g., Refs.~\cite{Chen:2019lcm,Alexandrou:2020zbe,Alexandrou:2021bbo,Bhattacharya:2022aob,Cichy:2023dgk,Bhattacharya:2023ays,Bhattacharya:2023nmv,Bhattacharya:2023jsc,Bhattacharya:2024qpp,HadStruc:2024rix,Ding:2024hkz}.
In particular, the recent important development is to access GPDs in asymmetric frames of reference \cite{Bhattacharya:2022aob}, which enables computations for several momentum transfer values at once, leading to much improved computational efficiency.
Nevertheless, the lattice calculations of GPDs are still in their exploratory stage, and several sources of systematic uncertainties remain to be quantified.
Hence, it is somewhat unclear how lattice results can be meaningfully combined with experimental data in the phenomenology of GPDs, given significant sources of unquantified systematics related to them, such as higher-twist and higher-order corrections.

In this work, we address this question by exploring the possibility of combining lattice and experimental data. We focus specifically on the zero skewness case, that is, we only utilize elastic scattering experimental data. Combining these with lattice results allows us to address nucleon tomography and to extract the total angular momentum of partons, however, both only for the valence quarks. In addition to accessing this information, we are also interested in studying the systematics of nucleon tomography using the concept of shadow GPDs. We note that the limitation to only the valence quarks comes solely from the use of data for elastic scattering, as lattice-QCD results provide information on both the valence and sea components. The latter will be discussed in this work but will not be used in the extraction of the aforementioned information. For similar reasons, this work is limited to the so-called unpolarized parton distributions, despite lattice QCD also providing access to helicity and transversity cases~\cite{Alexandrou:2020zbe,Alexandrou:2021bbo,Bhattacharya:2023jsc}.

This article is organized as follows. In Sect.~\ref{sec:basics}, we remind the basics of the GPD formalism and define the nomenclature used throughout this text, while in Sect.~\ref{sec:lattice} we briefly summarize the extraction of lattice observables that are used in this work. In Sect.~\ref{sec:comparison}, we present a comparison between lattice-QCD results and parameterizations of PDFs, elastic form factors (FFs), and GPDs. In the same section, we also present the double ratio, a quantity that allows the cancellation of some sources of systematic uncertainties associated with lattice-QCD computations. In Sect.~\ref{sec:extraction}, we describe the extraction of tomography information, including the presentation of data for elastic scattering, and the introduction of the shadow term used to study model dependence. Finally, in Sect.~\ref{sec:results}, we present the obtained results, and in Sect.~\ref{sec:summary}, we provide a concise summary of this work.

%% file: sec_basics.tex
\section{Basics of GPD formalism}
\label{sec:basics}

\begin{figure}[!htp]
  \centering
  \caption{Momenta of the relevant particles and their plus components: four-momenta $p$ and $p'$ describe the initial and final-state protons, $k-\Delta/2$ and $k+\Delta/2$ describe the emission and reabsorption of a single quark, while $x+\xi$ and $x-\xi$ represent the fractional quark momenta, with $t$ being the Mandelstam variable describing the four-momentum transfer.} 
  \label{fig:gpd}
  \subfloat[Momenta of particles in the symmetric reference frame]{\label{fig:gpd1}\includegraphics[height=0.13\textwidth]{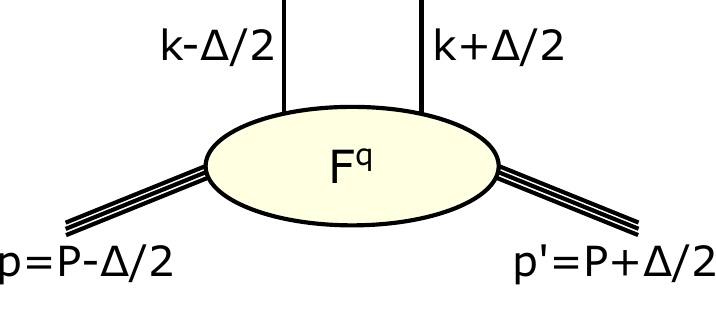}}\quad\quad\quad\quad
  \subfloat[Plus components of momenta]{\label{fig:gpd2}\includegraphics[height=0.13\textwidth]{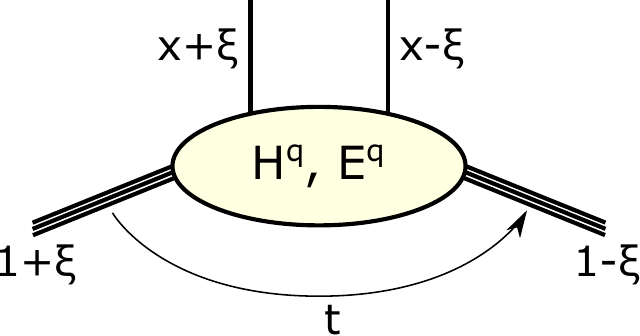}}
\end{figure}

For this investigation, we focus on the unpolarized case for quark GPDs for the proton. The formal definition of proton GPDs $H^q$ and $E^q$ for quarks, expressed through matrix elements of quark operators on the light-cone is as follows:
\begin{align}
\frac{1}{2} \int \frac{d z^-}{2\pi}\, e^{ix P^+ z^-}
  \langle p'|\, \bar{q}(-{\textstyle\frac{1}{2}} z)\, \gamma^+ q({\textstyle\frac{1}{2}} z) 
  \,|p \rangle \Big|_{z^+=0,\, \mathbf{z}=0}
\label{eq:gpd_definition_q}
= \frac{1}{2P^+} \bar{u}(p') \left[
  H^q(x,\xi,t) \gamma^+ +
  E^q(x,\xi,t) \frac{i \sigma^{+\alpha} \Delta_\alpha}{2M} 
  \, \right] u(p) \,,
\end{align}
where $M$ is the proton mass, and, for brevity, we omit the Wilson line. The light-cone vectors are given by $n_{\pm}=(1,0,0,\pm 1)/\sqrt{2}$, such that any four-momentum can be expressed as:
\begin{equation}
v^{\mu} = v^{+}n_{+}^{\mu} + v^{-}n_{-}^{\mu}+v_{\perp}^{\mu} \,,
\end{equation}
where $v^{\pm}= v \cdot n_{\mp} = (v^0 \pm v^{3})/\sqrt{2}$ and $v_{\perp}=(0,\mathbf{v},0)$. Specific four-momenta and variables are defined with the help of Fig.~\ref{fig:gpd}. GPDs depend on three kinematic variables: $x$, $\xi$, and $t$. The variable $x=k^{+}/P^{+}$ describes the average longitudinal momentum of the active parton, $\xi=(p^{+}-p'^{+})/(p^{+}+p'^{+})=-\Delta^{+}/(2P^{+})$ the change of this momentum, while $t=\Delta^2$ (one of the Mandelstam variables) characterizes the change of the proton's four-momentum. Additionally, GPDs also depend on the factorization scale, $\mu$, but this dependence will be suppressed throughout the text for brevity.

The so-called forward limits of $H^q$ and $E^q$ are:
\begin{align}
    H^q(x, 0, 0) &= q(x) \,, \\
    E^q(x, 0, 0) &= e_{q}(x) \,,
\end{align}
where $q(x)$ is the unpolarized PDF, while $e_q(x)$ has no correspondence in physics of (semi-)inclusive scattering and is subject to modeling or a lattice-QCD calculation. One of the most striking features of GPDs, being a consequence of the Lorentz invariance of these objects, is the so-called polynomiality, expressed by:
\begin{align}
\int_{-1}^1 \dd x\, x^n H^q(x,\xi,t) &=
  \sum_{i=0 \atop \scriptstyle{\rm even}}^n (2\xi)^i A^q_{n+1,i}(t)
  + \mbox{mod}(n,2)\, (2\xi)^{n+1} C_{n+1}^q(t) \,, \\
\int_{-1}^1 \dd x\, x^n E^q(x,\xi,t) &=
  \sum_{i=0 \atop \scriptstyle{\rm even}}^n (2\xi)^i B^q_{n+1,i}(t)
  - \mbox{mod}(n,2)\, (2\xi)^{n+1} C_{n+1}^q(t)\,.
\label{eq:polynomiality}
\end{align}
The polynomiality states that a given Mellin moment of GPD is a polynomial in even powers of $\xi$, where $A^{q}_{n+1,i}(t)$, $B^{q}_{n+1,i}(t)$ and $C^{q}_{n+1}(t)$ are generalized form factors. For $n=0$, they are equivalent to the contributions of a given quark to Dirac and Pauli~FFs:
\begin{align}
\int_{-1}^1 \dd x\,H^q(x,\xi,t) = A_{1,0}^q(t) \equiv F_{1}^{q}(t) \,, \label{eq:bacis_F1}\\
\int_{-1}^1 \dd x\,E^q(x,\xi,t) = B_{1,0}^q(t) \equiv F_{2}^{q}(t) \,. \label{eq:bacis_F2}
\end{align}

Nucleon tomography requires no knowledge of GPD at $\xi\neq0$, and for an unpolarized proton is given by
\begin{equation}
q(x,{\bf b_\perp}) =
\int \frac{\dd^2{\bf \Delta}_\perp}{(2\pi)^2}  
e^{-i{\bf b_\perp}\cdot {\bf \Delta}_\perp}
H_q(x,0,-{\bf \Delta}_\perp^2) \,.
\end{equation}
The impact parameter, ${\bf b_\perp}=(b_x, b_y)$, is defined in a coordinate system whose origin is set by the center of momentum of all proton constituents~\cite{Burkardt:2002hr}. In the limiting case where the entire proton's momentum is carried by a single parton, $x=1$, the origin of the coordinate system coincides with its position. Therefore, $q(x=1, {\bf b_\perp}) = \delta({\bf b_\perp})$, which requires a vanishing $t$-dependence of the GPD $H^q$ at $x=1$:
\begin{equation}
\lim_{x \to 1} \frac{\dd}{\dd t} H(x,0,t) = 0\,.
\label{eq:bacics_xEq1}
\end{equation}
Because of the coordinate system in which nucleon tomography is defined, one can easily find the average relative distance between the active parton and the spectator system~\cite{Diehl:2004cx}:
\begin{equation}
d_{q} = \frac{\sqrt{\langle b^2 \rangle_x^q}}{1-x} \,, 
\label{eq:bacics_spectators}
\end{equation}
where
\begin{equation}
\langle b^2 \rangle_x^q = \frac{\int \dd^2 {\bf b_\perp} {\bf b}_\perp^2 q(x,{\bf b_\perp})}{\int \dd^2 {\bf b_\perp}q(x,{\bf b_\perp})} \,.
\end{equation}
The requirement of keeping this distance finite at $x \to 1$ results in the additional constraint on the parameters of the GPD model used by us to describe the experimental and lattice data (see Eq.~\eqref{eq:H_profile}). If the proton is not polarized longitudinally, i.e. along the direction of its motion setting the $Z$-axis, but rather transversely, say, along the $X$-axis, the density of partons will be distorted and given by:
\begin{equation}
q_X(x,{\bf b_\perp}) = 
q(x,{\bf b_\perp})
-\frac{1}{2M} \frac{\partial}{\partial b_y}
e_q(x,{\bf b_\perp})\,,
\end{equation}
where
\begin{equation}
e_q(x,{\bf b_\perp})  =
\int \frac{\dd^2{\bf \Delta_\perp}}{(2\pi)^2}
e^{-i{\bf b_\perp}\cdot {\bf \Delta}_\perp}
E_q(x,0,-{\bf \Delta}_\perp^2) \,.
\end{equation}

Additional information about the proton is provided by fully considering the connection between GPDs and elements of the energy-momentum tensor. For brevity, we refrain from explaining this broad subject in detail and instead refer to one of the available reviews, such as~\cite{Burkert:2023wzr}. However, from the point of view of this work, it is important to recall the so-called Ji's sum rule:
\begin{equation}
2J^{q} = A_{2,0}(0) + B_{2,0}(0) = \int_{-1}^{1}\dd x\, x \Big(H^q(x, \xi, 0) + E^q(x, \xi, 0)\Big)\,,
\label{eq:basics_ji}
\end{equation}
where $A_{2,0}(0)$ and $B_{2,0}(0)$ are the generalized form factors, see Eq.~\eqref{eq:polynomiality}. This sum rule is particularly important, as it allows for evaluating the total angular momentum carried by specific partons.

%% file: sec_lattice.tex
\section{Lattice QCD input}
\label{sec:lattice}

In this section, we briefly summarize the extraction of lattice observables that are used in this work.
For a more detailed account, we refer to Refs.~\cite{Bhattacharya:2022aob,Bhattacharya:2024qpp}.

We calculate Euclidean matrix elements (MEs) of the following non-local vector operator:
\begin{align}
\label{eq:MEs}
F^\mu (z, P_f, P_i) = \langle N(P_f) | \bar\psi(z) \gamma^\mu W(0, z) \psi (0) | N (P_i) \rangle,\,\,\,
\end{align}
with $|N(P_i)\rangle$, $|N(P_f)\rangle$ denoting the nucleon's initial and final states with the corresponding four-momenta $P_i$ or $P_f$, respectively.
The quark and antiquark fields of the operator are spatially separated by a distance $z$ along the 3-direction and connected by a Wilson line $W(0,z)$.
For the construction of $H$ and $E$ pseudo-GPD MEs, we use Dirac indices $\mu=0,\,1,2\,$ in the operator insertion.
We also define the momentum transfer, $\Delta=P_f-P_i$, with $t=-\Delta^2$, and the so-called Ioffe time, $\nu=P_3z$.

The MEs of Eq.~(\ref{eq:MEs}) are calculated in an asymmetric frame of reference, in which the final state has a fixed momentum, $P_f=(P_f^0,0,0,P_f^3)$ and the momentum transfer is embodied by the initial state momentum, i.e.\ $P_i=(P_f^0-\Delta^0, -\Delta^1,-\Delta^2, P_f^3)$.
In this work, we only consider momentum transfer along the transverse directions, i.e.\ we consider only zero skewness ($\xi=0$).
Given such a setup, we proceed by extracting Lorentz-invariant amplitudes, $A_i$, and construct the $H$ and $E$ pseudo-GPDs in terms of these amplitudes, see Ref.~\cite{Bhattacharya:2022aob} for explicit expressions.
Out of the two pseudo-GPD definitions proposed in Ref.~\cite{Bhattacharya:2022aob}, we use the Lorentz-invariant (LI) one. It can be interpreted that different pseudo-GPD definitions are characterized by different contamination from power-suppressed higher-twist effects (HTEs).
In the present case of unpolarized GPDs, the LI variant turns out to have a more favorable interplay of HTEs, with the convergence improvement observed particularly for the $E$ case~\cite{Cichy:2023dgk}.

In our work, we use MEs computed in Refs.~\cite{Bhattacharya:2022aob,Bhattacharya:2024qpp}, and we refer to these works for a more comprehensive discussion of the methodology.
The lattice setup consists of $N_f=2+1+1$ twisted mass clover-improved quarks (a degenerate up and down quark doublet and a non-degenerate heavier doublet of strange and charm quarks) and Iwasaki-improved gluons~\cite{Alexandrou:2018egz}.
The lattice parameters are: $a\approx0.093$ fm (lattice spacing), $L^3\times T=32^3\times64$ (lattice size, amounting to $L\approx3$ fm in physical units) and a pion mass of approx.~260 MeV.
All MEs are calculated at a source-sink separation of 10 lattice spacings, leading to negligible excited-state contamination at our current level of precision.

The use of the asymmetric kinematic frame discussed above allows us to access several momentum transfer vectors and values of $-t$.
In this work, we use $\vec{\Delta}=(\Delta_1,\Delta_2,0)$ with $\Delta_{1/2}=\{1,2,3,4\}(2\pi/L)$ and the other transverse component $\Delta_{2/1}$ chosen such that $-t<2.3$ GeV$^2$. 
This leads to the following combinations for the transverse momentum transfer (in units of $(2\pi/L)$): $(1,0)$, $(2,0)$, $(3,0)$, $(4,0)$, $(1,1)$, $(2,1)$, $(3,1)$, $(2,2)$ with permutations exchanging the role of $\Delta_{1/2}$ and sign changes of both $\Delta_{1/2}$.
In physical units, these combinations give $-t=\{0.17,\,0.34,\,0.65,\,0.81,\,1.24,\,1.38,\,1.52,\,2.29\}$ GeV$^2$.
The MEs are obtained at several momentum boosts $P_3$, corresponding to $P_3=(2\pi/L)n$, with $n=\{0,\,1,\,2,\,3,\,4\}$, amounting to $P_3=\{0,\,0.42,\,0.83,\,1.25,\,1.67\}$ GeV in physical units.
For more details of this setup, we refer to Ref.~\cite{Bhattacharya:2024qpp}, in particular to Table I in this publication, that documents the numbers of used gauge field configurations, source positions and the total numbers of measurements.

In Refs.~\cite{Bhattacharya:2022aob,Bhattacharya:2024qpp}, only the flavor non-singlet combination $u-d$ was considered, which profits from the cancellation of quark-disconnected contributions in a setup of degenerate light quarks.
However, it was shown in Ref.~\cite{Alexandrou:2021oih} that the disconnected contributions are strongly suppressed in the unpolarized case and, in particular, much smaller than our current statistical precision.
Hence, we neglect them and consider flavor-separated GPDs of up and down quarks.

After the extraction of bare pseudo-GPD MEs, we renormalize them in a ratio scheme \cite{Orginos:2017kos}, canceling the divergences related to the endpoints and the Wilson line with zero-momentum unpolarized PDFs.
The ensuing objects, often referred to as pseudo-ITDs (Ioffe time distributions) are renormalization group invariant, but are defined at particular kinematic scales given by $1/z$ and are still Euclidean objects.
Thus, they need to be related to physical ITDs, the $H$ and $E$ GPDs in coordinate space.
This proceeds by applying perturbative evolution and matching kernels \cite{Radyushkin:2018cvn,Zhang:2018ggy,Izubuchi:2018srq,Radyushkin:2018nbf,Li:2020xml}.
The former evolve the ITDs from the scales $1/z$ to a common scale $\mu$, chosen to be 2 GeV, while the latter convert the Euclidean objects to their light-cone counterparts in the $\MSb$ scheme.
We restrict ourselves to one-loop evolution and matching, with two-loop effect found to be beyond our current precision \cite{Bhat:2022zrw}.
The explicit formulas are given in Ref.~\cite{Bhattacharya:2024qpp}.
These formulas should be applied at short distances to avoid large contributions from $\mathcal{O}(z^2\LambdaQCD^2)$ HTEs.
We adopt the pragmatic criterion for the maximum value of $z$, $\zmax$, discussed in Ref.~\cite{Bhattacharya:2024qpp}.
According to it, $\zmax$ is chosen such that matched ITDs obtained from different combinations of $(P_3,z)$ but at a common Ioffe time are compatible among each other.
Since our combined analysis with experimental data uses only the real part of lattice-extracted ITDs, we choose $\zmax=9a\approx0.84$ fm~\cite{Bhattacharya:2024qpp}.
While this value seems rather large from the perturbative point of view, the practical size of HTEs is demonstrably suppressed for $z\leq\zmax$ at the current precision level, likely profiting from cancellations of HTEs in the ratio scheme and the convergence properties of the LI variant of pseudo-GPD definitions.

The final evolved and matched ITDs are input to our analysis presented in the remainder of this paper.
They are physical objects, directly related to the internal structure of the nucleon and are given for $\mu=2~\mathrm{GeV}$.
Nevertheless, one needs to keep in mind that at this stage, they are still contaminated by unquantified systematic effects, related to lattice-specific features (e.g.\ a single lattice spacing and a single lattice volume), the non-physical pion mass and other aspects (e.g.\ a limited nucleon boost).

%% file: sec_comparison.tex
\section{Comparison between lattice-QCD results and existing phenomenological results}
\label{sec:comparison}

In this section, we compare lattice-QCD results used in this work with models of GPDs and parameterizations of elastic FFs and unpolarized PDFs. The comparison is done in the space of Ioffe time, $\nu$, that can be related to the usual momentum space by the Fourier transform,
\begin{equation}
    \hat H^q(\nu,\xi,t) = \int_{-1}^{1}\mathrm{d}x\,e^{i x \nu}H^q(x,\xi,t)\,,
    \label{eq:hatH}
\end{equation}
which is shown here for a quark flavor $q$. This relation also holds for the GPD $E^q$, in which we are also interested in this study.

It is instructive to rewrite Eq.~\eqref{eq:hatH} with the help of charge-even, $H^{q(+)}$, and charge-odd, $H^{q(-)}$, combinations of GPDs, sometimes called ``singlet'' and ``non-singlet'' (or ``valence'') combinations, which are defined as follows:
\begin{align}
H^{q(+)}(x,\xi,t) & = H^{q}(x,\xi,t) - H^{q}(-x,\xi,t)\,, \\
H^{q(-)}(x,\xi,t) & = H^{q}(x,\xi,t) +
H^{q}(-x,\xi,t)\,.
\end{align}
Because of the obvious symmetry properties, this gives us:
\begin{align}
\re\hat H^{q}(\nu,\xi,t) & = \int_{0}^{1}\mathrm{d}x\,\cos(x\nu)H^{q(-)}(x,\xi,t)\,, \\
\im\hat H^{q}(\nu,\xi,t) & = \int_{0}^{1}\mathrm{d}x\,\sin(x\nu)H^{q(+)}(x,\xi,t)\,.
\end{align}
Since in this analysis we are only interested in the case $\xi=0$, which is relevant for nucleon tomography, in the following, we will omit the $\xi$-dependence in the formulae for brevity. At $\xi=0$,  further insight is provided by the following decomposition into valence and sea contributions: 
\begin{align}
H^{q(-)}(x,t) & = H^{q_\mathrm{val}}(|x|, t)\,, \\
H^{q(+)}(x,t) & = \left( H^{q_\mathrm{val}}(|x|, t) + 2H^{q_\mathrm{sea}}(|x|, t) \right) \mathrm{sgn}(x)\,, 
\end{align}
where we used the following:
\begin{align}
H^{q_\mathrm{val}}(x, t) & = 0\quad\mathrm{for}\quad x<0\,, \\
H^{q_\mathrm{sea}}(x, t) & = -H^{q_\mathrm{sea}}(-x, t)\,. 
\end{align}
We, therefore, see that the real part of $\hat H^q(\nu, t)$ is sensitive only to valence quarks, while the imaginary part of this quantity is also sensitive to the sea contribution. It is also important to note that knowing $H^{q(-)}(x,t)$ is sufficient to determine $H^{q_{\mathrm{val}}}(x,t)$. The latter can be used to evaluate any Mellin moment of this quantity.

The quantity $\hat H(\nu, t)$ must also fulfill the constraint \eqref{eq:bacics_xEq1}, which in Ioffe time space can be expressed with the help of the inverse Fourier transform as follows:
\begin{align}
\lim_{x \to 1}\frac{\dd}{\dd t}\int_{0}^{\infty} \dd \nu e^{-ix\nu} \hat H^{q}(\nu, t) = 0 \,.
\end{align}

\begin{figure}[tp!]
  \centering
  \caption{Comparison of lattice-QCD results (markers with uncertainty bars) with GK~\cite{Goloskokov:2006hr,Goloskokov:2008ib} (solid curve) and VGG~\cite{Vanderhaeghen:1999xj} (dashed curve) GPD models, and, only for (a) and (b), MSW analysis~\cite{Moutarde:2018kwr} (solid band). Left (right) plots are for real (imaginary) parts of $H^{q}(\nu, t)$, while top (bottom) rows are for up (down) quarks. Each subfigure is made for different kinematics specified in the subcaption.} 
  \label{fig:comp}
  \subfloat[As a function of $\nu$ for $t = 0$.]{\label{fig:compHnu0}\includegraphics[width=\textwidth]{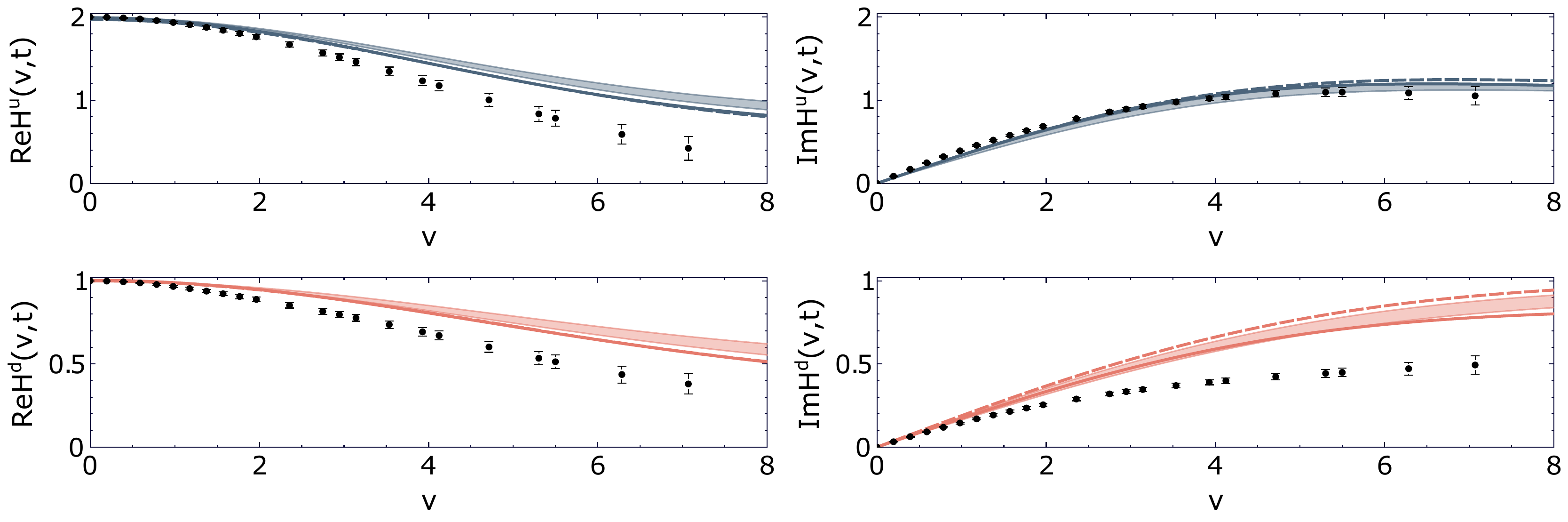}}\\
  \subfloat[As a function of $|t|$ for $\nu = 0$.]{\label{fig:compHt0}\includegraphics[width=\textwidth]{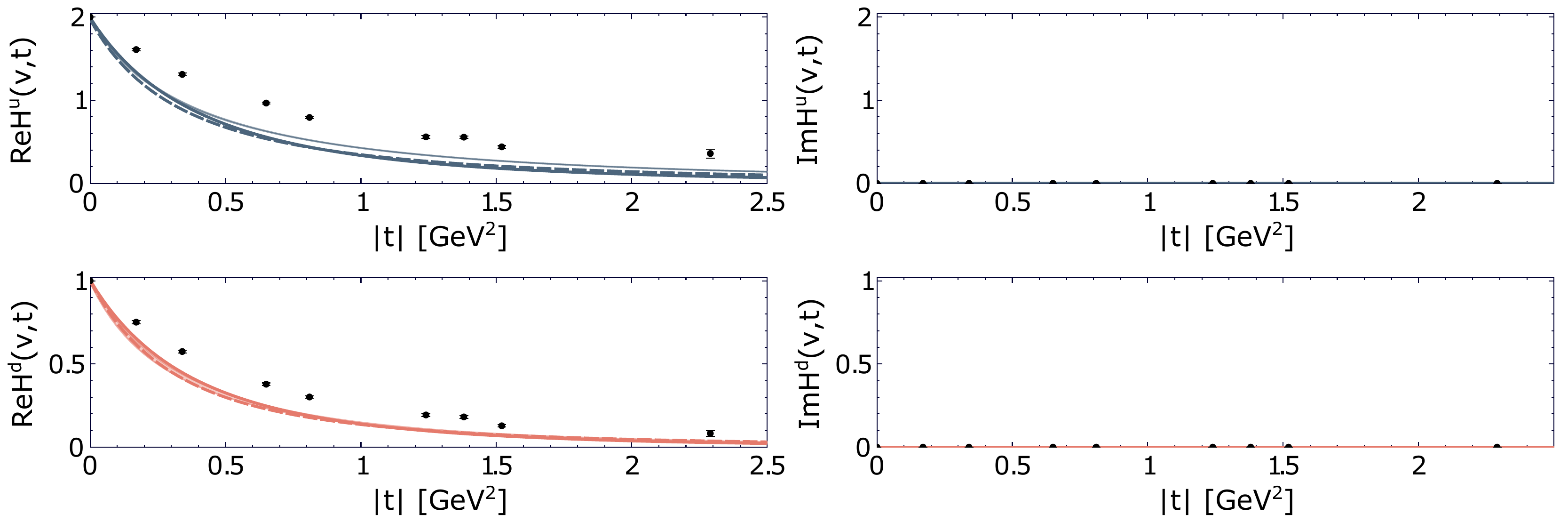}}\\
  \subfloat[As a function of $\nu$ for $|t| = 0.65~\mathrm{GeV}^2$.]{\label{fig:compHnu065}\includegraphics[width=\textwidth]{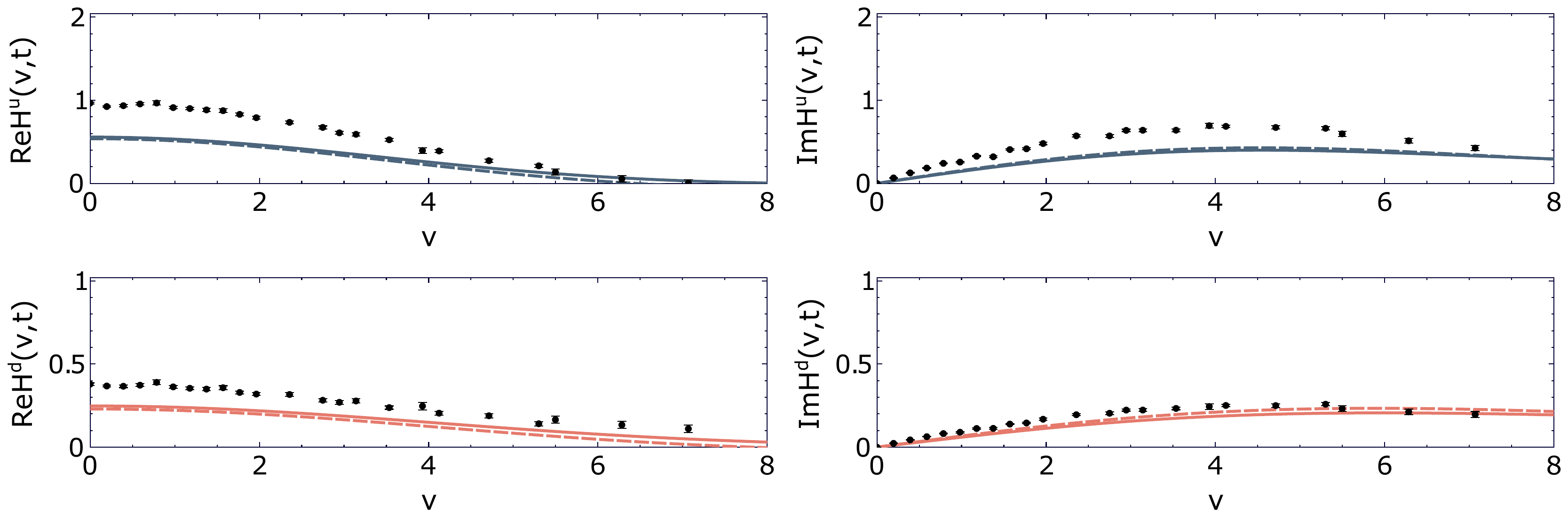}}
\end{figure}

The comparison between lattice-QCD results and several parameterizations of the up quark and down quarks of the unpolarized PDF ($t=0$ case) and elastic FFs ($\nu=0$ case) is shown in Figs.~\ref{fig:compHnu0} and~\ref{fig:compHt0}, respectively. The parametrizations utilized in this comparison come from the GK~\cite{Goloskokov:2006hr,Goloskokov:2008ib} and VGG~\cite{Vanderhaeghen:1999xj} GPD models (both implemented in the PARTONS framework~\cite{Berthou:2015oaw}), and the study presented in Ref.~\cite{Moutarde:2018kwr}, which we will refer to as MSW. That is, for PDFs, we show the comparison for a custom fit to the CTEQ6m set~\cite{Pumplin:2002vw} (GK), the original MSTW08 set~\cite{Martin:2009iq} available via the LHAPDF library~\cite{Buckley:2014ana} (VGG), and a custom fit to the NNPDF3.0 set~\cite{NNPDF:2014otw} (MSW). For elastic form factors, we have two simple parametrizations roughly reproducing the main features of the data (GK and VGG) and the result of the elaborate global analysis (MSW). The figures demonstrate a moderate agreement between lattice-QCD and parametrizations of PDFs and elastic FFs, which does not come as a surprise, taking into account that the current state of lattice-QCD computations is still very much exploratory. The discrepancy is also visible in $\nu, t \neq 0$ cases, as demonstrated in Fig.~\ref{fig:compHnu065}. In this case, however, one should keep in mind that the uncertainties of the two presented GPD models are unknown. In particular, these models are based on Radyushkin's double distribution Ansatz, but other modeling strategies, such as those based on the Mellin-Barnes framework~\cite{Kumericki:2015lhb}, dual parameterization~\cite{Polyakov:2002wz, Muller:2014wxa}, or artificial neural networks~\cite{Dutrieux:2021wll}, could yield different estimates.

\begin{figure}[!tp]
  \centering
  \caption{Comparison of lattice-QCD results (markers with uncertainty bars) with GK (solid curve) and VGG (dashed curve) GPD models for the single ratios $\im\hat H^u(\nu, t)/\im\hat H^u(\nu, 0)$.} 
  \label{fig:lim}
  \subfloat[$|t| = 0.34~\mathrm{GeV}^2$]{\label{fig:lim034}\includegraphics[width=0.33\textwidth]{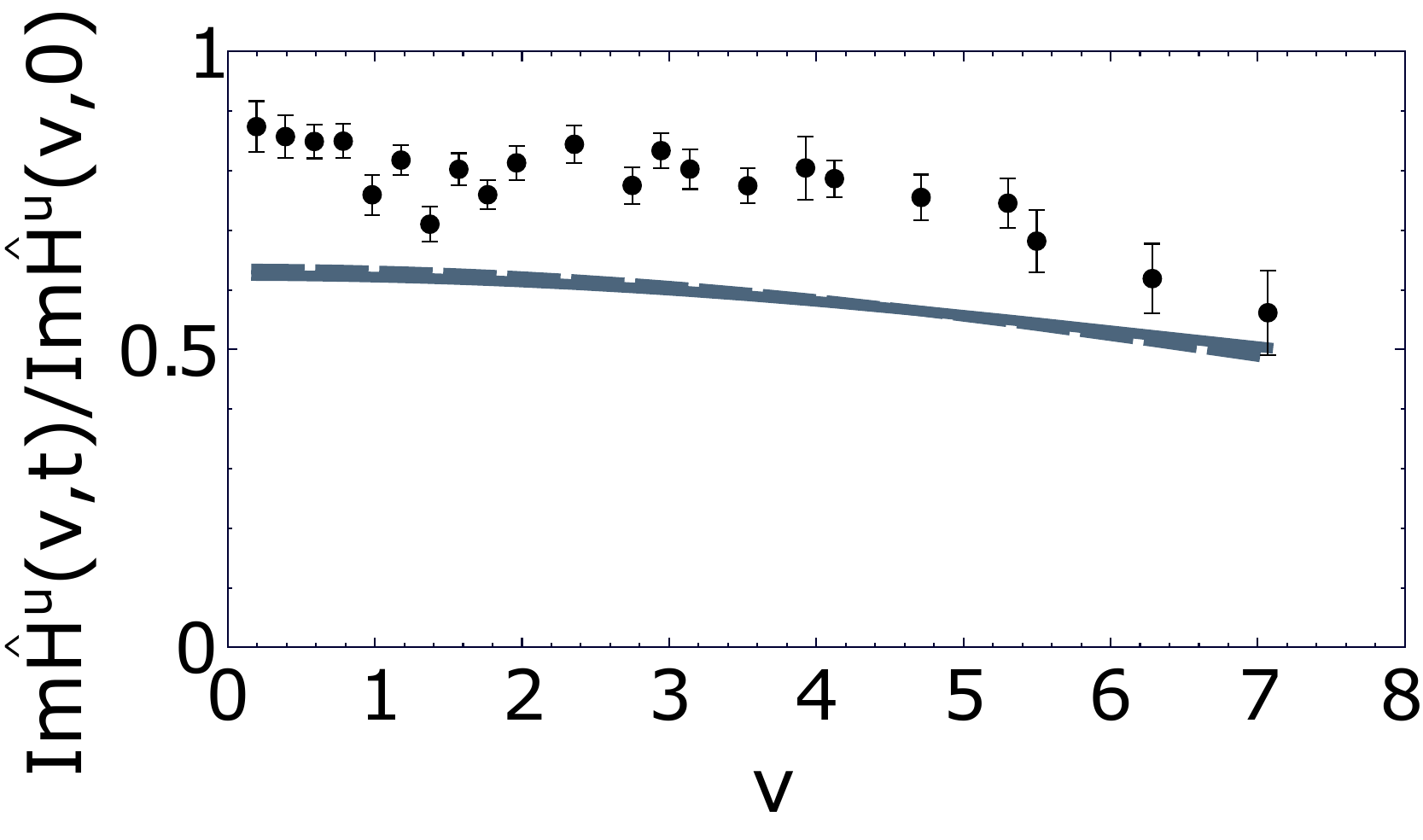}}
  \subfloat[$|t| = 0.81~\mathrm{GeV}^2$]{\label{fig:lim081}\includegraphics[width=0.33\textwidth]{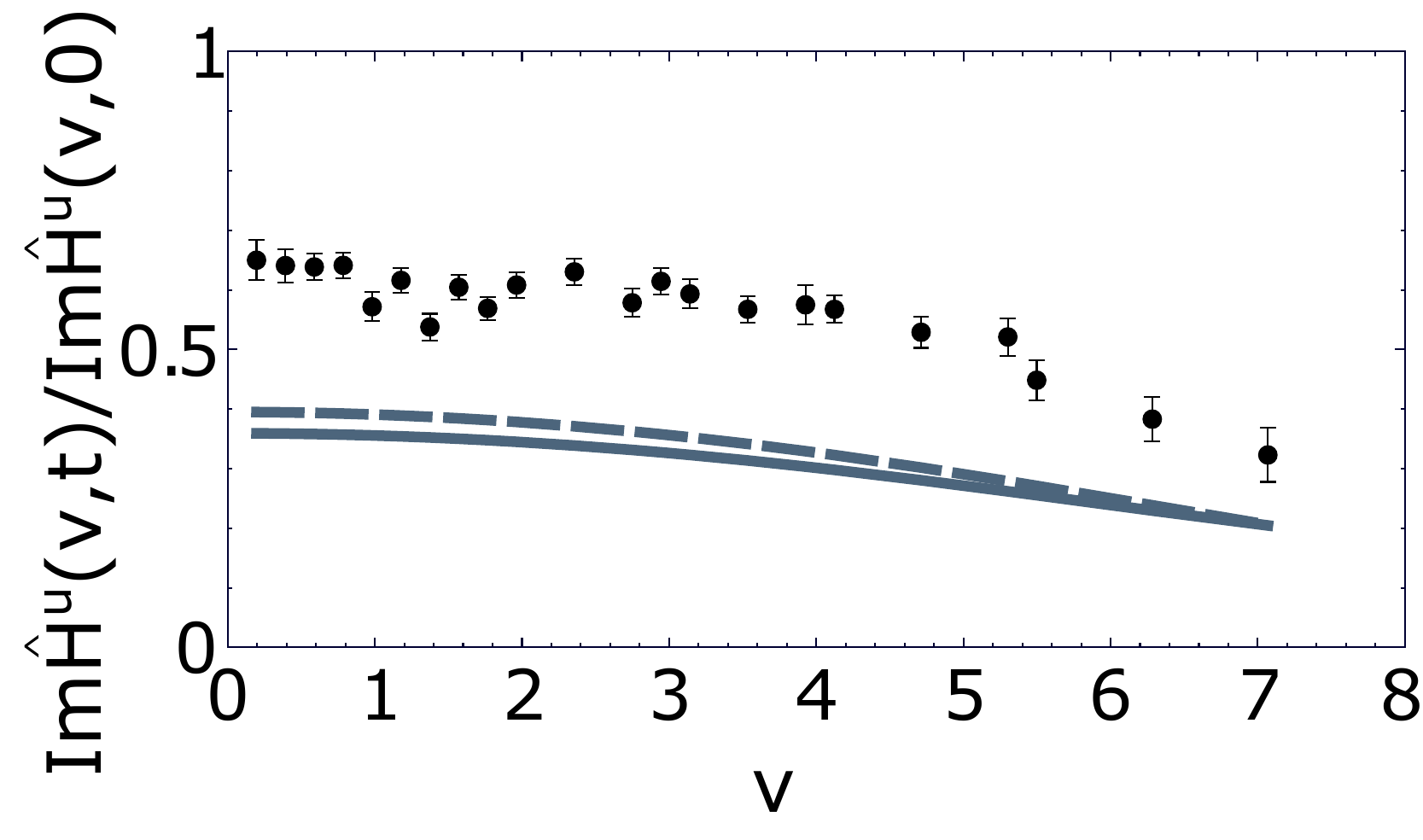}}
   \subfloat[$|t| = 1.52~\mathrm{GeV}^2$]{\label{fig:lim152}\includegraphics[width=0.33\textwidth]{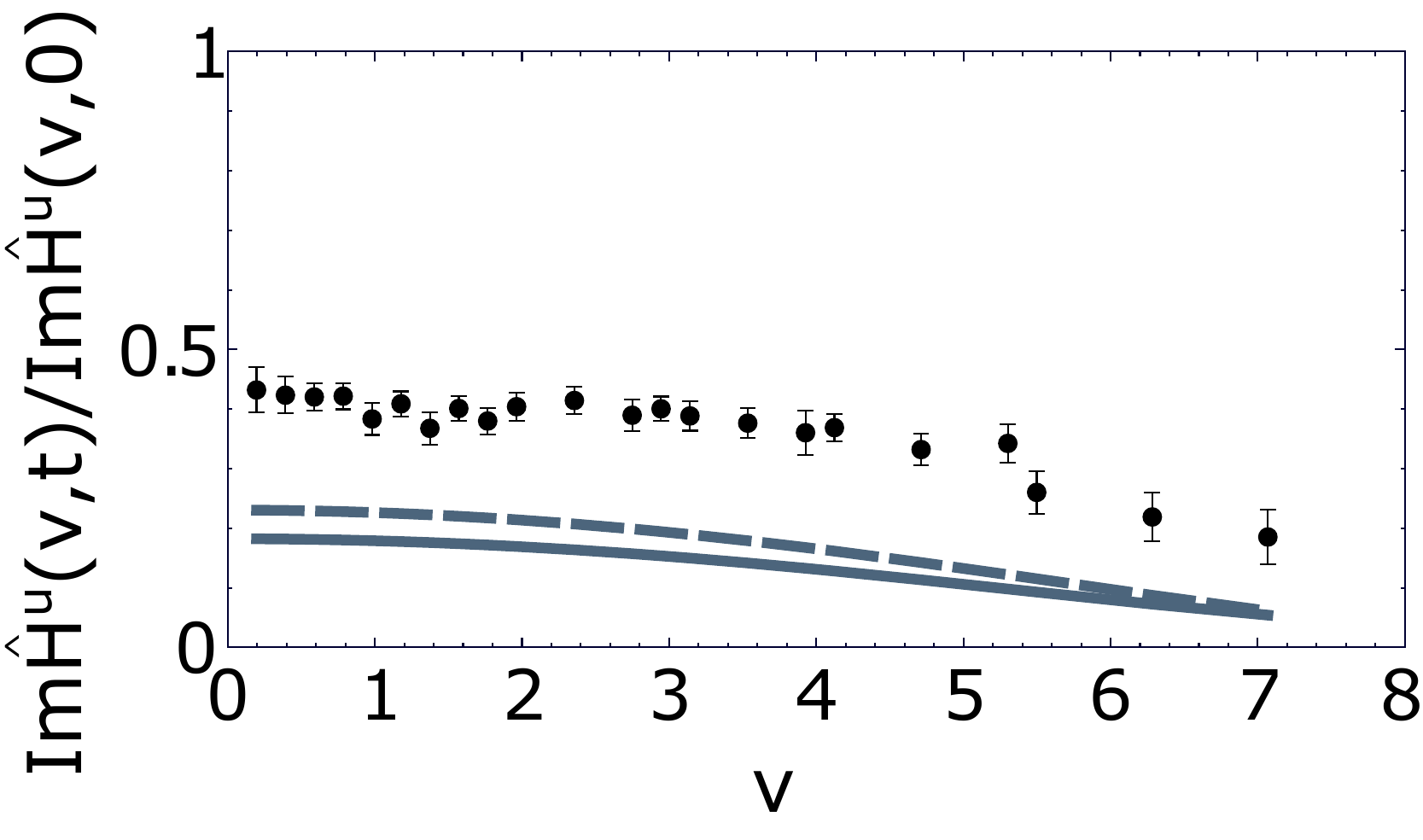}}
\end{figure}

The lack of agreement between lattice QCD results and parameterizations of PDFs and elastic FFs can be attributed to a variety of effects, in particular, to a nonphysical pion mass (for some discussion see, for instance,  Ref.~\cite{Alexandrou:2018pbm}). These effects will be corrected over time and, in fact, the comparison shown in Fig.~\ref{fig:comp} should be considered encouraging, given the current state-of-the-art in lattice QCD computations. The fundamental question is whether the presented lattice QCD results can be used \emph{now} in the phenomenological studies of GPDs without introducing significant bias and tension with better-known quantities. To make this possible, we introduce a new quantity called the double ratio. It is separately defined for the real and imaginary parts of~$\hat H^{q}(\nu,t)$:
\begin{align}
\dr_{\re}^{\hat H^q}(\nu, t) & = 
\frac{\re\hat H^q(\nu, t)}{\re\hat H^q(\nu, 0)}
\frac{\re\hat H^q(0, 0)}{\re\hat H^q(0, t)}
\,, \\
\dr_{\im}^{\hat H^q}(\nu, t) & = \lim_{\nu' \to 0}
\frac{\im\hat H^q(\nu, t)}{\im\hat H^q(\nu, 0)}
\frac{\im\hat H^q(\nu', 0)}{\im\hat H^q(\nu', t)}
\,.
\label{eq:dr}
\end{align}
The definitions of $\dr_{\re}^{\hat H^q}(\nu, t)$ and $\dr_{\im}^{\hat H^q}(\nu, t)$ must be different, because both $\im\hat H^q(0, 0)$ and $\im\hat H^q(0, t)$ vanish. The robustness of the $\dr_{\im}^{\hat H^q}(\nu, t)$ definition is demonstrated in Fig.~\ref{fig:lim}, which shows the single ratio $\im\hat H^q(\nu, t)/\im\hat H^q(\nu, 0)$ as a function of $\nu$ for a few values of $t$. In this figure, one can observe a plateau near $\nu=0$, indicating that $\lim_{\nu \to 0}(\im\hat H^q(\nu, t)/\im\hat H^q(\nu, t))$ can be reliably estimated by examining $\im\hat H^q(\nu, t)/\im\hat H^q(\nu, t)$ for small $\nu$. The plateau near $\nu=0$, also observed in Figs.~\ref{fig:compHnu0} and~\ref{fig:compHnu065} for the real parts, is not accidental. It is a consequence of the limited domain of integration in Eq.~\eqref{eq:hatH} compared to the usual definition of the Fourier transform. It appears whenever $x\nu$ frequencies are much smaller than the limiting $x=1$. In addition to the discussion presented in this paragraph, we note a poor agreement between lattice QCD results and GPD models for single ratios, as also shown in~Fig.~\ref{fig:lim}. 

For the real part, the double ratio $\dr_{\re}^{\hat H^q}(\nu, t)$ can be interpreted as a function describing the deviation of $\re\hat H^q(\nu, t)$ from the factorised Ansatz, being a product of the corresponding PDF and the elastic FF. Explicitly:
\begin{equation}
\re\hat H^{q}(\nu, t) = N_{H}^{q} \times \frac{q(\nu)}{N_{H}^{q}} \times \frac{F_{1}^{q}(t)}{N_{H}^{q}} \times \dr_{\re}^{\hat H^q}(\nu, t)\,,
\end{equation}
where, $N_{H}^{q} \equiv \re\hat H^{q}(0, 0)$ is $2$ for up quarks and $1$ for down quarks, $q(\nu) \equiv \re\hat H^{q}(\nu, 0)$ and $F_{1}^{q}(t) \equiv \re\hat H^{q}(0, t)$. A similar interpretation holds for the GPD $E^q$. However, its forward limit is not probed by (semi-)inclusive scattering and, therefore, is mostly unknown and a subject of modeling. The interpretation of $\dr_{\im}^{\hat H^q}(\nu, t)$ and $\dr_{\im}^{\hat E^q}(\nu, t)$ is, on the other hand, spoiled by the vanishing $\im\hat H^q(0, t) = \im\hat E^q(0, t) = 0$, that is, undefined elastic FFs.

\begin{figure}[!t]
  \centering
  \caption{Comparison of lattice-QCD results (markers with uncertainty bars) with GK (solid curve) and VGG (dashed curve) GPD models for the double ratios defined in Eq.~\eqref{eq:dr}. } 
  \label{fig:compDR}
  \subfloat[As a function of $\nu$ for $|t| = 0.65~\mathrm{GeV}^2$.]{\label{fig:compDRHnu065}\includegraphics[width=\textwidth]{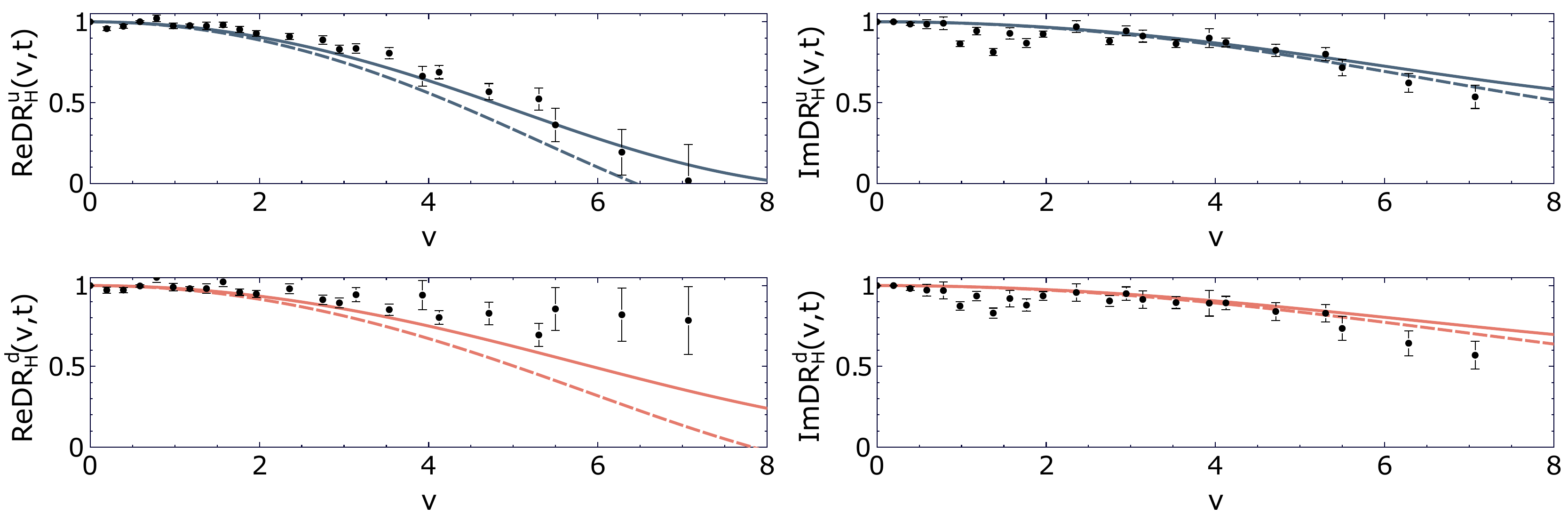}}\\
  \subfloat[As a function of $|t|$ for $\nu = 3.14$.]{\label{fig:compDRHt314}\includegraphics[width=\textwidth]{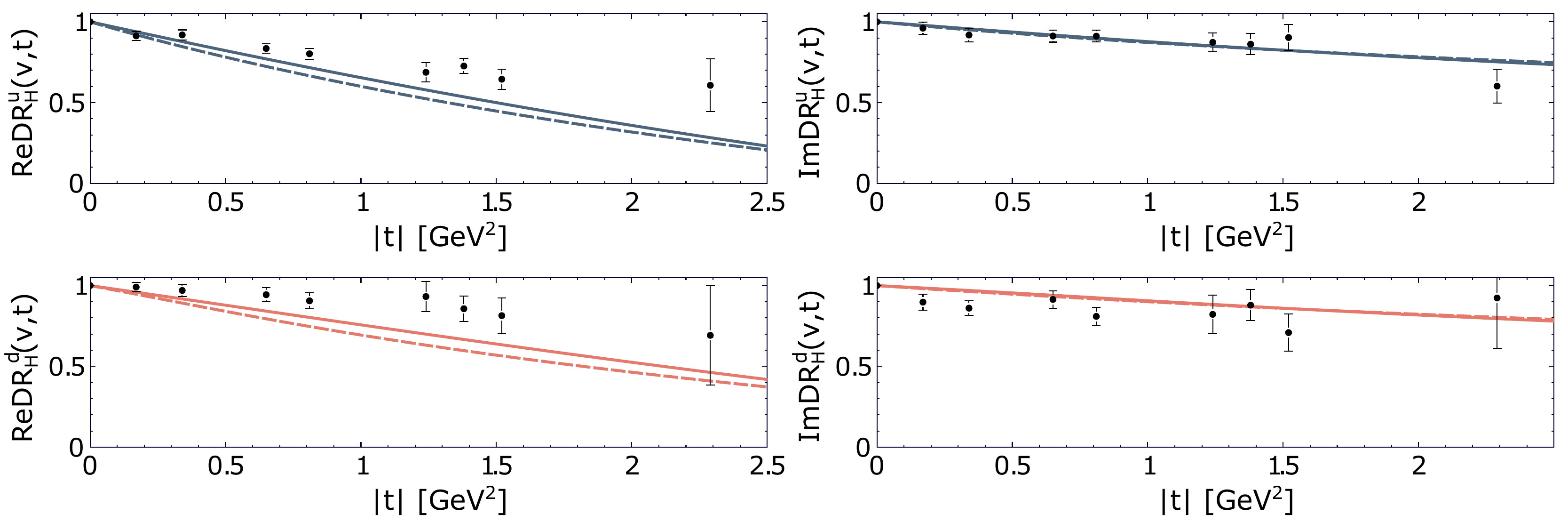}}
\end{figure}

The comparison between lattice-QCD data and GPD models for the double ratios is shown in Figs.~\ref{fig:compDRHnu065} and \ref{fig:compDRHt314}, as a function of $\nu$ and $t$, respectively. The comparison done for PDFs and elastic FFs is trivial, as, by definition, $\dr(\nu, 0) = \dr(0, t) = 1$ for both the real and imaginary parts and for all GPD types. Based on the presented figures, we find that the lattice-QCD results are now in much better agreement with the models. They either agree with them, or the discrepancy is within the spread between the models. It therefore seems that by removing explicit information on PDFs and elastic FFs, we have reduced many sources of systematic errors. This statement should be scrutinized with future lattice-QCD calculations, for instance, by studying the stability of double ratios with a more accurate reproduction of PDFs and elastic FFs. 

%% file: sec_extraction.tex
\section{Extraction of tomography information}
\label{sec:extraction}

In this section, we describe the extraction of tomography information from lattice-QCD results and data on elastic scattering, assuming a specific parameterization of PDFs. For the $H$ GPD, our Ansatz consists of two parts:
\begin{equation}
H^{q}(x,t) = H_C^{q}(x,t) + H_S^{q}(x,t)\,,
\label{eq:fitHAnsatz}
\end{equation}
where $H_C^{q}(x,t)$ is a ``classic'' term contributing to the description of both lattice-QCD results and data for elastic scattering, while $H_S^{q}(x,t)$ is a ``shadow'' term only sensitive to the double ratios.

For $H^{q}_C(x,t)$, we take the Ansatz previously used in Ref.~\cite{Moutarde:2018kwr}:
\begin{equation}
H^{q}_C(x,t) = q(x)\exp(f_{H}^{q}(x)t)\,,
\end{equation}
where the profile function containing free parameters $p_{H,i}^{q}$ is:
\begin{equation}
f_{H}^{q}(x) = p_{H,0}^{q} \log(1/x) + p_{H,1}^{q} (1-x)^2 - p_{H,0}^{q} (1-x)x\,.
\label{eq:H_profile}
\end{equation}
The opposite coefficients multiplying $\log(1/x)$ and $(1-x)x$ terms allow us to keep the distance between the active quark and the spectator system finite, see Eq.~\eqref{eq:bacics_spectators}. 

The shadow term must vanish both at $t=0$ (no contribution to PDFs) and upon integration over $x$ (no contribution to elastic FFs),
\begin{equation}
\int_{0}^{1}\dd x H^{q}_S(x,t) = 0\,.
\label{eq:shadow_x_integral}
\end{equation}
In addition, its contribution to $H^{q}(x, t)$ in \eqref{eq:fitHAnsatz} must not violate the positivity constraint and the requirement expressed by Eq.~\eqref{eq:bacics_xEq1}. Since the shadow term is not determined by first principles, it is subject to modeling. As an example, we propose the following Ansatz, fulfilling all the aforementioned requirements:
\begin{align}
H^{q}_S(x,t) = p_{H,2}^{q} \times
\Big((1 - x)^{b_{H}^{q}} - A(t) (1 - x)^{(b_{H}^{q} + 1)}\Big) \times
\Big(
\exp(p_{H,3}^{q} (1 - x) t) - 
\exp(p_{H,4}^{q} (1 - x) t)
\Big)\,.
\label{eq:shadow}
\end{align}
The function $A(t)$ is found by requiring \eqref{eq:shadow_x_integral} and is given by
\begin{align}
A(t) & = \nonumber \\
& -\Big(
    (b+1) p_{3} p_{4} t \big(-p_{3} (-p_{3} t)^b \Gamma
   (b+1,-p_{4} t)+p_{4} (-p_{4} t)^b \Gamma (b+1,-p_{3} t)+p_{3}
   \Gamma (b+1) (-p_{3} t)^b \nonumber \\
   & -p_{4} \Gamma (b+1) (-p_{4} t)^b\big)
   \Big)/
   \Big(
   b^2
   p_{3}^2 \Gamma (b+1) (-p_{3} t)^b-b^2 p_{4}^2 \Gamma (b+1) (-p_{4}
   t)^b+p_{3}^2 p_{4}^2 t^2 e^{p_{3} t} (-p_{3} t)^b (-p_{4}
   t)^b \nonumber \\
   & -p_{3}^2 p_{4}^2 t^2 e^{p_{4} t} (-p_{3} t)^b (-p_{4} t)^b+b
   p_{3}^2 p_{4} t (-p_{3} t)^b \Gamma (b+1,-p_{4} t)+p_{3}^2
   p_{4} t (-p_{3} t)^b \Gamma (b+1,-p_{4} t) \nonumber \\
   & -b p_{3}^2 (-p_{3} t)^b
   \Gamma (b+2,-p_{4} t)-p_{3}^2 (-p_{3} t)^b \Gamma (b+2,-p_{4}
   t)-p_{3}^2 p_{4} t (-p_{3} t)^b \Gamma (b+2,-p_{4} t)+2 b p_{3}^2
   \Gamma (b+1) (-p_{3} t)^b \nonumber \\
   & +p_{3}^2 \Gamma (b+1) (-p_{3} t)^b-b p_{3}
   p_{4}^2 t (-p_{4} t)^b \Gamma (b+1,-p_{3} t)-p_{3} p_{4}^2 t
   (-p_{4} t)^b \Gamma (b+1,-p_{3} t)+b p_{4}^2 (-p_{4} t)^b \Gamma
   (b+2,-p_{3} t) \nonumber \\
   & +p_{4}^2 (-p_{4} t)^b \Gamma (b+2,-p_{3} t)+p_{3}
   p_{4}^2 t (-p_{4} t)^b \Gamma (b+2,-p_{3} t)-2 b p_{4}^2 \Gamma (b+1)
   (-p_{4} t)^b-p_{4}^2 \Gamma (b+1) (-p_{4} t)^b
   \Big)\,,
   \label{eq:shadow_A}
\end{align}
where $p_{i} \equiv p_{H,i}^{q}$, $b \equiv b_{H}^{q}$, $\Gamma(a,z) = \int_z^{\infty}\dd t\, t^{a-1}e^{-t}$ is the incomplete gamma function, and $\Gamma(a) = \Gamma(a,0)$. The evaluation of $H^q_S$ is typically unstable for $|t| \lesssim 0.1~\mathrm{GeV}^2$, as both the numerator and denominator of Eq.~\eqref{eq:shadow_A} become very small. This issue does not pose a threat to our fit, as the lattice-QCD data used do not cover this region. However, it may be relevant for the later extraction of tomography information. Since $A(t)$ is flat in the problematic region, it is safe to approximate it there with the help of the Taylor expansion done at, say, $|t_0| = 0.2~\mathrm{GeV}^2$. The coefficient $b_{H}^{q}$ controls the power behavior of the shadow term at $x \to 1$ and is set to match the corresponding coefficient used in the parameterization of PDFs:
\begin{align}
q(x) = N_{q}^{H} q_0(x) / \left( \int_0^1 \dd x\, q_0(x) \right) \,, \quad
q_0(x) = x^{-\delta_{H}^{q}}(1-x)^{b_{H}^{q}}\sum_{i=0}^4 c_{H,i}^{q} x^i \,,
\end{align}
where the coefficients $\delta_{H}^{q}$, $b_{H}^{q}$ and $c_{H,i}^{q}$ have been fixed in a fit to the NNPDF3.0 set in Ref.~\cite{Moutarde:2018kwr}, and where $N_{u}^{H}=2$ for up quarks and $N_{d}^{H}=1$ for down quarks. For brevity, we omit the expression for the normalization integral, as it can be easily computed analytically. The matching of $b_{H}^{q}$ coefficients helps keep the parton densities positive in impact parameter space. In summary, in addition to the feature described in the last sentence, the shadow term will not contribute to PDFs, as $H_S^q(x,0)=0$, nor to elastic FFs, because of Eq.~\eqref{eq:shadow_x_integral}, and it can only be constrained by lattice-QCD results. Moreover, because of the use of $(1-x)$ terms in the profile function, see Eq.~\eqref{eq:shadow}, it fulfills the requirement given by~Eq.~\eqref{eq:bacics_xEq1}.

For the $E$ GPD, we have:
\begin{equation}
E^{q}(x,t) = e_{q}(x)\exp(f_{E}^{q}(x)t)\,.
\end{equation}
The profile function is:
\begin{equation}
f_{E}^{q}(x) = p_{H,0}^{q} \log(1 / x) + p_{E,0}^{q} (1 - x)^2 + p_{E,1}^{q}  x (1 - x)\,,
\end{equation}
where the coefficient $p_{H,0}^{q}$ is the same as in Eq.~\eqref{eq:H_profile}, helping to keep the parton densities positive in the impact parameter space. For the $E$ GPD in the current analysis, we refrain from using the shadow term, as the number of free parameters is already inflated by the need to constrain the forward limit in the fit:
\begin{equation}
e_q(x) = N_{q}^{E} e_{q0}(x) / \left( \int_0^1 \dd x\, e_{q0}(x) \right) \,, \quad
e_{q0}(x) =  x^{-\delta_{q}^{E}} (1 - x)^{b_{q}^{E}} (1 + \gamma_{q}^{E}\sqrt{x})\,,
\end{equation}
where $N_{q}^{E} = \kappa_q$ are the ``partonic'' anomalous magnetic moments, $b_{u}^{E}=4.65$, $b_{d}^{E}=5.25$, $\gamma_{u}^{E} = 4$, $\gamma_{d}^{E} = 0$ and $\delta_{q}^{E}$ is a free parameter. This Ansatz for $e_q(x)$ has been originally proposed in Ref.~\cite{Diehl:2013xca}. Releasing some of the fixed $e_q(x)$ coefficients in the fit is possible, however, this is left for future, more elaborate analyses.

\begin{table}[!t]
\centering
\caption{Elastic form factor and lattice-QCD data used in this analysis. Data for $R^{n}$ are taken from Ref.~\cite{Diehl:2013xca}, evaluated from the original material specified in this table. The two last columns indicate the goodness of the fit for the set of central replicas.}
\label{tab:EFFdata}
\begin{tabular}{C{0.19\linewidth}C{0.19\linewidth}C{0.19\linewidth}C{0.19\linewidth}C{0.19\linewidth}}
\toprule
Observable			& Reference					& Number of points ($\#$)& $\chi^2$ & $\chi^2 /\#$\\ \midrule
$G_{M, N}^{p}$ 	& \cite{Arrington:2007ux}	& $54$ & $46$ & $0.86$\\    
$R^{p}$           & \cite{Arrington:2007ux, Milbrath:1997de, Pospischil:2001pp, Gayou:2001qt, Gayou:2001qd, Punjabi:2005wq, MacLachlan:2006vw, Puckett:2010ac, Paolone:2010qc, Ron:2011rd, Zhan:2011ji} & $54$ & $88$ & $1.63$\\
$G_{M, N}^{n}$    & \cite{Anklin:1998ae, Kubon:2001rj, Anklin:1994ae, Lachniet:2008qf, Anderson:2006jp} & $36$ & $22$ & $0.63$\\   
$R^{n}$           & \cite{Herberg:1999ud, Glazier:2004ny, Plaster:2005cx, Passchier:1999cj, Zhu:2001md, Warren:2003ma, Geis:2008aa, Bermuth:2003qh, rohe_pc, Riordan:2010id} & $21$ & $26$ & $1.23$\\    
$G_{E}^{n}$       & \cite{Schiavilla:2001qe} & $12$ & $2.5$ & $0.21$\\   
$r_{nE}^{2}$      & \cite{Beringer:1900zz} & $1$ & $7.3$ & $7.3$\\
$\re\dr_{H}^{u}$      & \cite{Bhattacharya:2024qpp} & $176$ & $245$ & $1.39$\\
$\re\dr_{H}^{d}$      & \cite{Bhattacharya:2024qpp} & $176$ & $253$ & $1.44$\\
$\re\dr_{E}^{u}$      & \cite{Bhattacharya:2024qpp} & $176$ & $180$ & $1.02$\\
$\re\dr_{E}^{d}$      & \cite{Bhattacharya:2024qpp} & $176$ & $324$ & $1.84$\\ \midrule
TOTAL & & $882$ & $1185$ & $1.34$\\ \bottomrule
\end{tabular}
\end{table}
\begin{table}[!t]
\centering
\caption{Mean values of fitted parameters with uncertainties. The values presented in this table are for tentative orientation only. Directly using them may lead to incorrect conclusions, as it would neglect the correlations.}
\label{tab:fitResult}
\begin{tabular}{C{0.11\linewidth}C{0.11\linewidth}C{0.11\linewidth}C{0.11\linewidth}C{0.11\linewidth}C{0.11\linewidth}C{0.11\linewidth}C{0.11\linewidth}}
\toprule
$p_{H,0}^u$ & $p_{H,1}^u$ & $p_{H,2}^u$ & $p_{H,3}^u$ & $p_{H,4}^u$ & $p_{H,0}^d$ & $p_{H,1}^d$ & $p_{H,2}^d$\\ 
\midrule
$0.876 \pm 0.076$ & 
$-0.32 \pm 0.12$ & 
$0.7 \pm 4.1$ & 
$1.07 \pm 0.85$ & 
$0.82 \pm 0.80$ & 
$0.500 \pm 0.068$ & 
$0.72 \pm 0.18$ & 
$0.0 \pm 2.9$\\
\midrule
\\
\midrule
$p_{H,3}^d$ & $p_{H,4}^d$ & $p_{E,0}^u$ & $p_{E,1}^u$ & $\delta_u^E$ & $p_{E,0}^d$ & $p_{E,1}^d$ & $\delta_d^E$\\ 
\midrule
$0.95 \pm 0.76$ & 
$0.80 \pm 0.74$ & 
$-0.44 \pm 0.25$ & 
$-0.93 \pm 0.17$ & 
$0.710 \pm 0.039$ & 
$-0.36 \pm 0.22$ & 
$-0.37 \pm 0.31$ & 
$0.806 \pm 0.034$\\ 
\bottomrule
\end{tabular}
\end{table}
\begin{figure}[t!]
    \centering
    \caption{Lattice-QCD and elastic FF data (markers with uncertainty bars) compared with the fit results (dashed bands). The data coming from lattice-QCD computations (double ratios) are shown in the first two rows for only $|t|=0.65~\mathrm{GeV}^2$.}
    \label{fig:fit}\includegraphics[width=\linewidth]{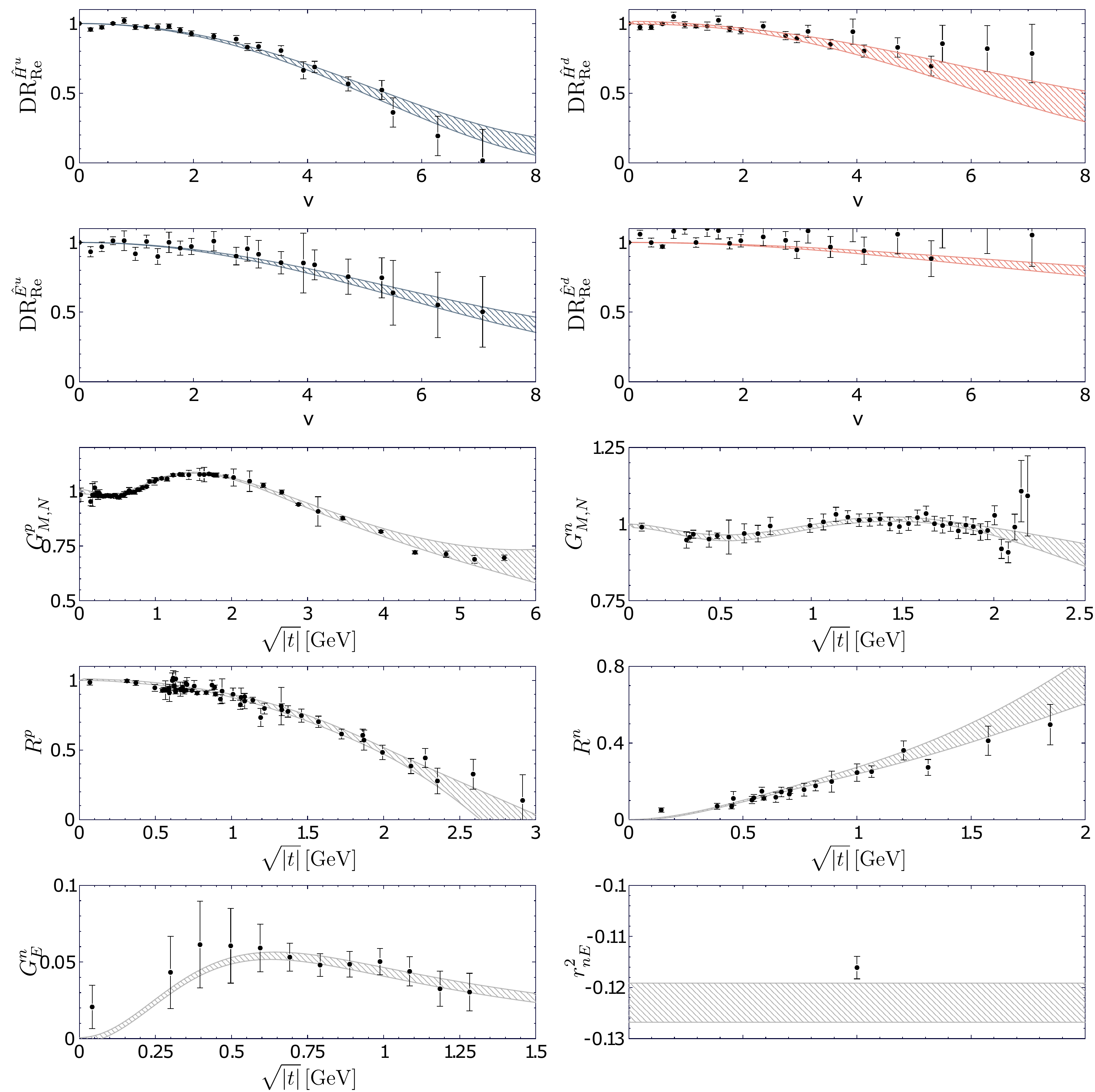} 
\end{figure}

Both lattice-QCD and elastic data used in our fit are specified in Table~\ref{tab:EFFdata}. The latter were initially selected in Ref.~\cite{Diehl:2013xca}, and they consist of the following observables:
\begin{itemize}[label=\small$\bullet$]
\item the normalized magnetic form factors:
\begin{equation}
G_{M, N}^{i}(t) = \frac{G_{M}^{i}(t)}{\mu_{i}G_{D}(t)} \,,
\end{equation}
where $\mu_{i}$ are magnetic moments, $i=p, n$, and
\begin{equation}
G_{D}(t) = \frac{1}{\left( 1 - t/M_D^2 \right)^{2}}
\end{equation}
is the dipole form factor with $M_D^2 = 0.71~\mathrm{GeV}^2$.
\item the normalized ratios of electric and magnetic form factors:
\begin{equation}
R^{i}(t) = \frac{\mu_{i}G_{E}^{i}(t)}{G_{M}^{i}(t)} \,,
\end{equation}
where $i=p, n$. 
\item the squared charge radius of neutron:
\begin{equation}
r_{nE}^{2} = 6 \frac{dG_{E}^{n}(t)}{dt}\bigg\rvert_{t = 0} \;.
\end{equation}
\end{itemize}
The relationships between the Sachs, Dirac, and Pauli form factors for the proton and neutron are as follows: 
\begin{align}
G_{M}^{i} &= F_{1}^{i} + F_{2}^{i} \,,  \nonumber \\
G_{E}^{i} &= F_{1}^{i} + \frac{t}{4m^{2}}F_{2}^{i}\,,\quad\mathrm{where}~i=p, n\,.
\end{align}
Finally, the ``partonic'' elastic FFs directly related to GPDs via Eqs.~\eqref{eq:bacis_F1} and~\eqref{eq:bacis_F2} are: 
\begin{align}
F_{i}^{p} &= e_{u} F_{i}^{u} + e_{d} F_{i}^{d} \,, \nonumber \\
F_{i}^{n} &= e_{u} F_{i}^{d} + e_{d} F_{i}^{u}\,,\quad\mathrm{where}~i=1, 2 \,.
\end{align}

The fit is performed using both Minuit~\cite{James:1975dr} (for the initial minimization without the shadow term) and a genetic algorithm~\cite{ga} (for the final approach). Data sensitive to all considered GPDs and flavors are fitted together. We therefore simultaneously constrain $16$ parameters: $10$ for the GPD $H$ and $6$ for the GPD $E$, see Table~\ref{tab:fitResult}. The uncertainties are propagated using replicas of PDF, lattice-QCD, and elastic FF data. The latter are generated by randomly smearing the central values of data points according to their associated uncertainties, see Ref.~\cite{Moutarde:2018kwr}. The positivity is enforced numerically by performing a simple numerical test for each potential solution of the fit to determine if it satisfies the positivity constraint. If it does not, the solution is discarded, which is particularly straightforward when using genetic algorithm minimization. The obtained value of $\chi^2$ normalized to the number of elastic FF and lattice-QCD data is $1185/882 \approx 1.34$ for the set of central replicas.  If the shadow term is not included, this value becomes $1488/882 \approx 1.68$, indicating that lattice-QCD results have a constraining power on this term. The quality of the fit can be inspected visually in~Fig.~\ref{fig:fit}.  

%% file: sec_results.tex
\section{Results}
\label{sec:results}

\begin{figure}[!tp]
  \centering
  \caption{Nucleon tomography for up quarks in an unpolarized proton at two values of $x$. The plots are obtained using the set of central replicas for PDFs, elastic FFs and lattice-QCD data, and separately show the contributions of the classic (dashed lines) and shadow terms (dotted lines), cf. Eq.~\eqref{eq:fitHAnsatz}, as well as their sum (solid lines).} 
  \label{fig:nt_1D_central}
  \subfloat[$x=0.2$]{\label{fig:nt_1D_central_02}\includegraphics[width=0.33\textwidth]{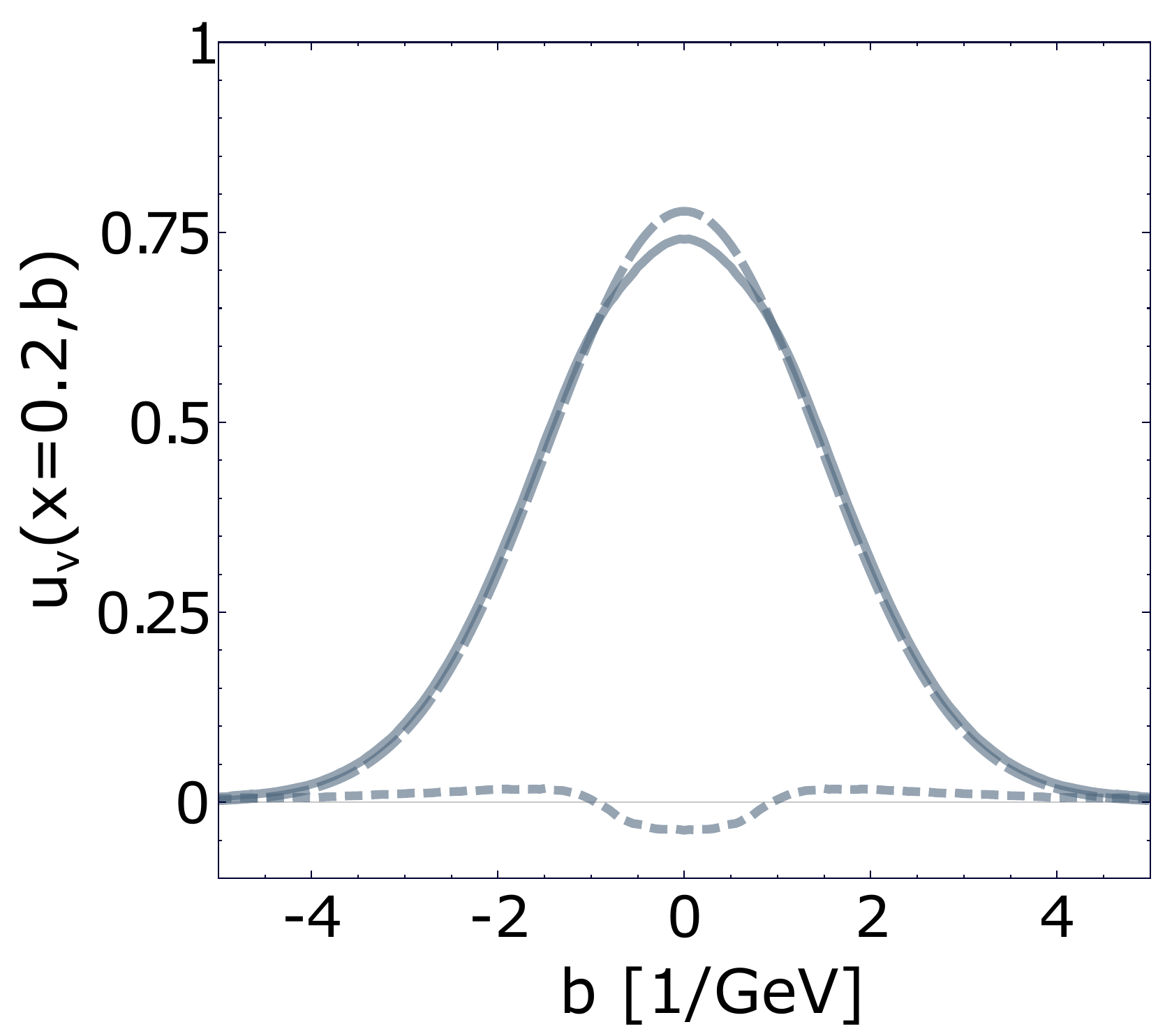}}
  \subfloat[$x=0.5$]{\label{fig:nt_1D_central_05}\includegraphics[width=0.33\textwidth]{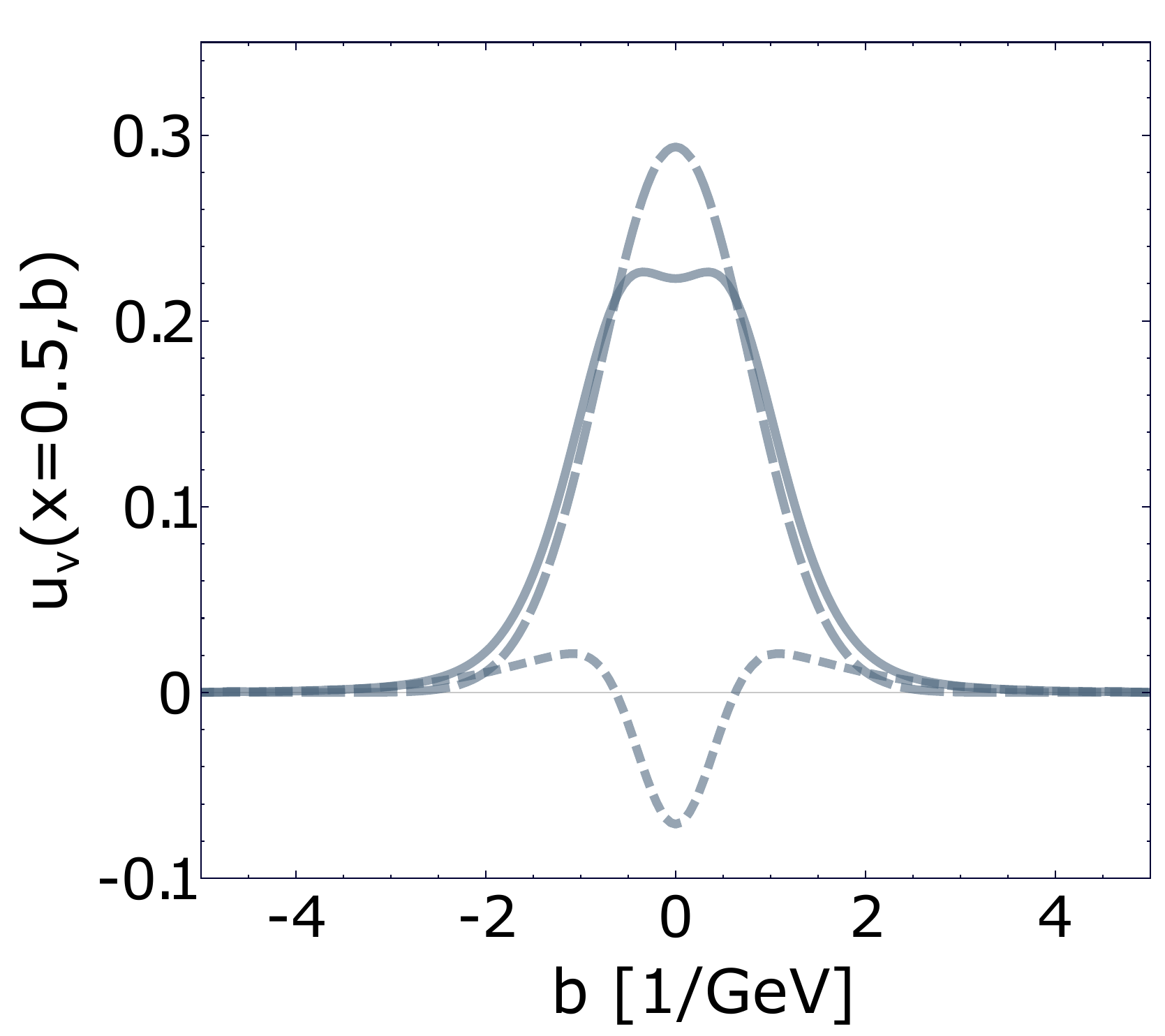}}
\end{figure}

We begin the discussion of our results by illustrating the impact of the shadow term on the tomographic images, as shown in Fig.~\ref{fig:nt_1D_central}. Several interesting features can be observed, some of which are anticipated based on the construction of this term. First, the shadow term contributes both positively and negatively, which is necessary to cancel out its overall contribution in $\int\dd b\, q(x, b)$. However, the shadow term does not make $q(x, b)$ negative at any point, a result of careful parameter selection and the numerical enforcement of positivity. Second, the shadow term modifies the bell shape imposed by the classic Ansatz utilizing the $\exp(f(x)t)$ function. This makes it a useful tool for studying more peculiar geometries of the proton. Finally, the shadow term primarily contributes in the region of high $x$. At low $x$, it does not compete with the classic term that includes $x^{-\delta}$ in the PDF parameterization. While it seems possible to modify the shadow term to resemble the PDF more closely, this is left for future, more systematic studies and is beyond the scope of this exploratory work.

\begin{figure}[!tp]
  \centering
  \caption{Nucleon tomography for unpolarized and transversely polarized protons at two values of $x$. The left column shows 1D profiles as a function of $b=(b_x^2 + b_y^2)^{\nicefrac{1}{2}}$ (for unpolarized proton) or $b_y$ (for transversely polarized proton) for up (blue) and down (red) valence quarks. The fit results (dashed bands) are compared to GK (solid lines) and VGG (dashed lines) models. The middle and right columns show $xu_{v}(x,b_x,b_y)/2$ and $xd_{v}(x,b_x,b_y)$ distributions, respectively, with the same color scales. The origins of the coordinate systems are marked by white~crosses.} 
  \label{fig:nt}
  \subfloat[Unpolarized proton for $x=0.2$]{\label{fig:nt_un02}
  \includegraphics[height=0.28\textwidth]{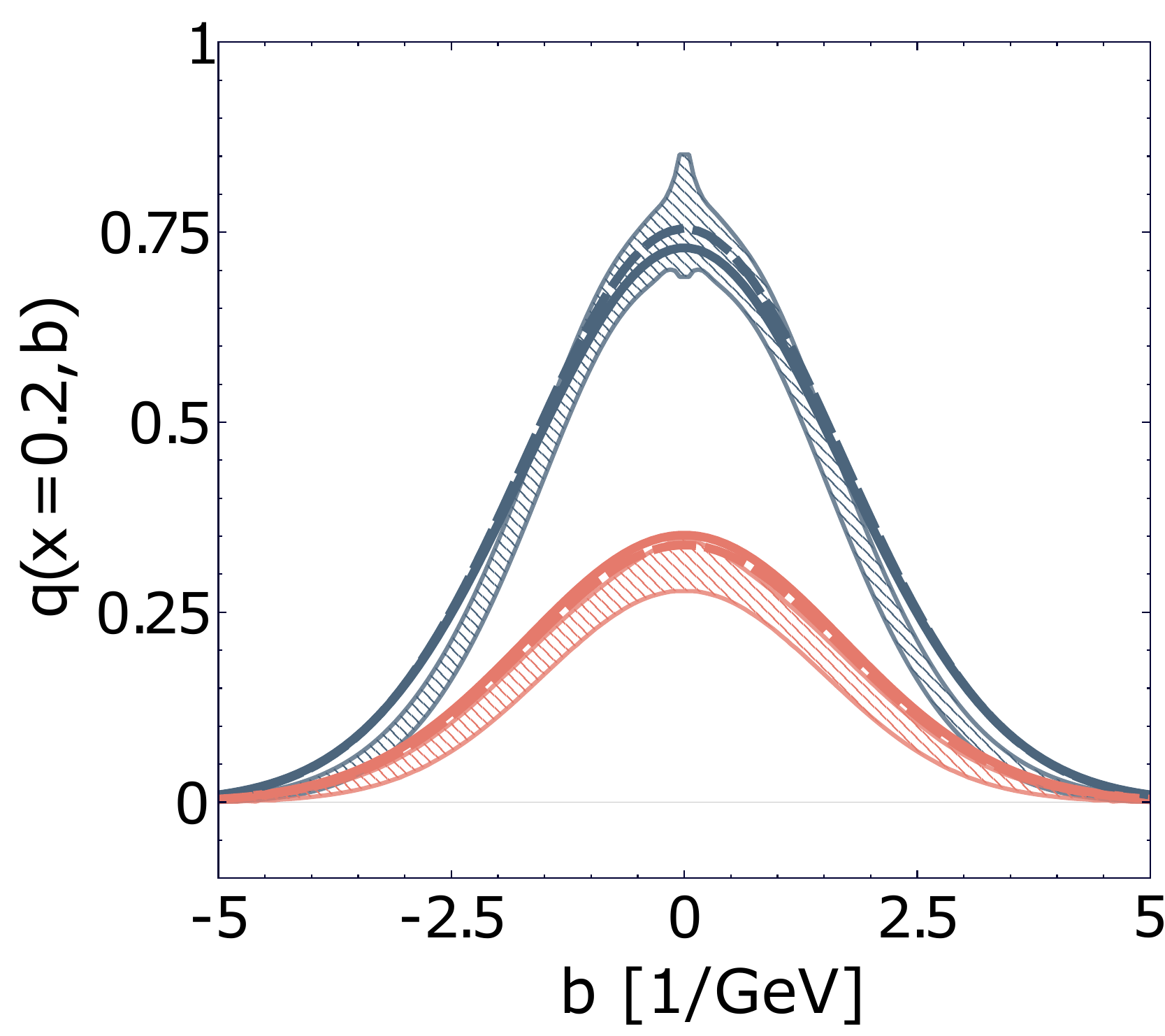}
  \includegraphics[height=0.28\textwidth]{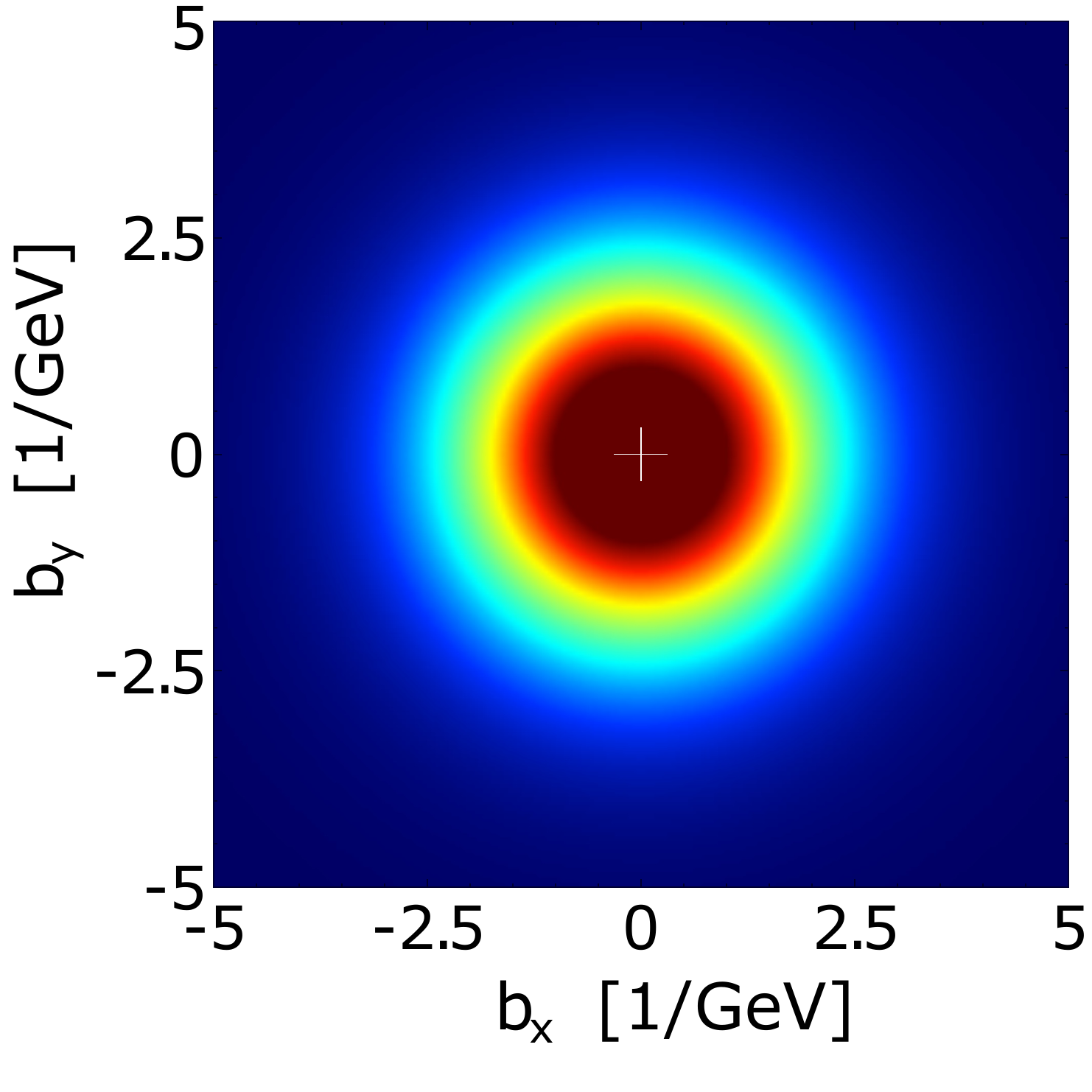}
   \includegraphics[height=0.28\textwidth]{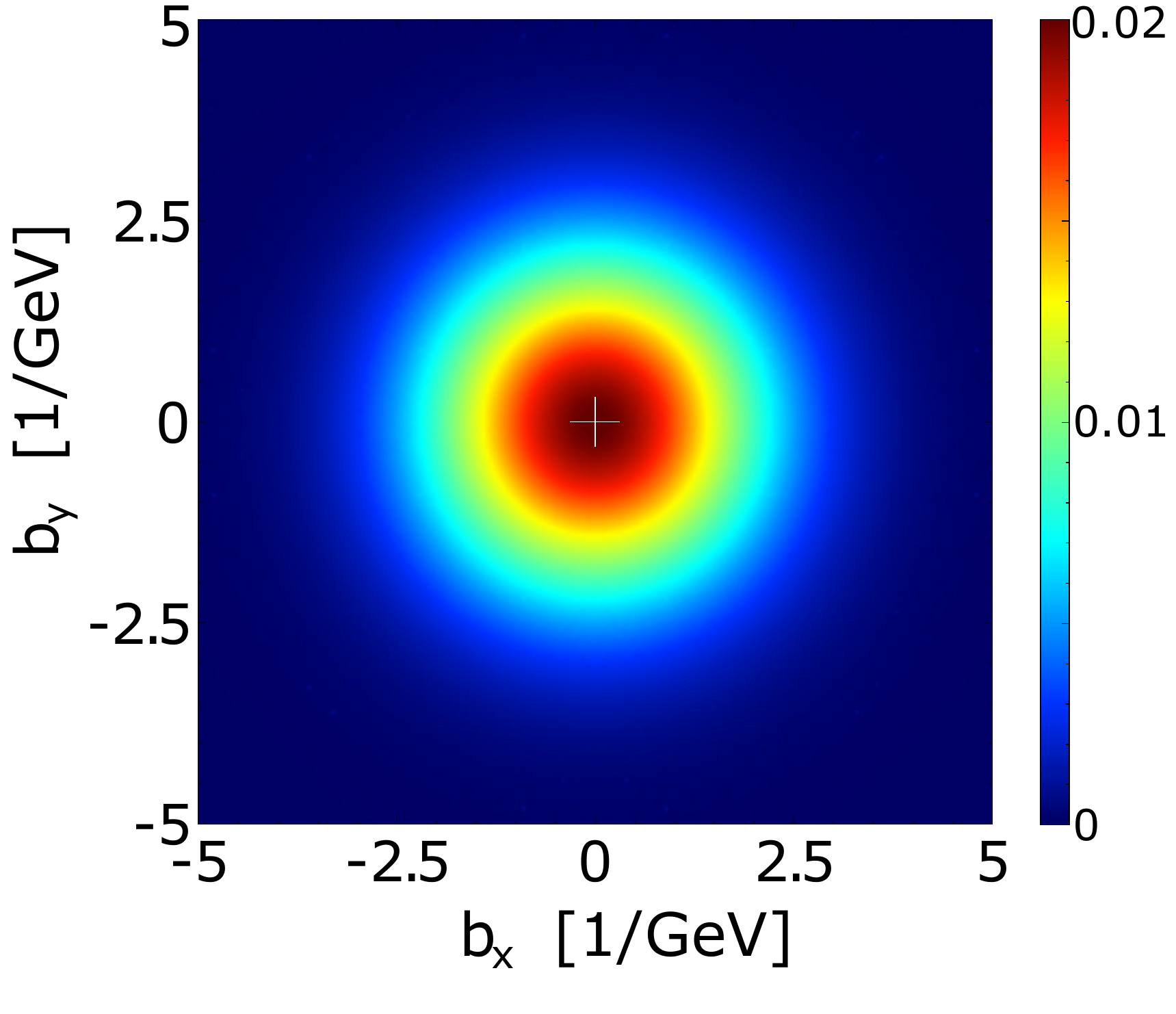}
  }\\
  \subfloat[Transversely polarized proton for $x=0.2$]{\label{fig:nt_pol02}
  \includegraphics[height=0.28\textwidth]{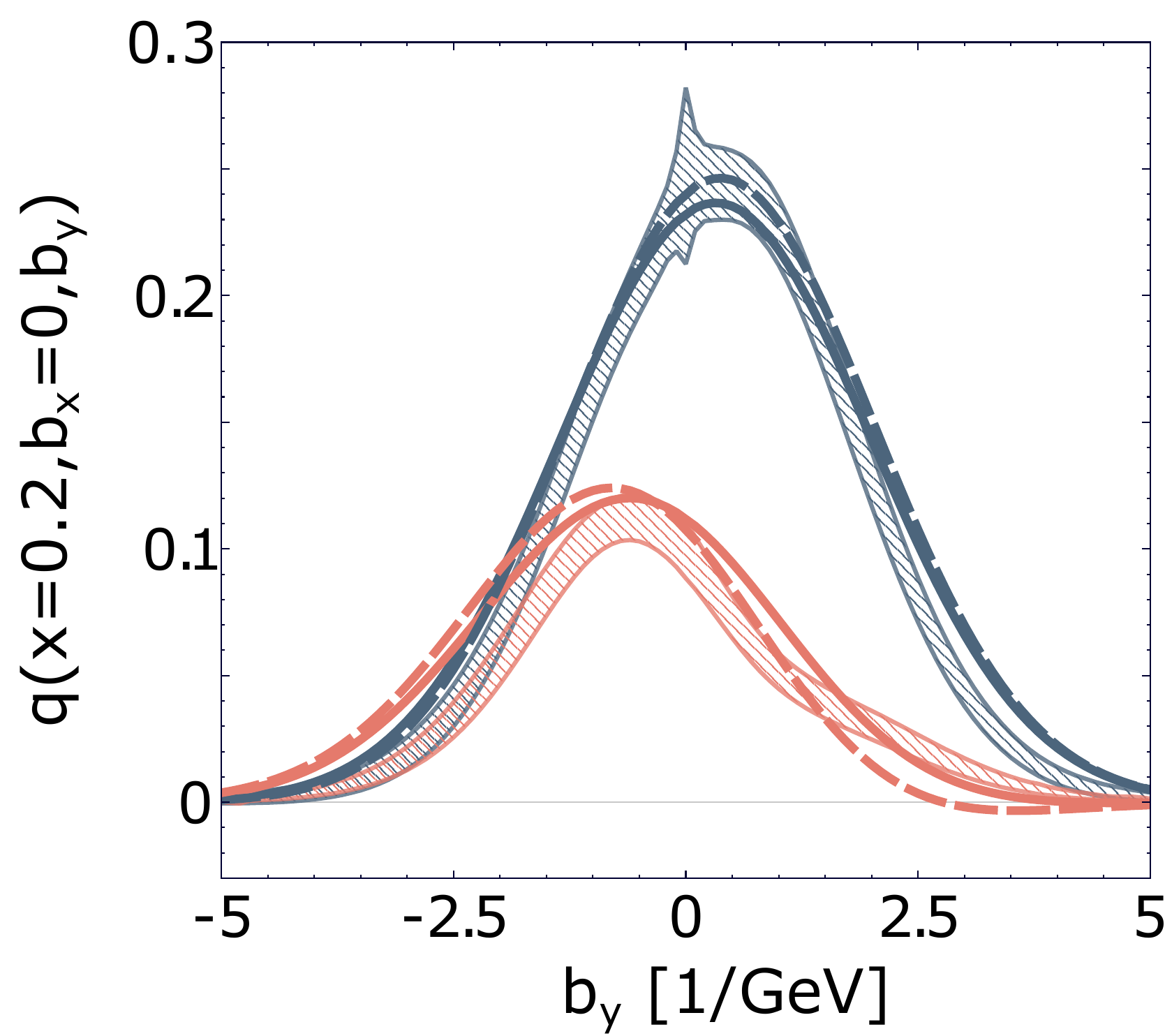}
  \includegraphics[height=0.28\textwidth]{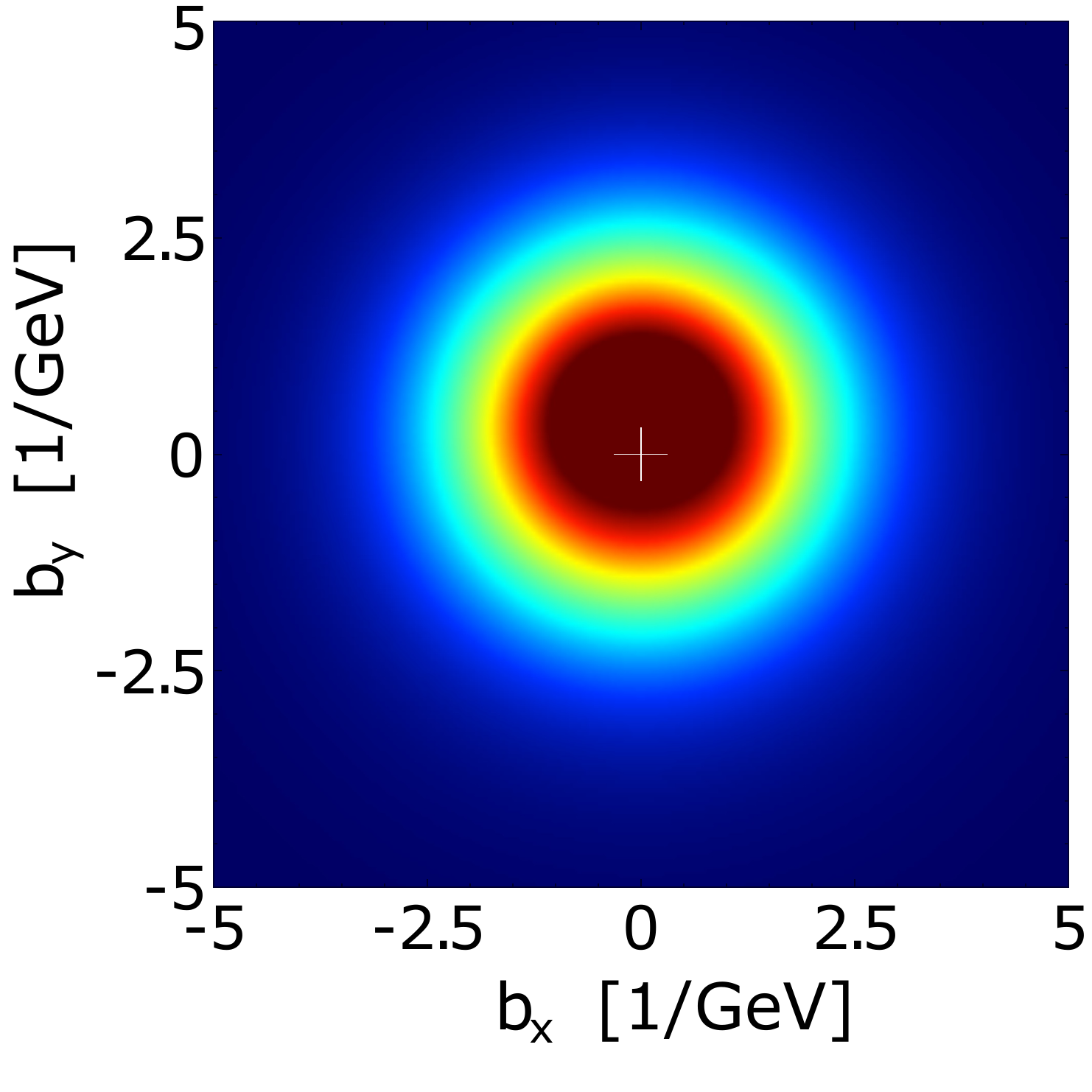}
   \includegraphics[height=0.28\textwidth]{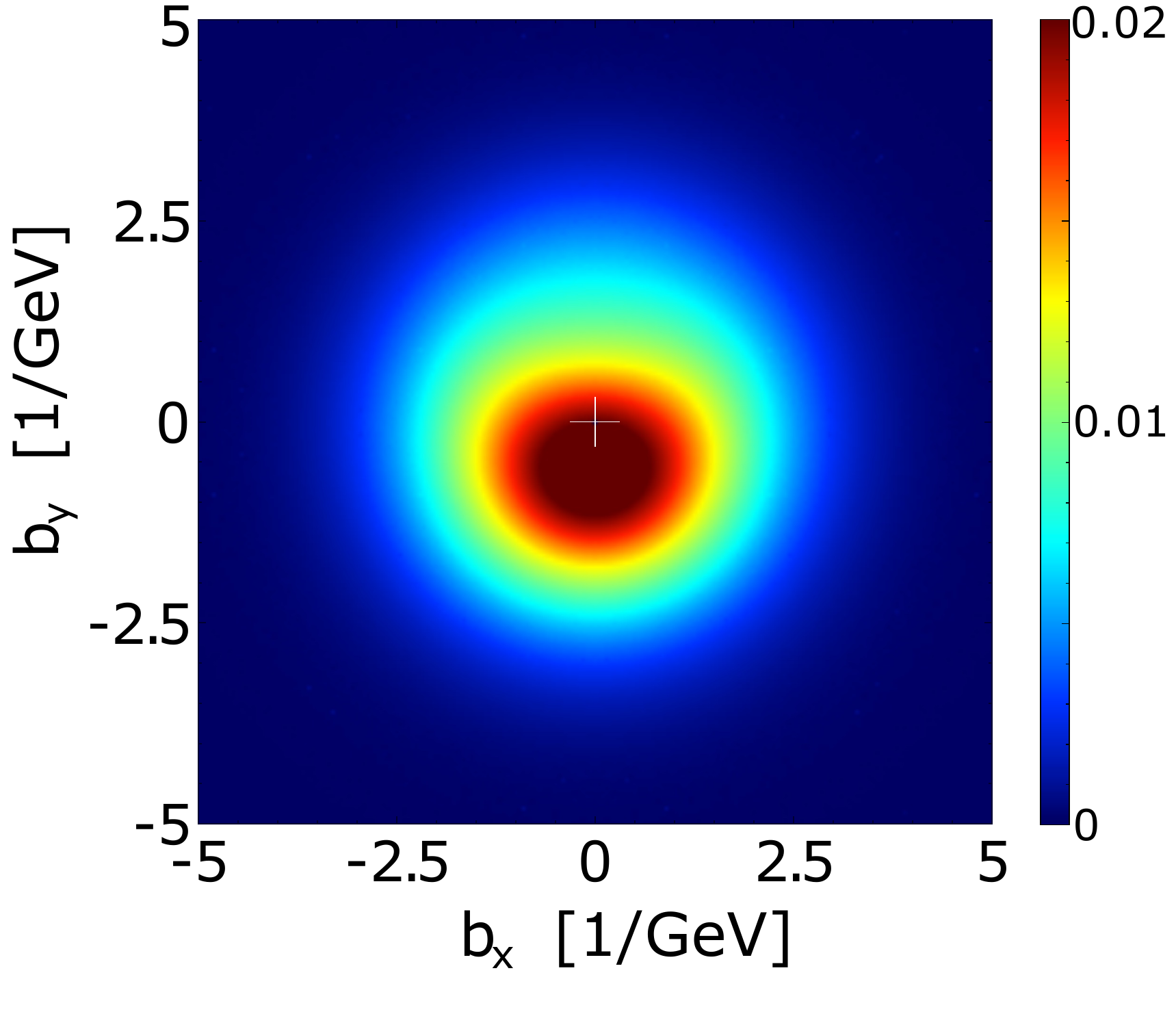}
  }\\
  \subfloat[Unpolarized proton for $x=0.5$]{\label{fig:nt_un05}
  \includegraphics[height=0.28\textwidth]{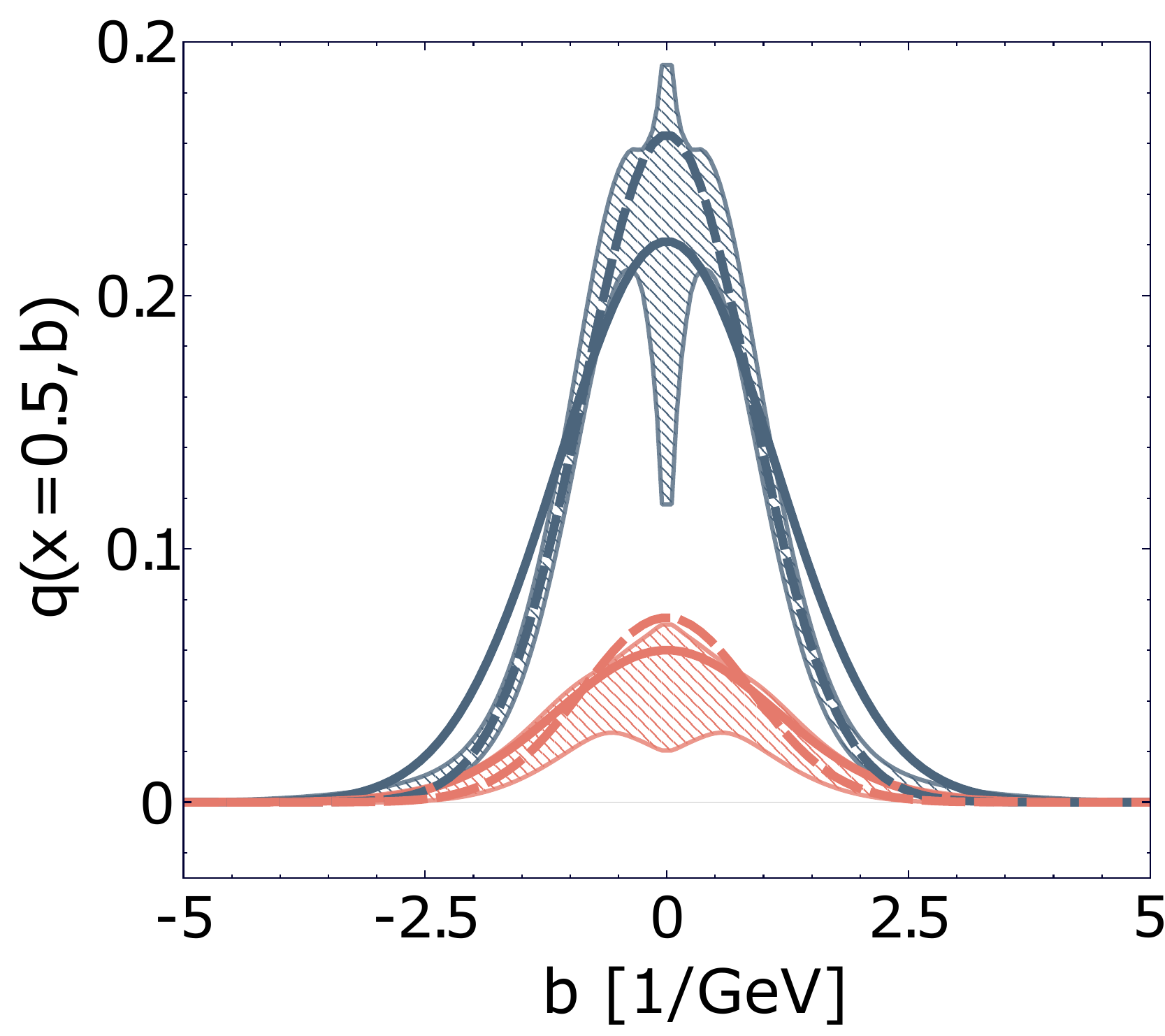}
  \includegraphics[height=0.28\textwidth]{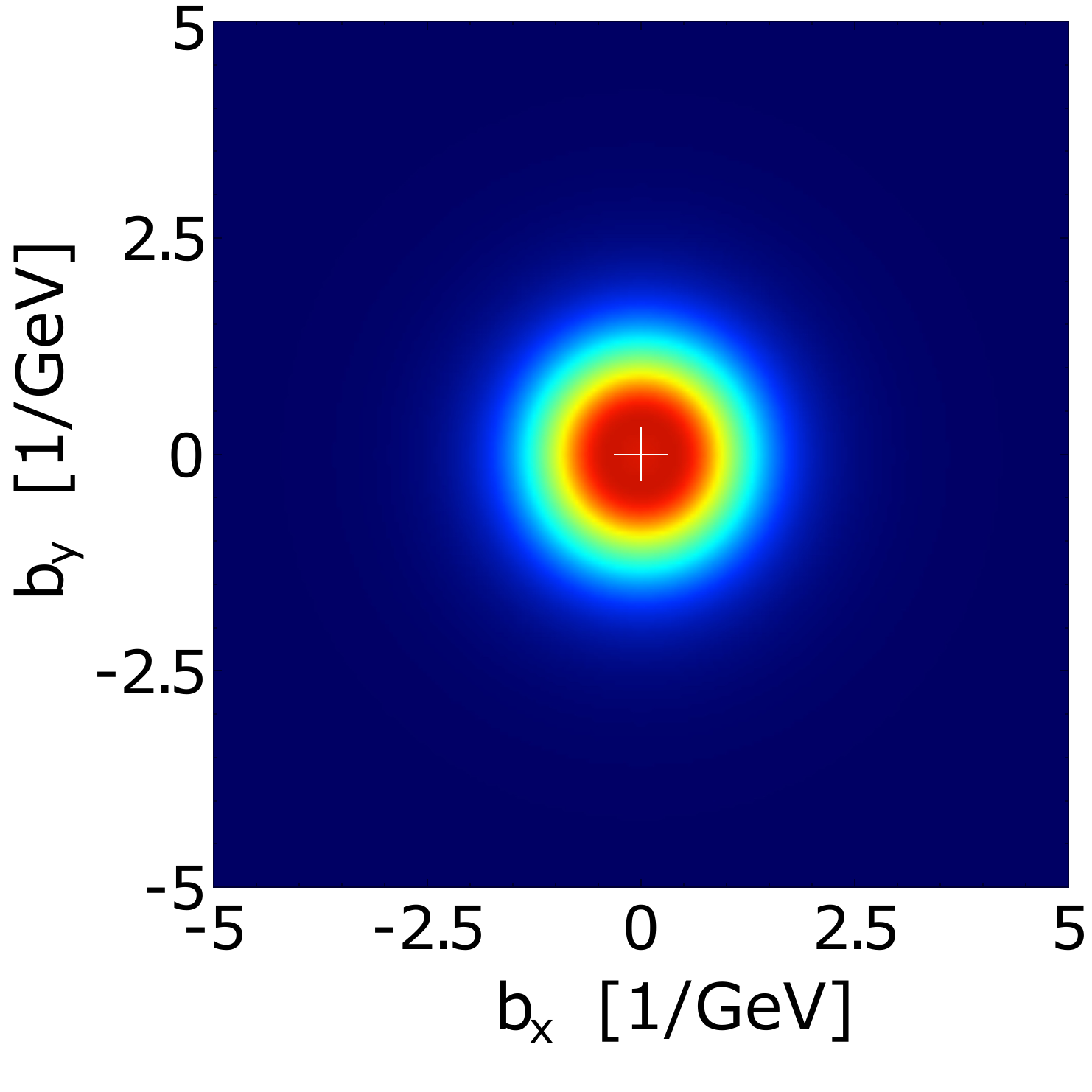}
   \includegraphics[height=0.28\textwidth]{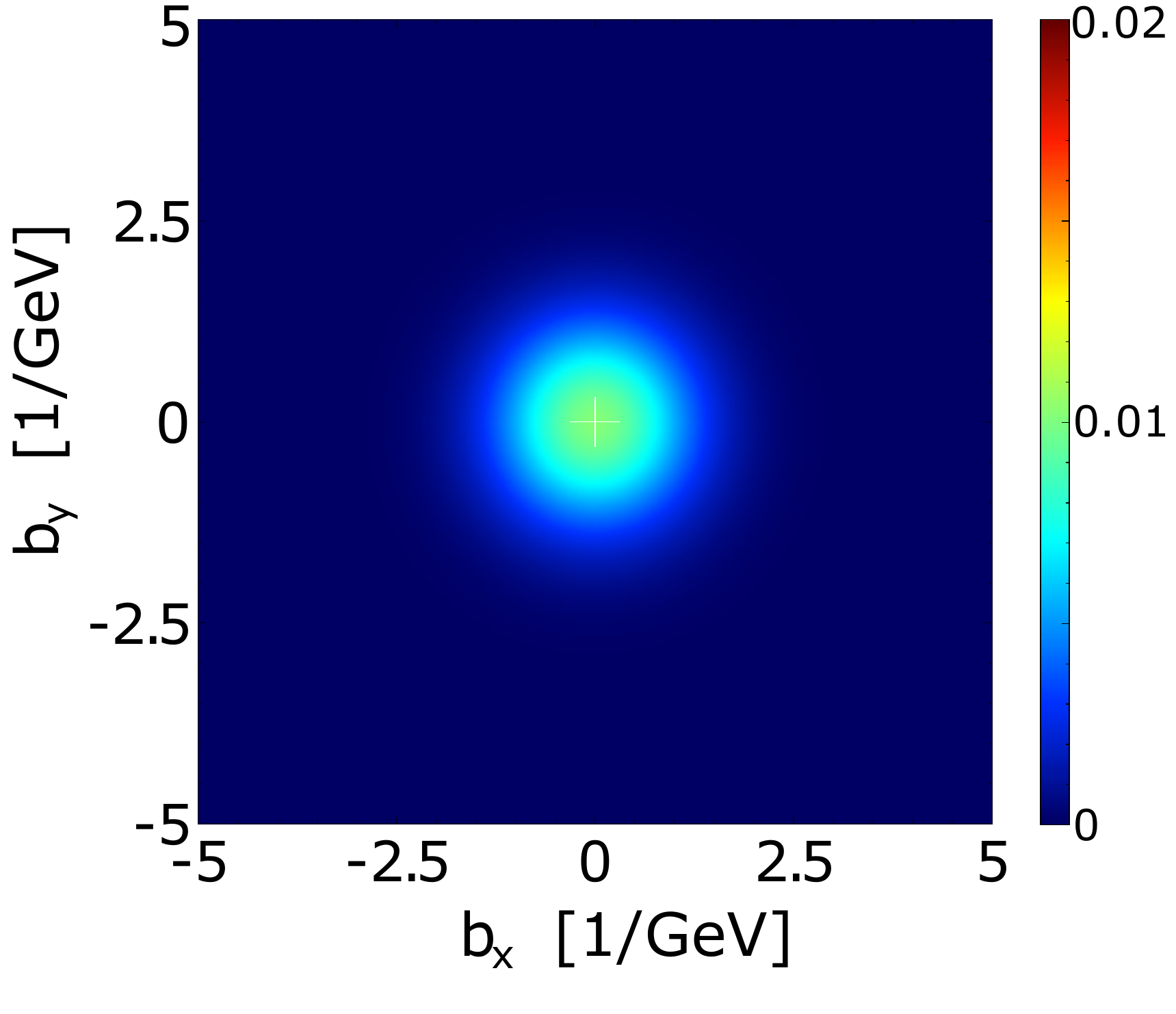}
  }\\
  \subfloat[Transversely polarized  proton for $x=0.5$]{\label{fig:nt_pol05}
  \includegraphics[height=0.28\textwidth]{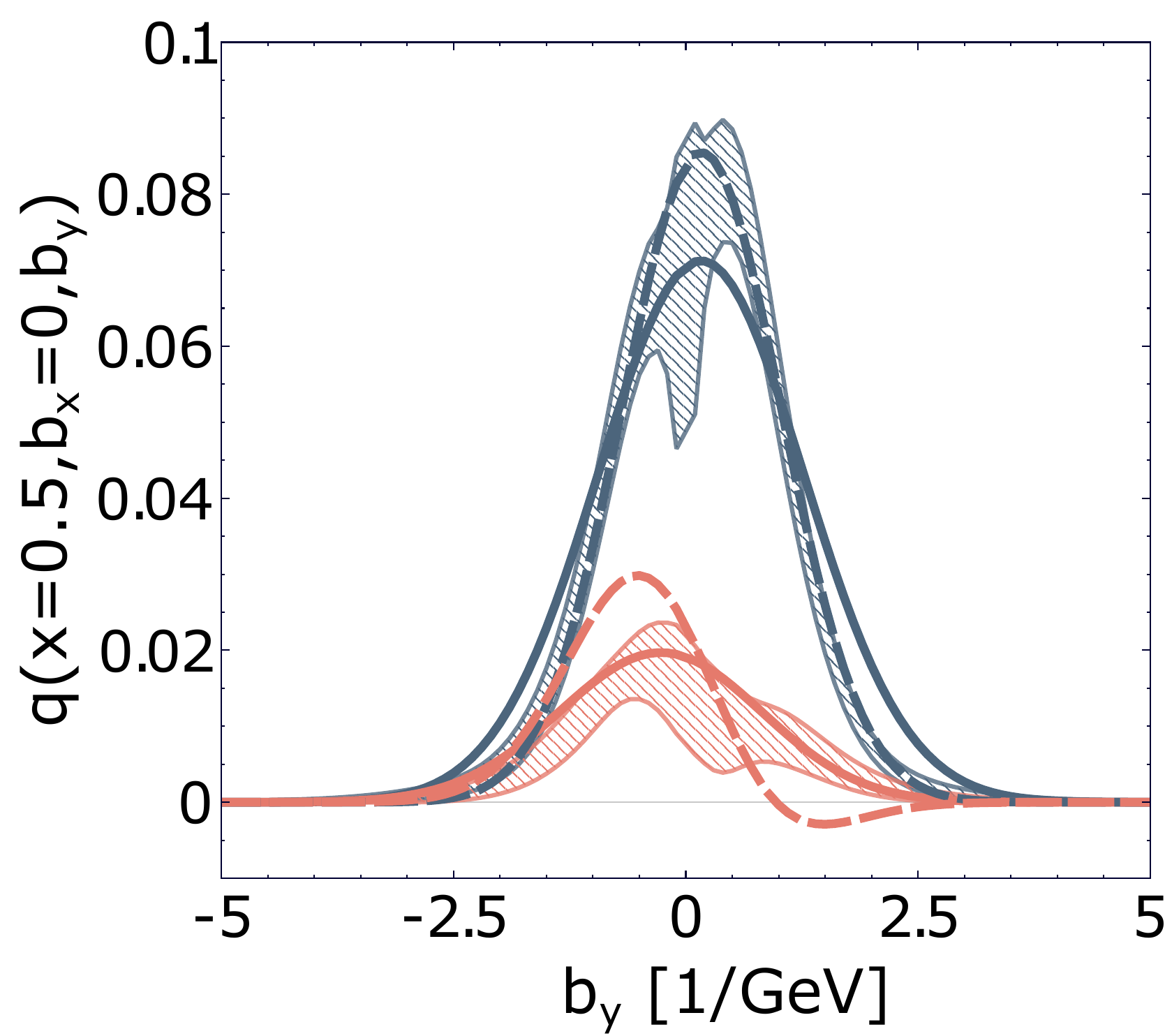}
  \includegraphics[height=0.28\textwidth]{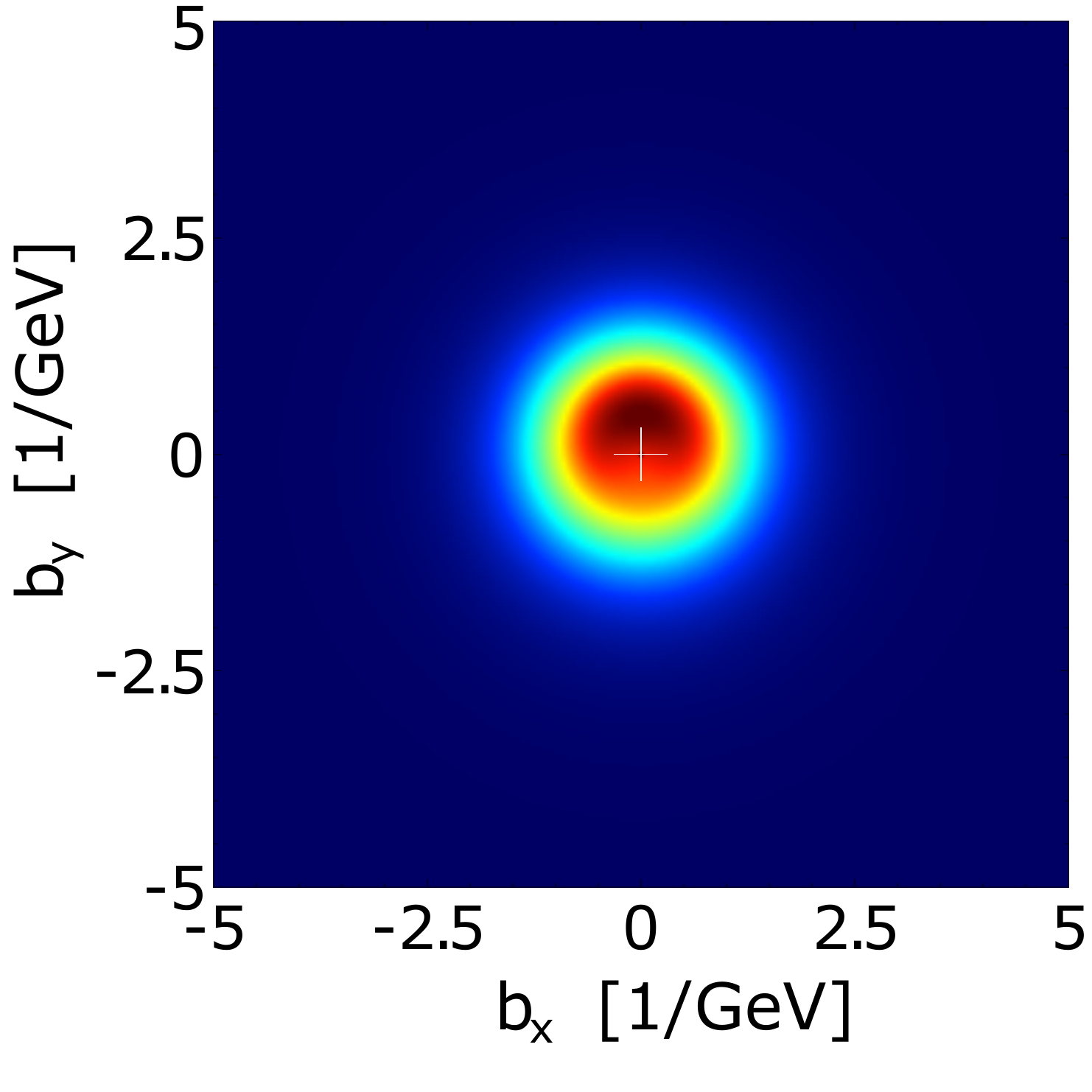}
   \includegraphics[height=0.28\textwidth]{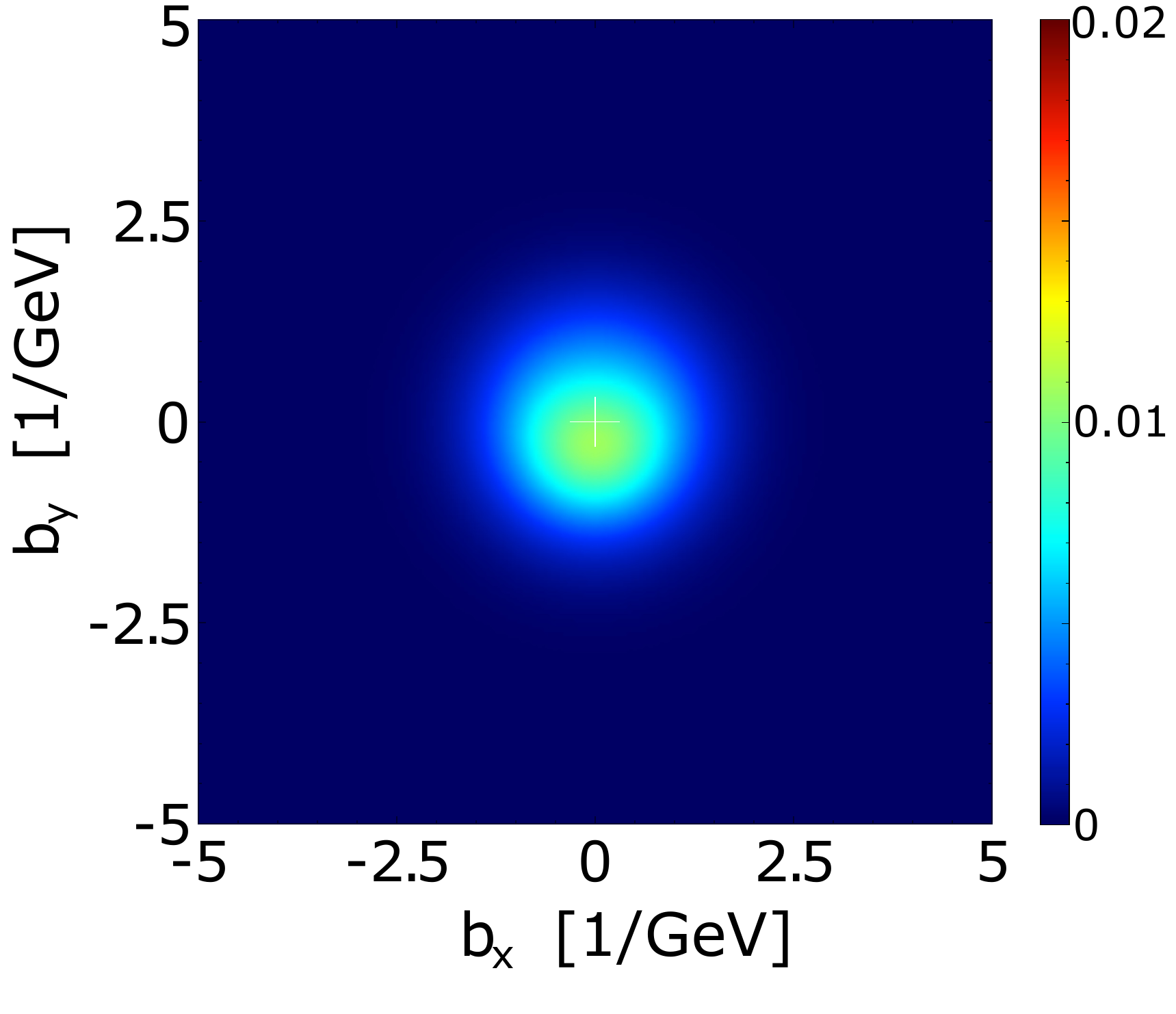}
  }

\end{figure}

\begin{figure}[!tp]
  \centering
  \caption{First Mellin moments of GPDs $H$ and $E$, see Eq.~\eqref{eq:polynomiality} for the definition, at $\xi = 0$ for up (blue) and down (red) valence quarks.  The fit results (dashed bands) are compared to GK (solid lines) and VGG (dashed lines) models.} 
  \label{fig:mom}
  \subfloat[For GPD $H$]{\label{fig:momH}
  \includegraphics[height=0.28\textwidth]{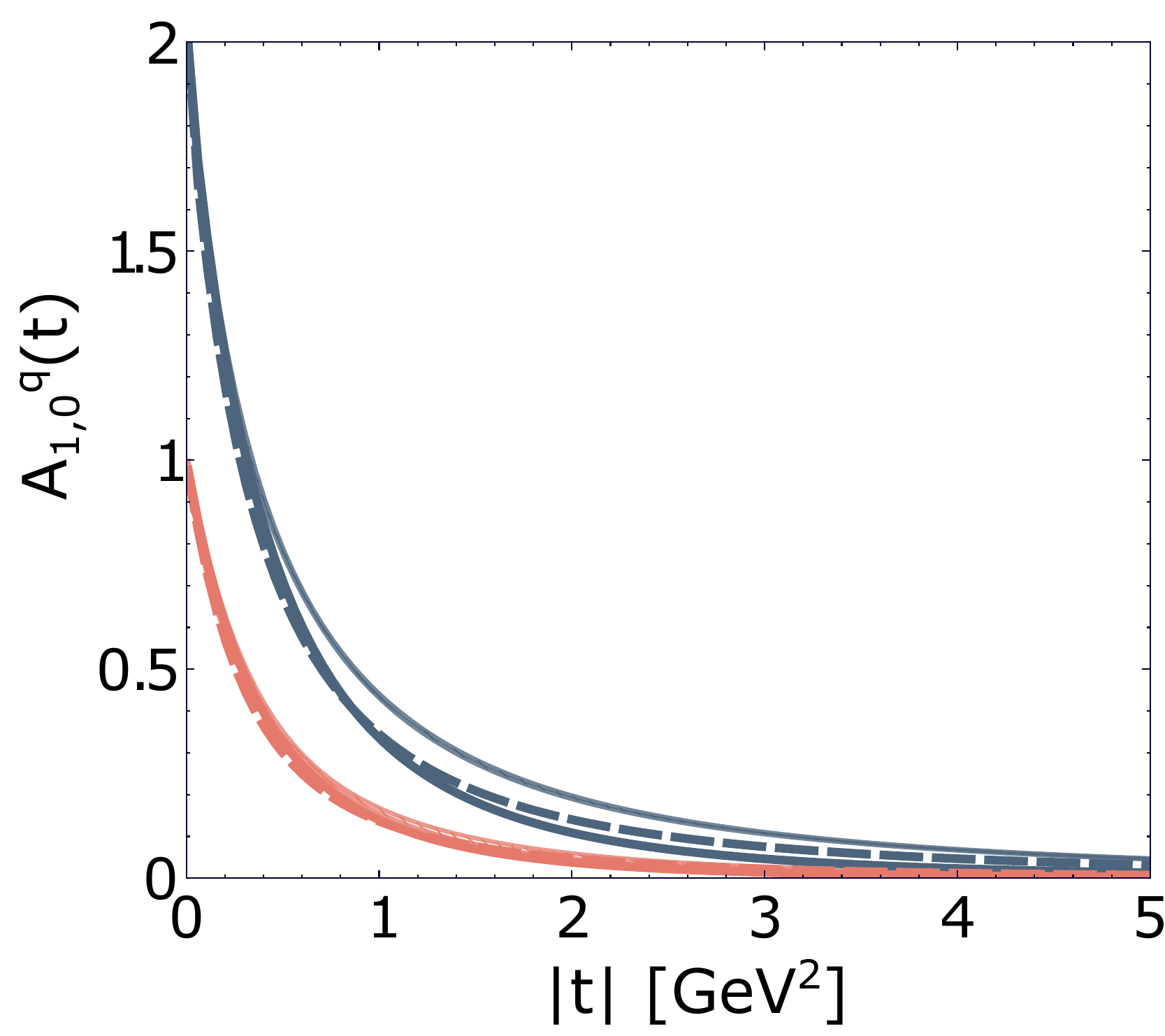}
  \includegraphics[height=0.28\textwidth]{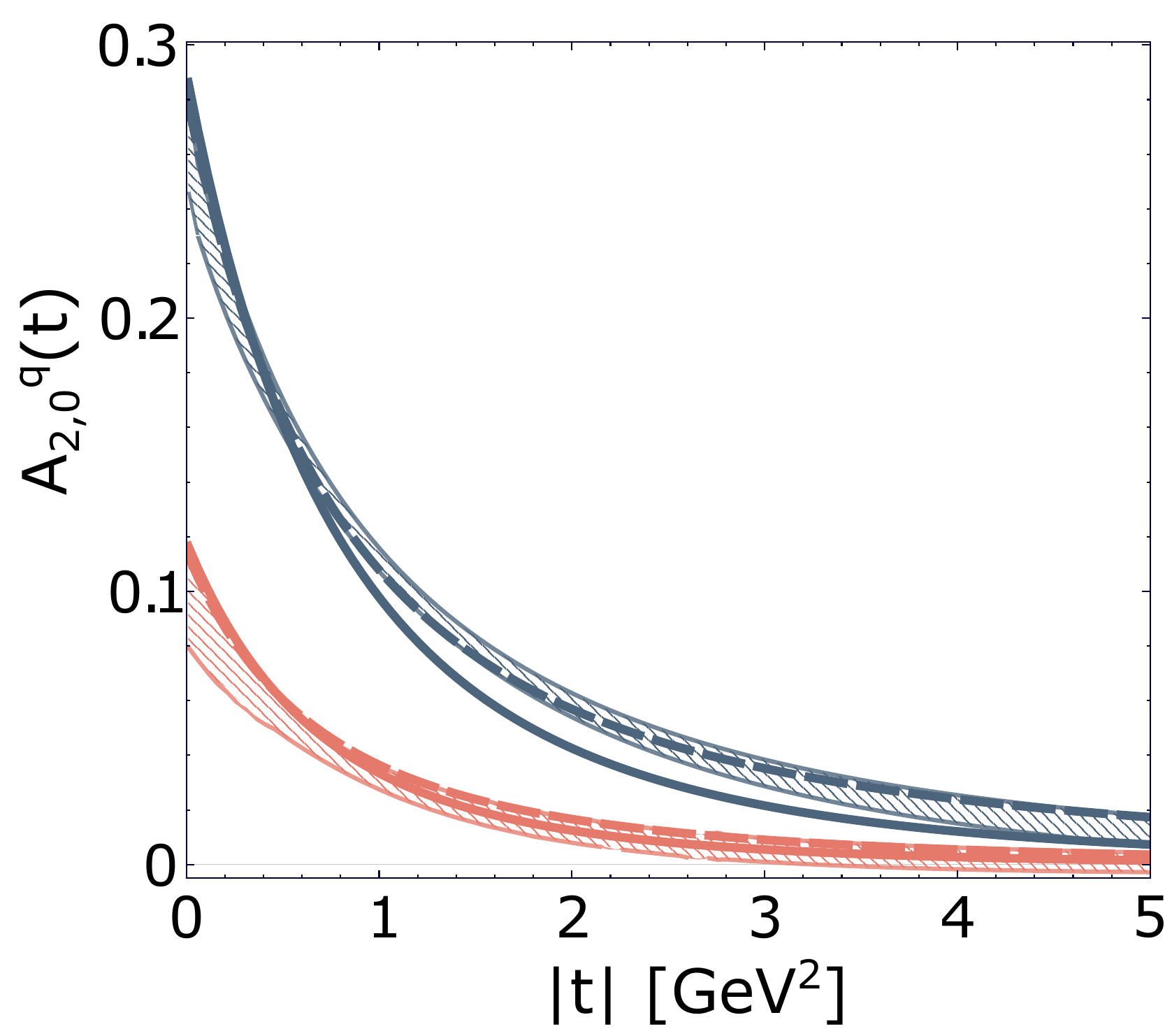}
   \includegraphics[height=0.28\textwidth]{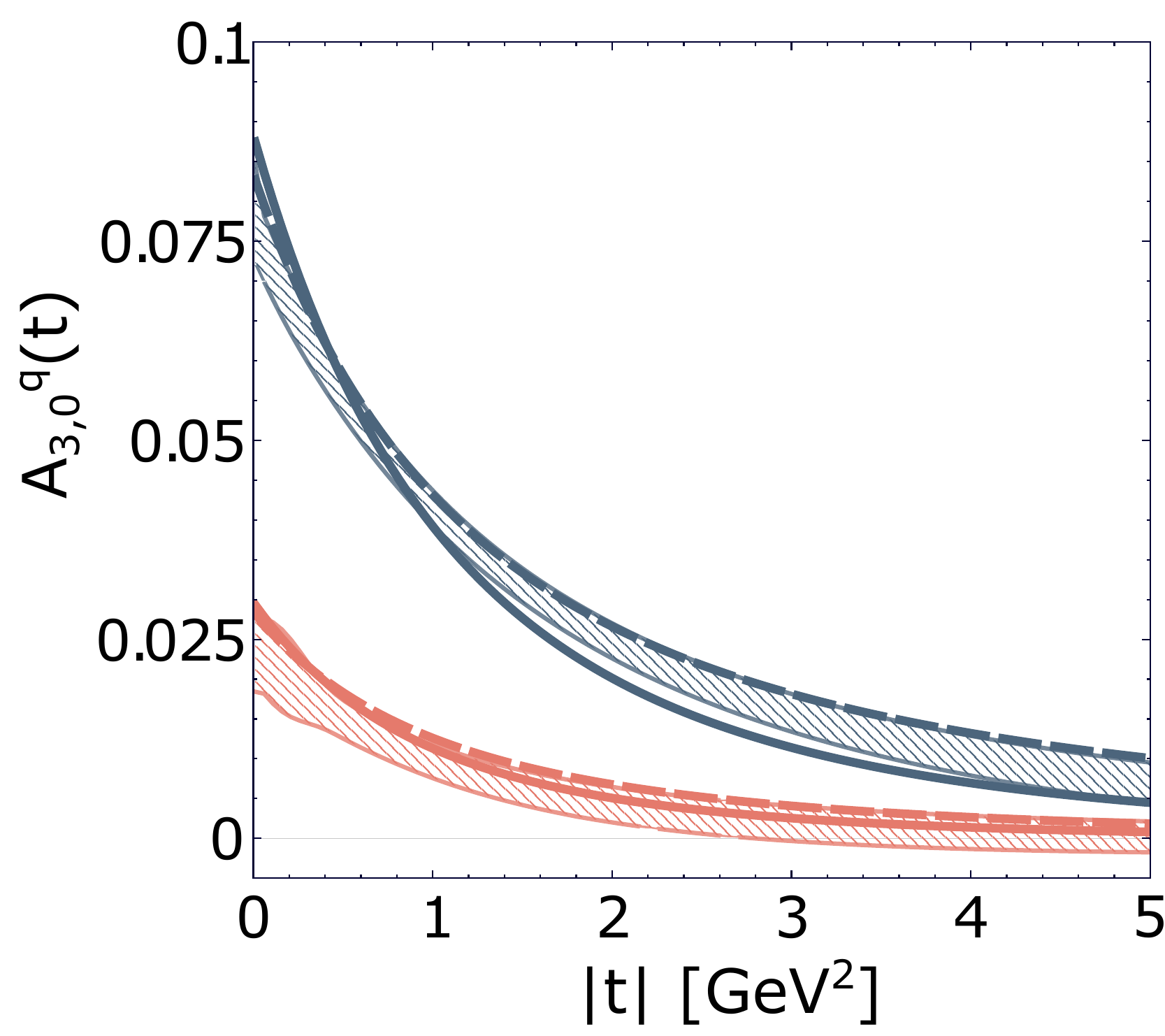}
   }
  \\
   \subfloat{
  \includegraphics[height=0.28\textwidth]{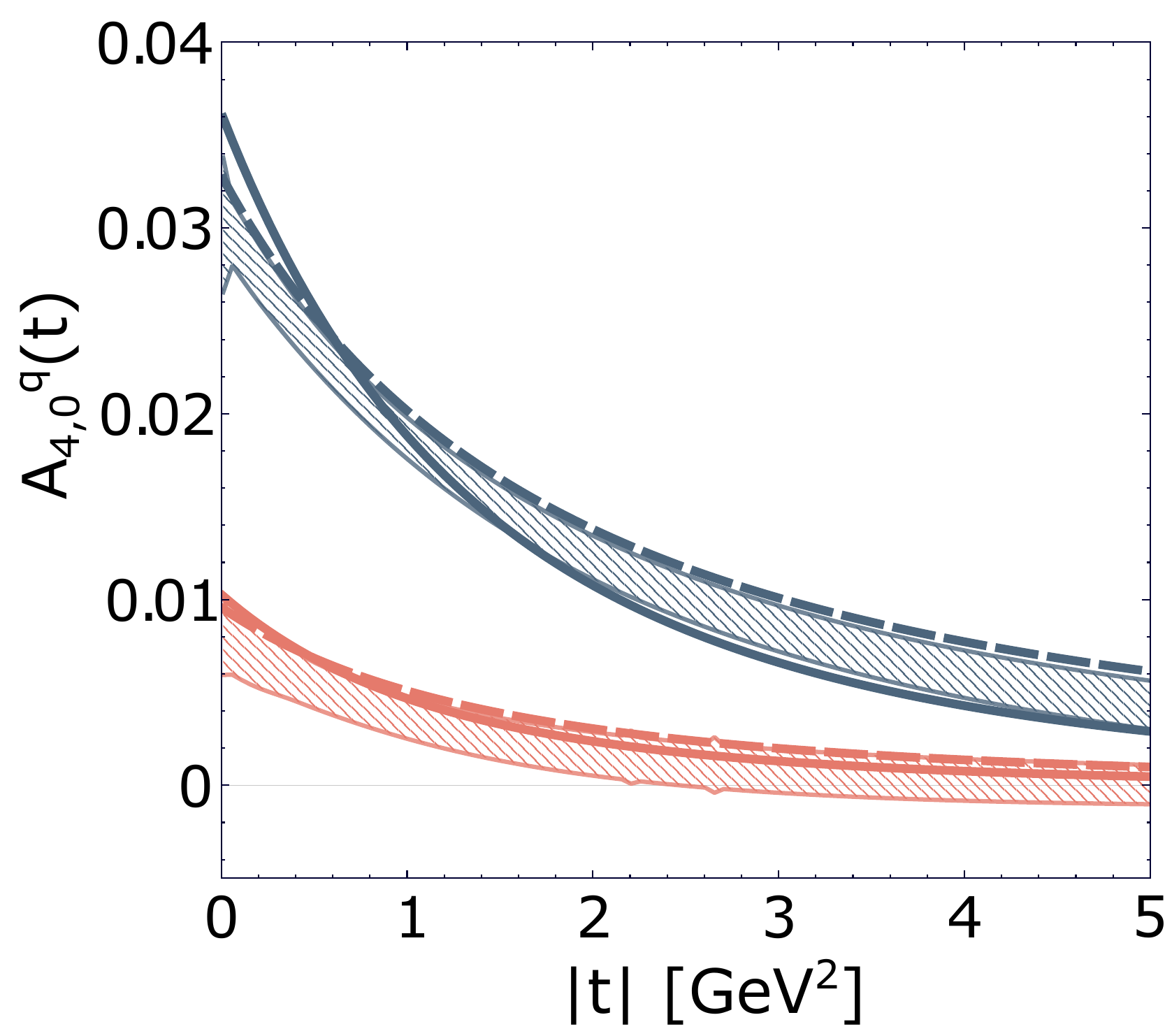}
  \includegraphics[height=0.28\textwidth]{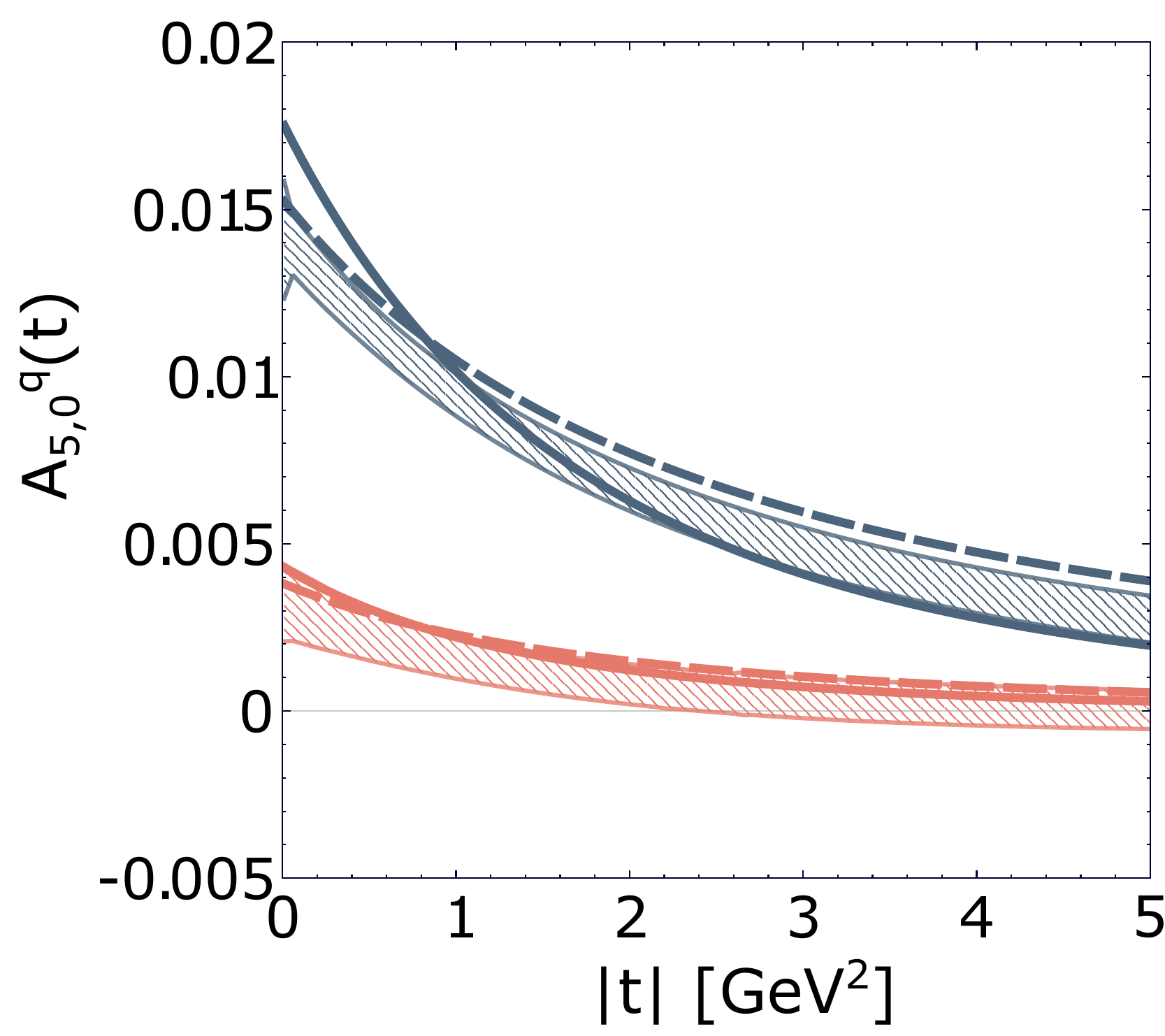}
   \includegraphics[height=0.28\textwidth]{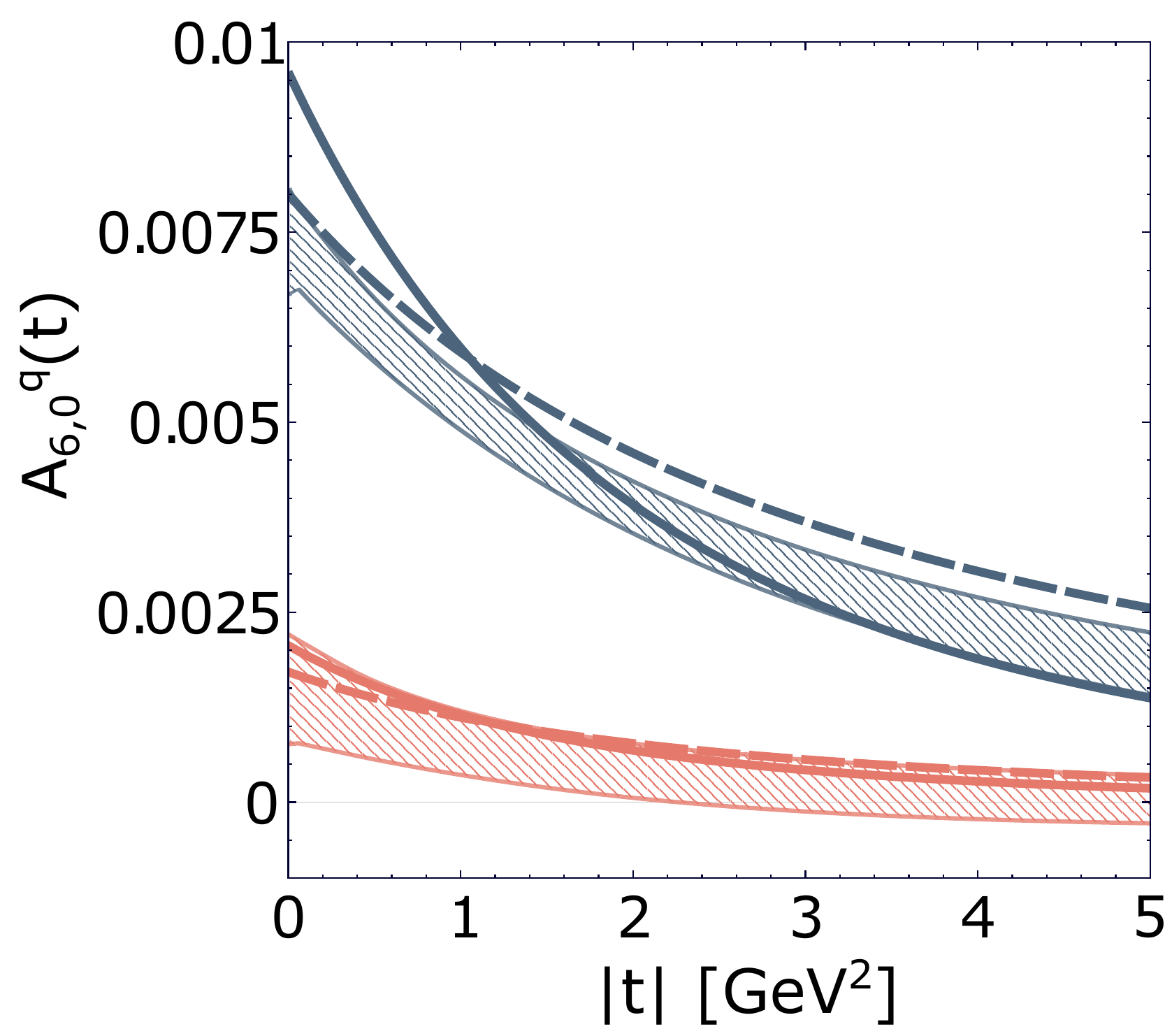}
  }\\
  \subfloat[For GPD $E$]{\label{fig:momE}
  \includegraphics[height=0.28\textwidth]{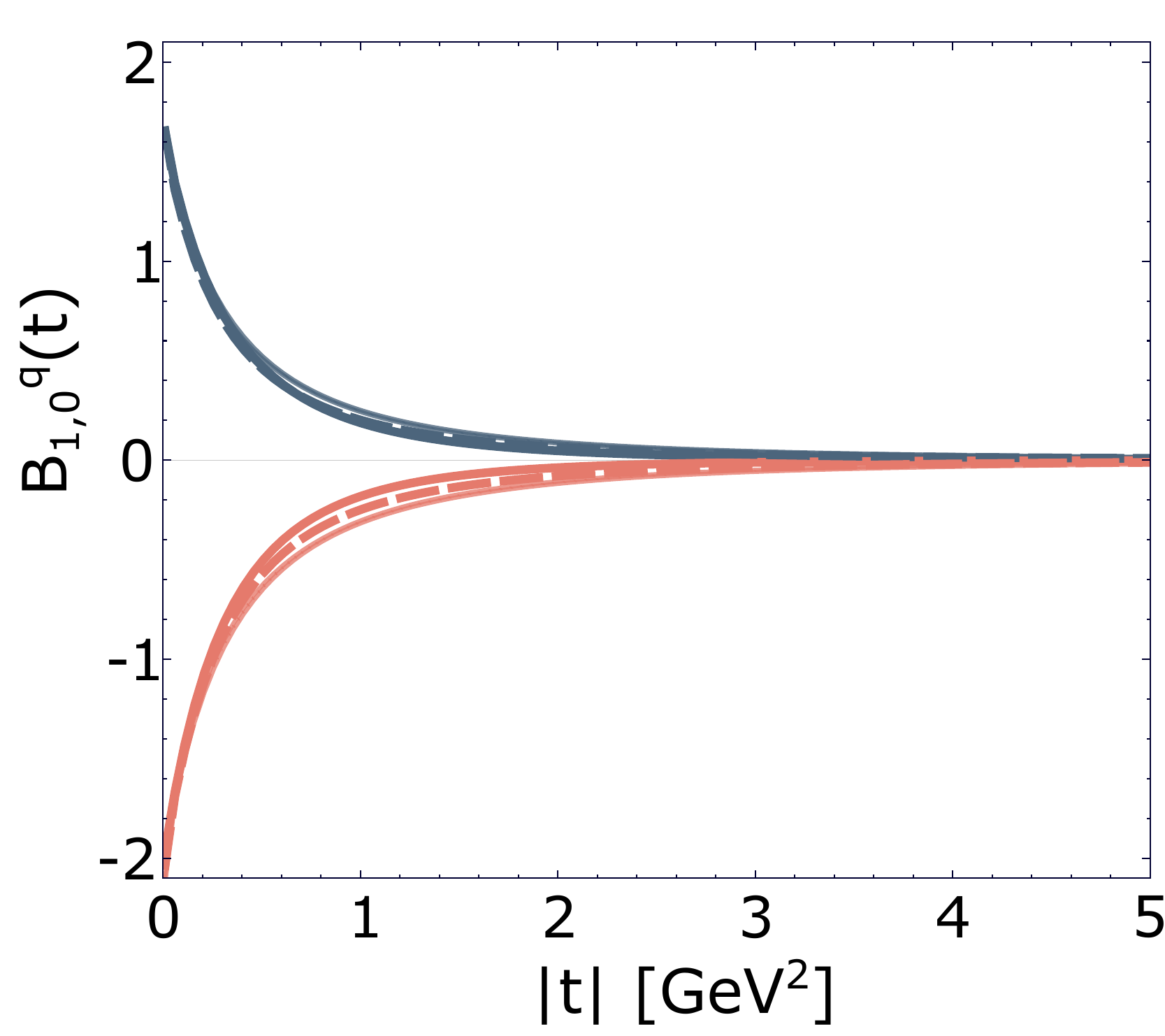}
  \includegraphics[height=0.28\textwidth]{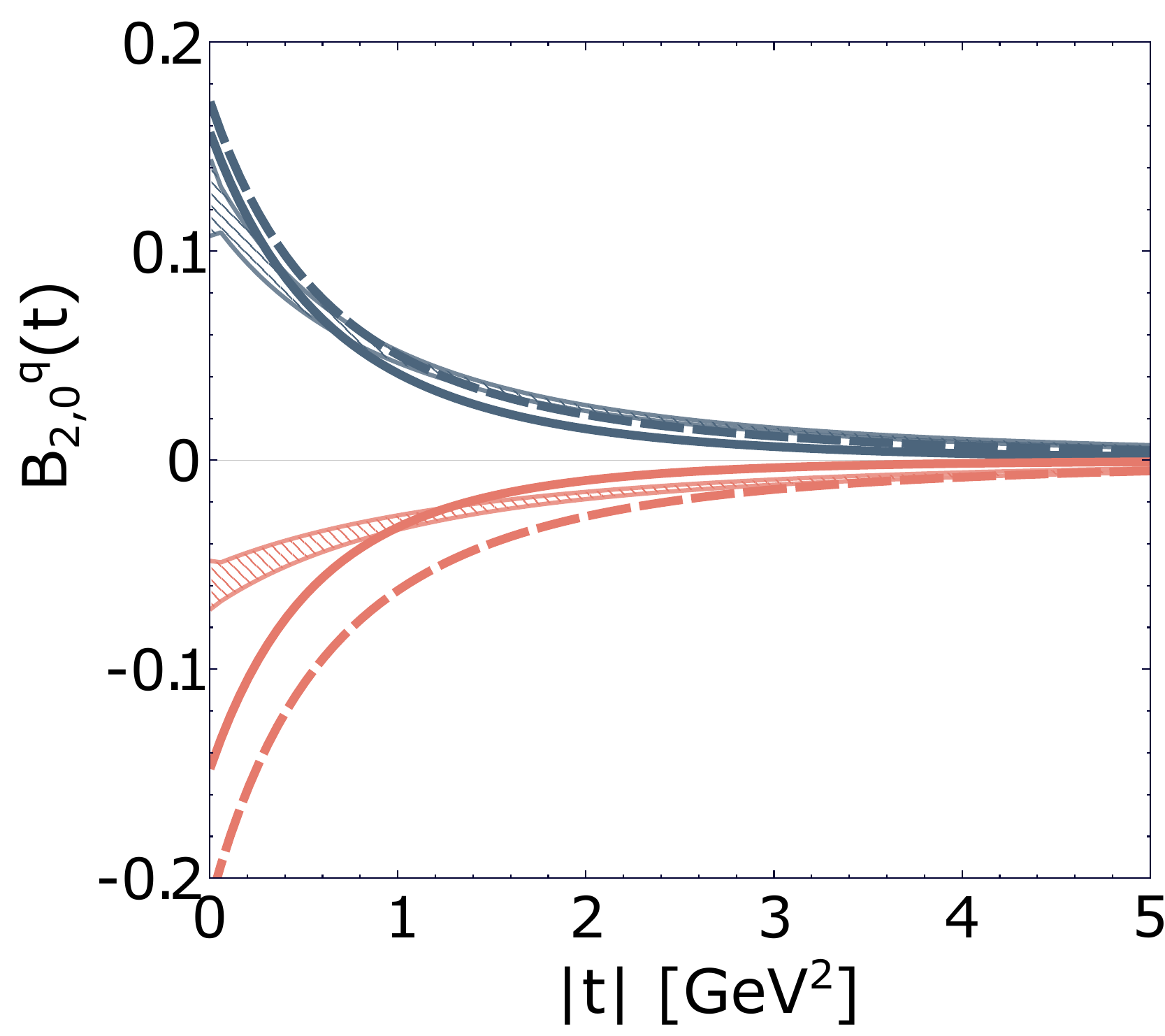}
   \includegraphics[height=0.28\textwidth]{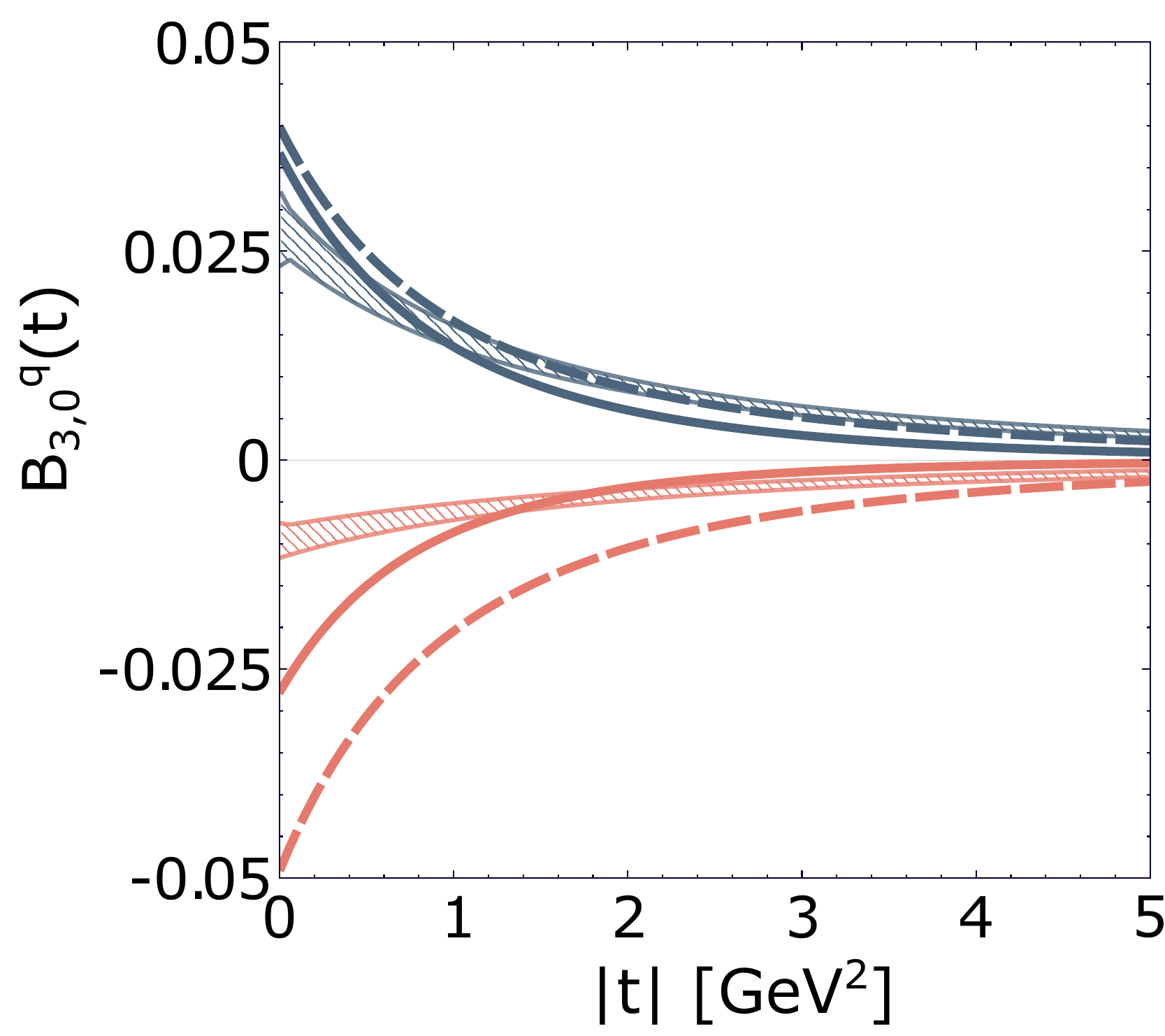}
  }\\
    \subfloat{
  \includegraphics[height=0.28\textwidth]{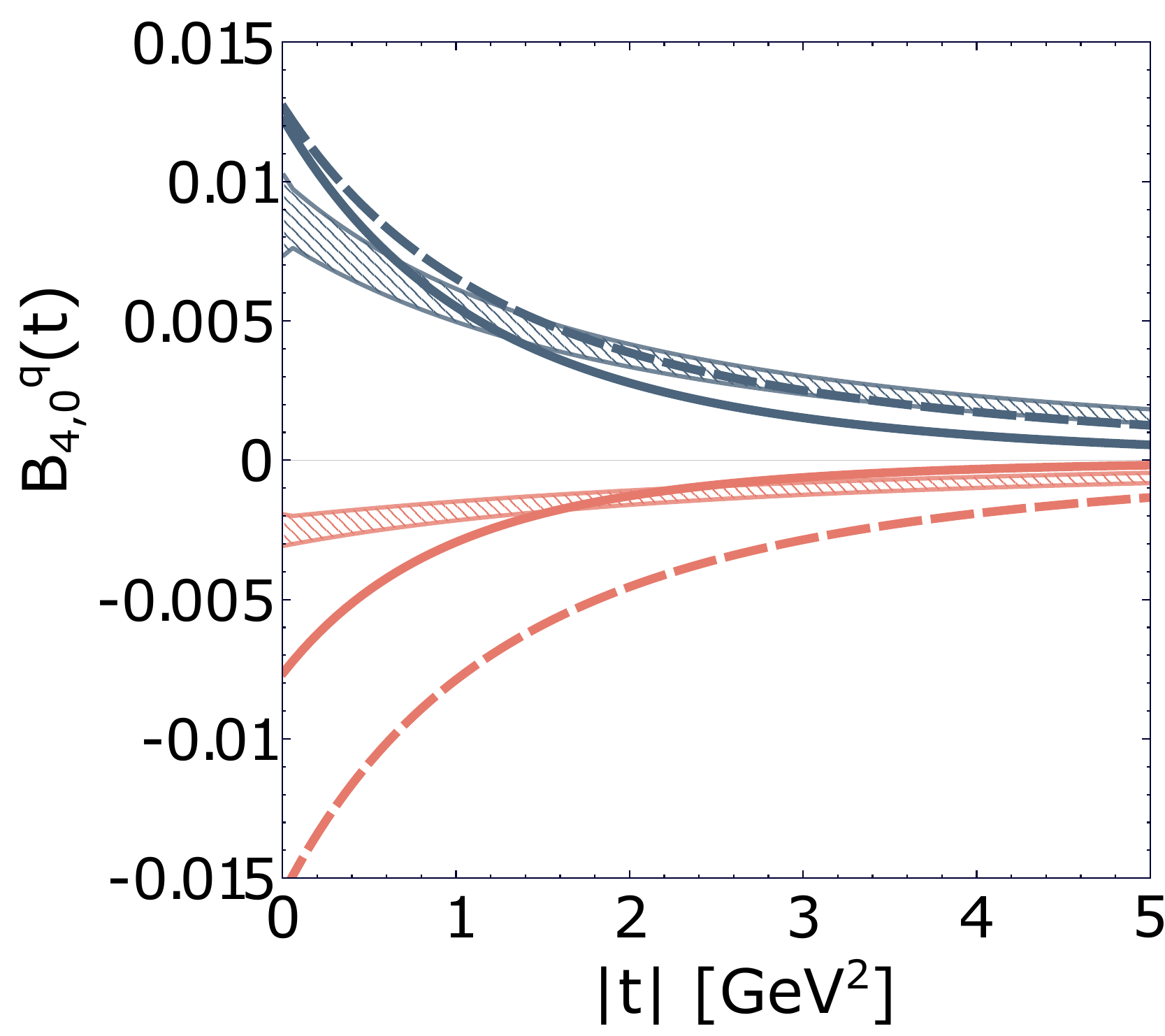}
  \includegraphics[height=0.28\textwidth]{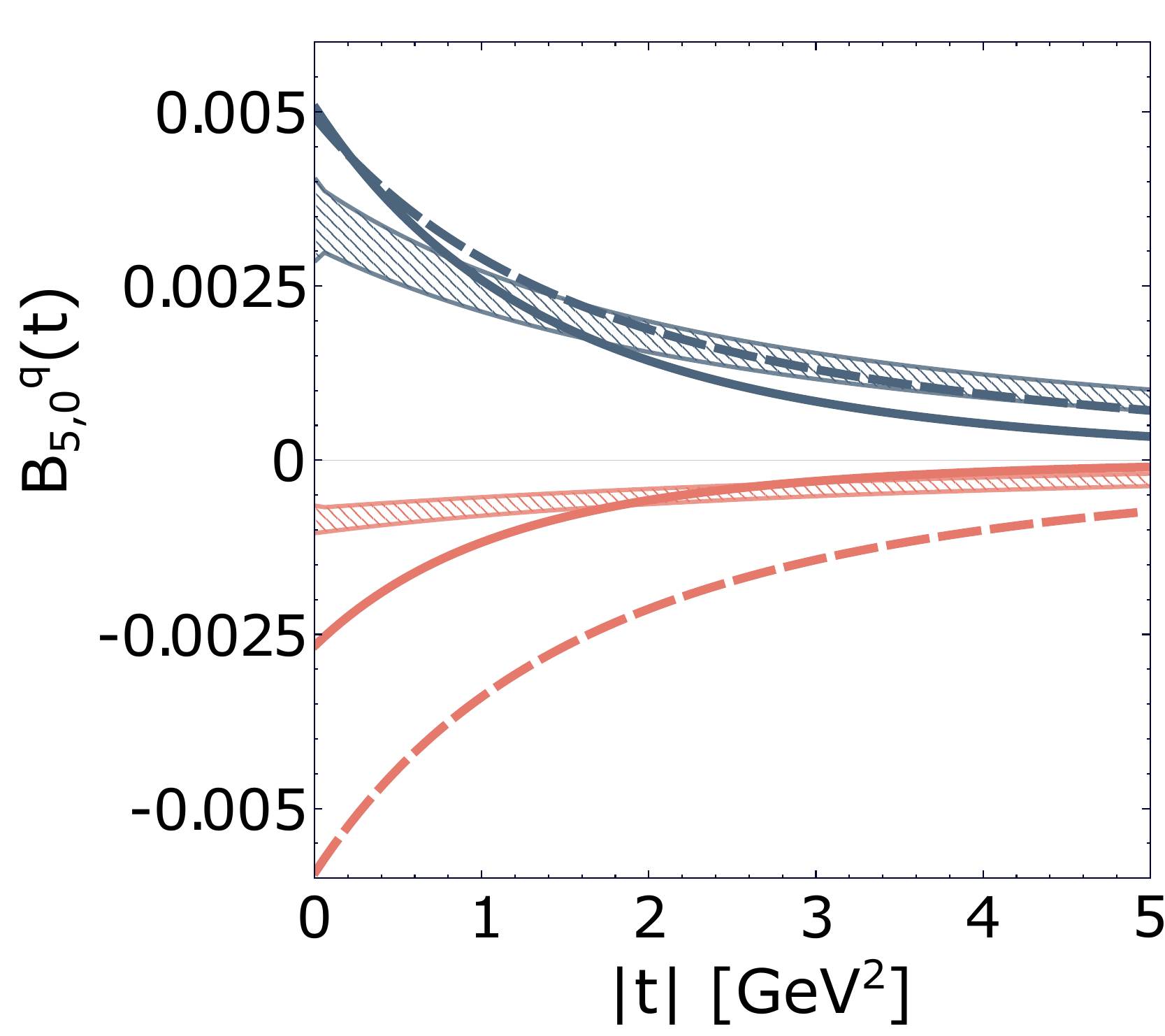}
   \includegraphics[height=0.28\textwidth]{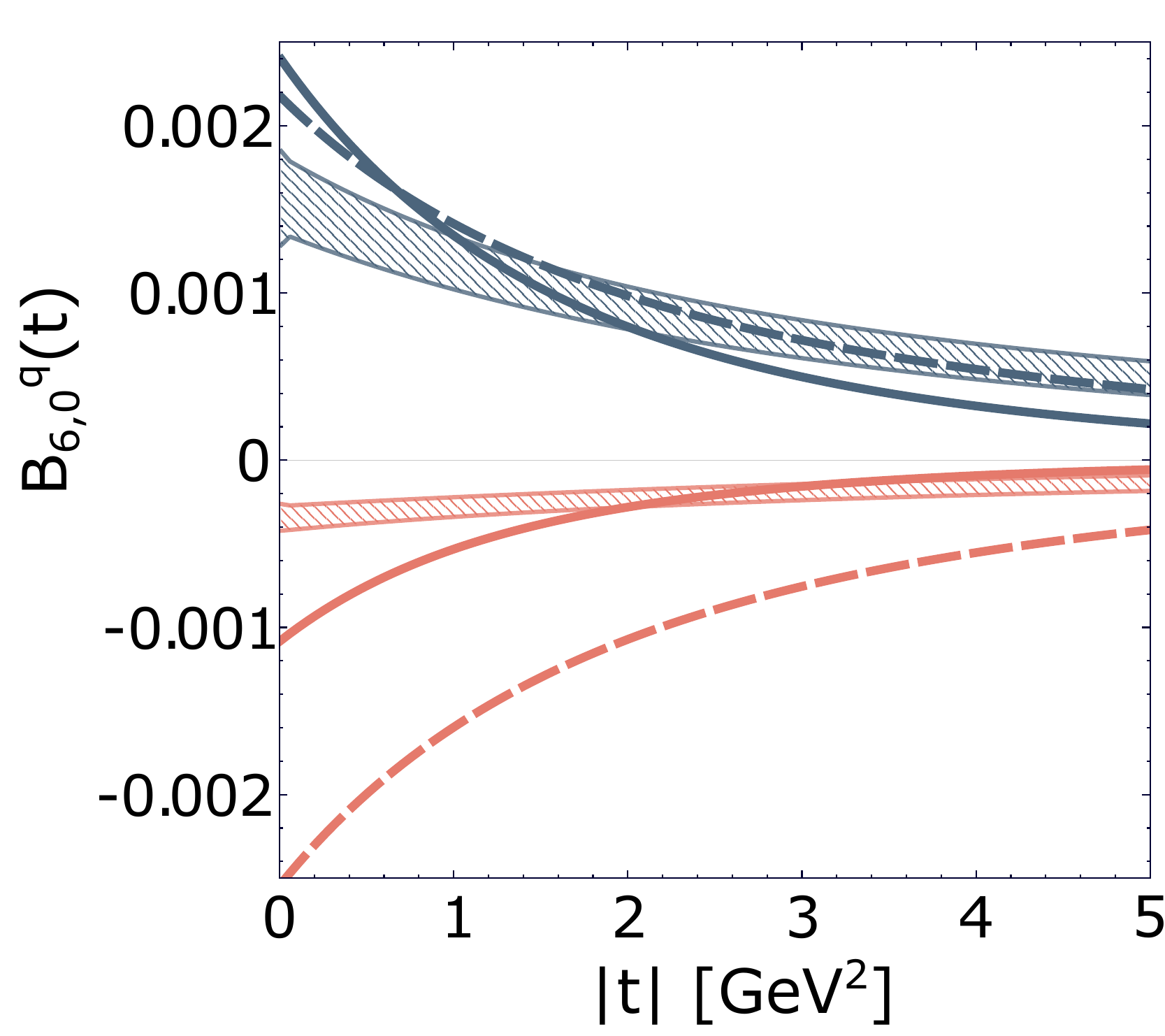}
  }

\end{figure}

\begin{figure}[!tp]
  \centering
  \caption{Total angular momentum of up and down valence quarks. Each point represents a single replica set used in this analysis. The green ellipse indicates the confidence level corresponding to $3\sigma$. Open markers are used for points outside the ellipse.} 
  \label{fig:ji}
  \includegraphics[width=0.33\textwidth]{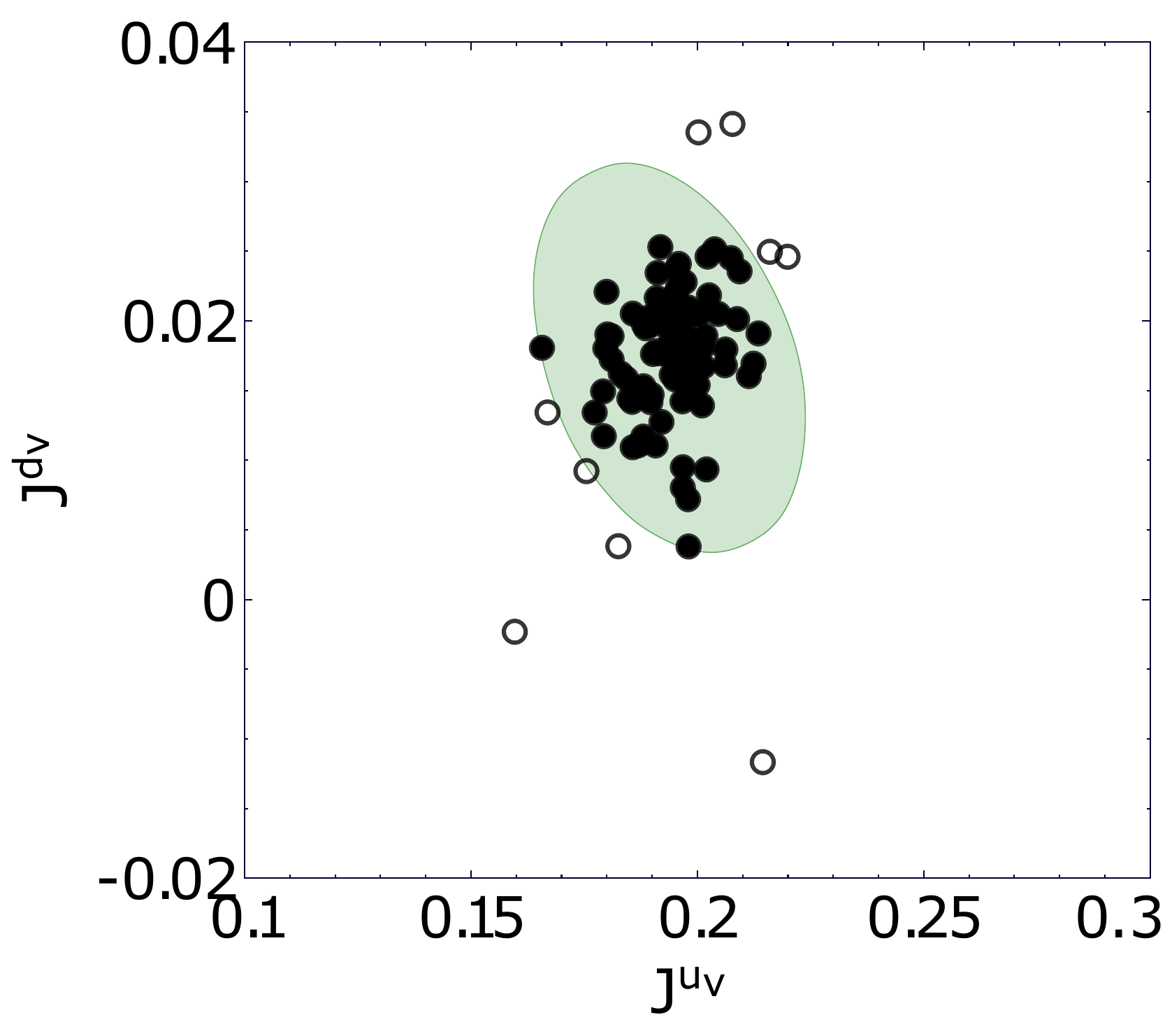}
\end{figure}
The tomographic images obtained from lattice QCD and elastic FF data are shown in Fig.~\ref{fig:nt} for both unpolarized and transversely polarized (along the $x$-axis) protons. This figure includes one-dimensional profiles, which allow us to present the uncertainties, as well as 2D distributions normalized so that a common color scale can be used for all images. The inflation of uncertainties near $b=0$ is due to the somewhat moderate constraint of the shadow term, primarily caused by the limited range in $t$ provided by the lattice-QCD data. The deviation from the bell shape is not observed within the estimated uncertainties. The shift in parton densities induced in a transversely polarized proton is clearly visible and, as expected, is opposite for up and down quarks. 

Finally, in Fig.~\ref{fig:mom}, we present the first six Mellin moments for both GPDs, $H^{q}$ and $E^{q}$. These plots may serve as a convenient tool for comparison between this work and the calculations of $A_{i,0}^{q}(t)$ and $B_{i,0}^{q}(t)$ quantities by other lattice QCD groups. Additionally, according to Eq.~\eqref{eq:basics_ji}, we evaluate total angular momenta of valence quarks from $A_{2,0}^{q}(0)$ and $B_{2,0}^{q}(0)$. The outcome is $J^{u_{v}} = 0.195 \pm 0.010$ and $J^{d_{v}} = 0.0173 \pm 0.0046$ evaluated at $\mu=2\,\mathrm{GeV}$. As proven by Fig.~\ref{fig:ji}, we observe only a small correlation between these two quantities in the extraction. The extracted values are similar to those obtained in Ref.~\cite{Diehl:2013xca}, i.e. $J^{u_{v}} = 0.230^{+0.009}_{-0.024}$ and $J^{d_{v}} = -0.004^{+0.010}_{-0.016}$, which is not entirely surprising, as our analysis shares several components with that one, in particular a similar Ansatz for the GPD $e_{q}(x)$ and the same selection of elastic data. Comparison with other analyses based on lattice QCD, such as Ref.~\cite{Bhattacharya:2023ays}, is difficult, because results specific to valence quarks are not provided.

%% file: sec_summary.tex
\section{Summary}
\label{sec:summary}

In this article, we explored the possibility of combining lattice QCD and elastic scattering data in a single analysis. Using the framework of GPDs, we extracted information on nucleon tomography and evaluated the total angular momentum of partons, however, only for valence quarks. Full information about sea quarks can be obtained by combining lattice QCD with exclusive scattering data, a possibility we plan to explore in the future. It should be noted, however, that at low-$\xi$, nucleon tomography information can be directly extracted from experimental data. This technique has been employed by HERA and CERN experiments (see~\cite{COMPASS:2018pup} and references therein), and will be further explored at the EIC~\cite{Aschenauer:2013hhw}. This highlights the strong complementarity between the current work on valence quarks and future analyses focusing on the sea component.

From the comparison between popular models of GPDs and parameterizations of PDFs and elastic FFs with lattice-QCD data, it is clear that it is not advantageous to straightforwardly use them in their original form in phenomenological applications. This is not surprising, and one can certainly expect the situation to improve over time. However, it seems that many sources of systematic uncertainty can be reduced by using the double ratios introduced in this work, though at the cost of losing direct information on PDFs and elastic FFs. This opens up the possibility of using lattice-QCD data now, which, among many obvious benefits, also provides much-needed feedback to lattice-QCD groups, stimulating improvements in future computations.
Clearly, the systematic uncertainties in lattice-extracted observables can be rigorously quantified or eliminated with additional calculations performed at multiple lattice spacings and volumes, directly at the physical pion mass, with increased nucleon boosts etc.

Finally, in this work, we introduced a new type of shadow GPDs, this time in the $(x,0,t)$ space, whereas previous applications of this concept focused on $(x,\xi)$ while neglecting the $t$-dependence. We demonstrated the usefulness of shadow terms in assessing model uncertainties and showed that they can be used to study nucleon tomography beyond the bell shape. The obtained results certainly still contain unknown model uncertainties, but this analysis is an important step toward a precise phenomenology of GPDs augmented with lattice QCD computations.

%% file: main.bbl
\begin{thebibliography}{108}%
\makeatletter
\providecommand \@ifxundefined [1]{%
 \@ifx{#1\undefined}
}%
\providecommand \@ifnum [1]{%
 \ifnum #1\expandafter \@firstoftwo
 \else \expandafter \@secondoftwo
 \fi
}%
\providecommand \@ifx [1]{%
 \ifx #1\expandafter \@firstoftwo
 \else \expandafter \@secondoftwo
 \fi
}%
\providecommand \natexlab [1]{#1}%
\providecommand \enquote  [1]{``#1''}%
\providecommand \bibnamefont  [1]{#1}%
\providecommand \bibfnamefont [1]{#1}%
\providecommand \citenamefont [1]{#1}%
\providecommand \href@noop [0]{\@secondoftwo}%
\providecommand \href [0]{\begingroup \@sanitize@url \@href}%
\providecommand \@href[1]{\@@startlink{#1}\@@href}%
\providecommand \@@href[1]{\endgroup#1\@@endlink}%
\providecommand \@sanitize@url [0]{\catcode `\\12\catcode `\$12\catcode `\&12\catcode `\#12\catcode `\^12\catcode `\_12\catcode `\%12\relax}%
\providecommand \@@startlink[1]{}%
\providecommand \@@endlink[0]{}%
\providecommand \url  [0]{\begingroup\@sanitize@url \@url }%
\providecommand \@url [1]{\endgroup\@href {#1}{\urlprefix }}%
\providecommand \urlprefix  [0]{URL }%
\providecommand \Eprint [0]{\href }%
\providecommand \doibase [0]{https://doi.org/}%
\providecommand \selectlanguage [0]{\@gobble}%
\providecommand \bibinfo  [0]{\@secondoftwo}%
\providecommand \bibfield  [0]{\@secondoftwo}%
\providecommand \translation [1]{[#1]}%
\providecommand \BibitemOpen [0]{}%
\providecommand \bibitemStop [0]{}%
\providecommand \bibitemNoStop [0]{.\EOS\space}%
\providecommand \EOS [0]{\spacefactor3000\relax}%
\providecommand \BibitemShut  [1]{\csname bibitem#1\endcsname}%
\let\auto@bib@innerbib\@empty
\bibitem [{\citenamefont {M\"uller}\ \emph {et~al.}(1994)\citenamefont {M\"uller}, \citenamefont {Robaschik}, \citenamefont {Geyer}, \citenamefont {Dittes},\ and\ \citenamefont {Ho\v{r}ej\v{s}i}}]{Muller:1994ses}%
  \BibitemOpen
  \bibfield  {author} {\bibinfo {author} {\bibfnamefont {D.}~\bibnamefont {M\"uller}}, \bibinfo {author} {\bibfnamefont {D.}~\bibnamefont {Robaschik}}, \bibinfo {author} {\bibfnamefont {B.}~\bibnamefont {Geyer}}, \bibinfo {author} {\bibfnamefont {F.~M.}\ \bibnamefont {Dittes}},\ and\ \bibinfo {author} {\bibfnamefont {J.}~\bibnamefont {Ho\v{r}ej\v{s}i}},\ }\bibfield  {title} {\bibinfo {title} {{Wave functions, evolution equations and evolution kernels from light ray operators of QCD}},\ }\href {https://doi.org/10.1002/prop.2190420202} {\bibfield  {journal} {\bibinfo  {journal} {Fortsch. Phys.}\ }\textbf {\bibinfo {volume} {42}},\ \bibinfo {pages} {101} (\bibinfo {year} {1994})},\ \Eprint {https://arxiv.org/abs/hep-ph/9812448} {arXiv:hep-ph/9812448} \BibitemShut {NoStop}%
\bibitem [{\citenamefont {Ji}(1997{\natexlab{a}})}]{Ji:1996ek}%
  \BibitemOpen
  \bibfield  {author} {\bibinfo {author} {\bibfnamefont {X.-D.}\ \bibnamefont {Ji}},\ }\bibfield  {title} {\bibinfo {title} {{Gauge-Invariant Decomposition of Nucleon Spin}},\ }\href {https://doi.org/10.1103/PhysRevLett.78.610} {\bibfield  {journal} {\bibinfo  {journal} {Phys. Rev. Lett.}\ }\textbf {\bibinfo {volume} {78}},\ \bibinfo {pages} {610} (\bibinfo {year} {1997}{\natexlab{a}})},\ \Eprint {https://arxiv.org/abs/hep-ph/9603249} {arXiv:hep-ph/9603249} \BibitemShut {NoStop}%
\bibitem [{\citenamefont {Ji}(1997{\natexlab{b}})}]{Ji:1996nm}%
  \BibitemOpen
  \bibfield  {author} {\bibinfo {author} {\bibfnamefont {X.-D.}\ \bibnamefont {Ji}},\ }\bibfield  {title} {\bibinfo {title} {{Deeply virtual Compton scattering}},\ }\href {https://doi.org/10.1103/PhysRevD.55.7114} {\bibfield  {journal} {\bibinfo  {journal} {Phys. Rev. D}\ }\textbf {\bibinfo {volume} {55}},\ \bibinfo {pages} {7114} (\bibinfo {year} {1997}{\natexlab{b}})},\ \Eprint {https://arxiv.org/abs/hep-ph/9609381} {arXiv:hep-ph/9609381} \BibitemShut {NoStop}%
\bibitem [{\citenamefont {Radyushkin}(1996)}]{Radyushkin:1996ru}%
  \BibitemOpen
  \bibfield  {author} {\bibinfo {author} {\bibfnamefont {A.~V.}\ \bibnamefont {Radyushkin}},\ }\bibfield  {title} {\bibinfo {title} {{Asymmetric gluon distributions and hard diffractive electroproduction}},\ }\href {https://doi.org/10.1016/0370-2693(96)00844-1} {\bibfield  {journal} {\bibinfo  {journal} {Phys. Lett. B}\ }\textbf {\bibinfo {volume} {385}},\ \bibinfo {pages} {333} (\bibinfo {year} {1996})},\ \Eprint {https://arxiv.org/abs/hep-ph/9605431} {arXiv:hep-ph/9605431} \BibitemShut {NoStop}%
\bibitem [{\citenamefont {Radyushkin}(1997)}]{Radyushkin:1997ki}%
  \BibitemOpen
  \bibfield  {author} {\bibinfo {author} {\bibfnamefont {A.~V.}\ \bibnamefont {Radyushkin}},\ }\bibfield  {title} {\bibinfo {title} {{Nonforward parton distributions}},\ }\href {https://doi.org/10.1103/PhysRevD.56.5524} {\bibfield  {journal} {\bibinfo  {journal} {Phys. Rev. D}\ }\textbf {\bibinfo {volume} {56}},\ \bibinfo {pages} {5524} (\bibinfo {year} {1997})},\ \Eprint {https://arxiv.org/abs/hep-ph/9704207} {arXiv:hep-ph/9704207} \BibitemShut {NoStop}%
\bibitem [{\citenamefont {Burkardt}(2000)}]{Burkardt:2000za}%
  \BibitemOpen
  \bibfield  {author} {\bibinfo {author} {\bibfnamefont {M.}~\bibnamefont {Burkardt}},\ }\bibfield  {title} {\bibinfo {title} {{Impact parameter dependent parton distributions and off forward parton distributions for $\zeta \to 0$}},\ }\href {https://doi.org/10.1103/PhysRevD.62.071503} {\bibfield  {journal} {\bibinfo  {journal} {Phys. Rev. D}\ }\textbf {\bibinfo {volume} {62}},\ \bibinfo {pages} {071503} (\bibinfo {year} {2000})},\ \bibinfo {note} {[Erratum: Phys. Rev. D 66, 119903 (2002)]},\ \Eprint {https://arxiv.org/abs/hep-ph/0005108} {arXiv:hep-ph/0005108} \BibitemShut {NoStop}%
\bibitem [{\citenamefont {Burkardt}(2003)}]{Burkardt:2002hr}%
  \BibitemOpen
  \bibfield  {author} {\bibinfo {author} {\bibfnamefont {M.}~\bibnamefont {Burkardt}},\ }\bibfield  {title} {\bibinfo {title} {{Impact parameter space interpretation for generalized parton distributions}},\ }\href {https://doi.org/10.1142/S0217751X03012370} {\bibfield  {journal} {\bibinfo  {journal} {Int. J. Mod. Phys. A}\ }\textbf {\bibinfo {volume} {18}},\ \bibinfo {pages} {173} (\bibinfo {year} {2003})},\ \Eprint {https://arxiv.org/abs/hep-ph/0207047} {arXiv:hep-ph/0207047} \BibitemShut {NoStop}%
\bibitem [{\citenamefont {Burkardt}(2004)}]{Burkardt:2004bv}%
  \BibitemOpen
  \bibfield  {author} {\bibinfo {author} {\bibfnamefont {M.}~\bibnamefont {Burkardt}},\ }\bibfield  {title} {\bibinfo {title} {{Generalized parton distributions for large x}},\ }\href {https://doi.org/10.1016/j.physletb.2004.05.070} {\bibfield  {journal} {\bibinfo  {journal} {Phys. Lett. B}\ }\textbf {\bibinfo {volume} {595}},\ \bibinfo {pages} {245} (\bibinfo {year} {2004})},\ \Eprint {https://arxiv.org/abs/hep-ph/0401159} {arXiv:hep-ph/0401159} \BibitemShut {NoStop}%
\bibitem [{\citenamefont {Goeke}\ \emph {et~al.}(2007)\citenamefont {Goeke}, \citenamefont {Grabis}, \citenamefont {Ossmann}, \citenamefont {Polyakov}, \citenamefont {Schweitzer}, \citenamefont {Silva},\ and\ \citenamefont {Urbano}}]{Goeke:2007fp}%
  \BibitemOpen
  \bibfield  {author} {\bibinfo {author} {\bibfnamefont {K.}~\bibnamefont {Goeke}}, \bibinfo {author} {\bibfnamefont {J.}~\bibnamefont {Grabis}}, \bibinfo {author} {\bibfnamefont {J.}~\bibnamefont {Ossmann}}, \bibinfo {author} {\bibfnamefont {M.~V.}\ \bibnamefont {Polyakov}}, \bibinfo {author} {\bibfnamefont {P.}~\bibnamefont {Schweitzer}}, \bibinfo {author} {\bibfnamefont {A.}~\bibnamefont {Silva}},\ and\ \bibinfo {author} {\bibfnamefont {D.}~\bibnamefont {Urbano}},\ }\bibfield  {title} {\bibinfo {title} {{Nucleon form-factors of the energy momentum tensor in the chiral quark-soliton model}},\ }\href {https://doi.org/10.1103/PhysRevD.75.094021} {\bibfield  {journal} {\bibinfo  {journal} {Phys. Rev. D}\ }\textbf {\bibinfo {volume} {75}},\ \bibinfo {pages} {094021} (\bibinfo {year} {2007})},\ \Eprint {https://arxiv.org/abs/hep-ph/0702030} {arXiv:hep-ph/0702030} \BibitemShut {NoStop}%
\bibitem [{\citenamefont {Polyakov}\ and\ \citenamefont {Schweitzer}(2018)}]{Polyakov:2018zvc}%
  \BibitemOpen
  \bibfield  {author} {\bibinfo {author} {\bibfnamefont {M.~V.}\ \bibnamefont {Polyakov}}\ and\ \bibinfo {author} {\bibfnamefont {P.}~\bibnamefont {Schweitzer}},\ }\bibfield  {title} {\bibinfo {title} {{Forces inside hadrons: pressure, surface tension, mechanical radius, and all that}},\ }\href {https://doi.org/10.1142/S0217751X18300259} {\bibfield  {journal} {\bibinfo  {journal} {Int. J. Mod. Phys. A}\ }\textbf {\bibinfo {volume} {33}},\ \bibinfo {pages} {1830025} (\bibinfo {year} {2018})},\ \Eprint {https://arxiv.org/abs/1805.06596} {arXiv:1805.06596 [hep-ph]} \BibitemShut {NoStop}%
\bibitem [{\citenamefont {Ji}(2013)}]{Ji:2013dva}%
  \BibitemOpen
  \bibfield  {author} {\bibinfo {author} {\bibfnamefont {X.}~\bibnamefont {Ji}},\ }\bibfield  {title} {\bibinfo {title} {{Parton Physics on a Euclidean Lattice}},\ }\href {https://doi.org/10.1103/PhysRevLett.110.262002} {\bibfield  {journal} {\bibinfo  {journal} {Phys. Rev. Lett.}\ }\textbf {\bibinfo {volume} {110}},\ \bibinfo {pages} {262002} (\bibinfo {year} {2013})},\ \Eprint {https://arxiv.org/abs/1305.1539} {arXiv:1305.1539 [hep-ph]} \BibitemShut {NoStop}%
\bibitem [{\citenamefont {Radyushkin}(2017)}]{Radyushkin:2017cyf}%
  \BibitemOpen
  \bibfield  {author} {\bibinfo {author} {\bibfnamefont {A.~V.}\ \bibnamefont {Radyushkin}},\ }\bibfield  {title} {\bibinfo {title} {{Quasi-parton distribution functions, momentum distributions, and pseudo-parton distribution functions}},\ }\href {https://doi.org/10.1103/PhysRevD.96.034025} {\bibfield  {journal} {\bibinfo  {journal} {Phys. Rev. D}\ }\textbf {\bibinfo {volume} {96}},\ \bibinfo {pages} {034025} (\bibinfo {year} {2017})},\ \Eprint {https://arxiv.org/abs/1705.01488} {arXiv:1705.01488 [hep-ph]} \BibitemShut {NoStop}%
\bibitem [{\citenamefont {Chatagnon}\ \emph {et~al.}(2021)\citenamefont {Chatagnon} \emph {et~al.}}]{CLAS:2021lky}%
  \BibitemOpen
  \bibfield  {author} {\bibinfo {author} {\bibfnamefont {P.}~\bibnamefont {Chatagnon}} \emph {et~al.} (\bibinfo {collaboration} {CLAS}),\ }\bibfield  {title} {\bibinfo {title} {{First Measurement of Timelike Compton Scattering}},\ }\href {https://doi.org/10.1103/PhysRevLett.127.262501} {\bibfield  {journal} {\bibinfo  {journal} {Phys. Rev. Lett.}\ }\textbf {\bibinfo {volume} {127}},\ \bibinfo {pages} {262501} (\bibinfo {year} {2021})},\ \Eprint {https://arxiv.org/abs/2108.11746} {arXiv:2108.11746 [hep-ex]} \BibitemShut {NoStop}%
\bibitem [{\citenamefont {Accardi}\ \emph {et~al.}(2023)\citenamefont {Accardi} \emph {et~al.}}]{Accardi:2023chb}%
  \BibitemOpen
  \bibfield  {author} {\bibinfo {author} {\bibfnamefont {A.}~\bibnamefont {Accardi}} \emph {et~al.},\ }\href@noop {} {\bibinfo {title} {{Strong Interaction Physics at the Luminosity Frontier with 22 GeV Electrons at Jefferson Lab}}} (\bibinfo {year} {2023}),\ \Eprint {https://arxiv.org/abs/2306.09360} {arXiv:2306.09360 [nucl-ex]} \BibitemShut {NoStop}%
\bibitem [{\citenamefont {Abdul~Khalek}\ \emph {et~al.}(2022)\citenamefont {Abdul~Khalek} \emph {et~al.}}]{AbdulKhalek:2021gbh}%
  \BibitemOpen
  \bibfield  {author} {\bibinfo {author} {\bibfnamefont {R.}~\bibnamefont {Abdul~Khalek}} \emph {et~al.},\ }\bibfield  {title} {\bibinfo {title} {{Science Requirements and Detector Concepts for the Electron-Ion Collider}: {EIC Yellow Report}},\ }\href {https://doi.org/10.1016/j.nuclphysa.2022.122447} {\bibfield  {journal} {\bibinfo  {journal} {Nucl. Phys. A}\ }\textbf {\bibinfo {volume} {1026}},\ \bibinfo {pages} {122447} (\bibinfo {year} {2022})},\ \Eprint {https://arxiv.org/abs/2103.05419} {arXiv:2103.05419 [physics.ins-det]} \BibitemShut {NoStop}%
\bibitem [{\citenamefont {Anderle}\ \emph {et~al.}(2021)\citenamefont {Anderle} \emph {et~al.}}]{Anderle:2021wcy}%
  \BibitemOpen
  \bibfield  {author} {\bibinfo {author} {\bibfnamefont {D.~P.}\ \bibnamefont {Anderle}} \emph {et~al.},\ }\bibfield  {title} {\bibinfo {title} {{Electron-ion collider in China}},\ }\href {https://doi.org/10.1007/s11467-021-1062-0} {\bibfield  {journal} {\bibinfo  {journal} {Front. Phys. (Beijing)}\ }\textbf {\bibinfo {volume} {16}},\ \bibinfo {pages} {64701} (\bibinfo {year} {2021})},\ \Eprint {https://arxiv.org/abs/2102.09222} {arXiv:2102.09222 [nucl-ex]} \BibitemShut {NoStop}%
\bibitem [{\citenamefont {Bertone}\ \emph {et~al.}(2021)\citenamefont {Bertone}, \citenamefont {Dutrieux}, \citenamefont {Mezrag}, \citenamefont {Moutarde},\ and\ \citenamefont {Sznajder}}]{Bertone:2021yyz}%
  \BibitemOpen
  \bibfield  {author} {\bibinfo {author} {\bibfnamefont {V.}~\bibnamefont {Bertone}}, \bibinfo {author} {\bibfnamefont {H.}~\bibnamefont {Dutrieux}}, \bibinfo {author} {\bibfnamefont {C.}~\bibnamefont {Mezrag}}, \bibinfo {author} {\bibfnamefont {H.}~\bibnamefont {Moutarde}},\ and\ \bibinfo {author} {\bibfnamefont {P.}~\bibnamefont {Sznajder}},\ }\bibfield  {title} {\bibinfo {title} {{Deconvolution problem of deeply virtual Compton scattering}},\ }\href {https://doi.org/10.1103/PhysRevD.103.114019} {\bibfield  {journal} {\bibinfo  {journal} {Phys. Rev. D}\ }\textbf {\bibinfo {volume} {103}},\ \bibinfo {pages} {114019} (\bibinfo {year} {2021})},\ \Eprint {https://arxiv.org/abs/2104.03836} {arXiv:2104.03836 [hep-ph]} \BibitemShut {NoStop}%
\bibitem [{\citenamefont {Moffat}\ \emph {et~al.}(2023)\citenamefont {Moffat}, \citenamefont {Freese}, \citenamefont {Clo\"et}, \citenamefont {Donohoe}, \citenamefont {Gamberg}, \citenamefont {Melnitchouk}, \citenamefont {Metz}, \citenamefont {Prokudin},\ and\ \citenamefont {Sato}}]{Moffat:2023svr}%
  \BibitemOpen
  \bibfield  {author} {\bibinfo {author} {\bibfnamefont {E.}~\bibnamefont {Moffat}}, \bibinfo {author} {\bibfnamefont {A.}~\bibnamefont {Freese}}, \bibinfo {author} {\bibfnamefont {I.}~\bibnamefont {Clo\"et}}, \bibinfo {author} {\bibfnamefont {T.}~\bibnamefont {Donohoe}}, \bibinfo {author} {\bibfnamefont {L.}~\bibnamefont {Gamberg}}, \bibinfo {author} {\bibfnamefont {W.}~\bibnamefont {Melnitchouk}}, \bibinfo {author} {\bibfnamefont {A.}~\bibnamefont {Metz}}, \bibinfo {author} {\bibfnamefont {A.}~\bibnamefont {Prokudin}},\ and\ \bibinfo {author} {\bibfnamefont {N.}~\bibnamefont {Sato}},\ }\bibfield  {title} {\bibinfo {title} {{Shedding light on shadow generalized parton distributions}},\ }\href {https://doi.org/10.1103/PhysRevD.108.036027} {\bibfield  {journal} {\bibinfo  {journal} {Phys. Rev. D}\ }\textbf {\bibinfo {volume} {108}},\ \bibinfo {pages} {036027} (\bibinfo {year} {2023})},\ \Eprint {https://arxiv.org/abs/2303.12006} {arXiv:2303.12006 [hep-ph]} \BibitemShut {NoStop}%
\bibitem [{\citenamefont {Qiu}\ and\ \citenamefont {Yu}(2023)}]{Qiu_2023}%
  \BibitemOpen
  \bibfield  {author} {\bibinfo {author} {\bibfnamefont {J.-W.}\ \bibnamefont {Qiu}}\ and\ \bibinfo {author} {\bibfnamefont {Z.}~\bibnamefont {Yu}},\ }\bibfield  {title} {\bibinfo {title} {Extraction of the parton momentum-fraction dependence of generalized parton distributions from exclusive photoproduction},\ }\bibfield  {journal} {\bibinfo  {journal} {Physical Review Letters}\ }\textbf {\bibinfo {volume} {131}},\ \href {https://doi.org/10.1103/physrevlett.131.161902} {10.1103/physrevlett.131.161902} (\bibinfo {year} {2023})\BibitemShut {NoStop}%
\bibitem [{\citenamefont {Deja}\ \emph {et~al.}(2023)\citenamefont {Deja}, \citenamefont {Martinez-Fernandez}, \citenamefont {Pire}, \citenamefont {Sznajder},\ and\ \citenamefont {Wagner}}]{Deja:2023ahc}%
  \BibitemOpen
  \bibfield  {author} {\bibinfo {author} {\bibfnamefont {K.}~\bibnamefont {Deja}}, \bibinfo {author} {\bibfnamefont {V.}~\bibnamefont {Martinez-Fernandez}}, \bibinfo {author} {\bibfnamefont {B.}~\bibnamefont {Pire}}, \bibinfo {author} {\bibfnamefont {P.}~\bibnamefont {Sznajder}},\ and\ \bibinfo {author} {\bibfnamefont {J.}~\bibnamefont {Wagner}},\ }\bibfield  {title} {\bibinfo {title} {{Phenomenology of double deeply virtual Compton scattering in the era of new experiments}},\ }\href {https://doi.org/10.1103/PhysRevD.107.094035} {\bibfield  {journal} {\bibinfo  {journal} {Phys. Rev. D}\ }\textbf {\bibinfo {volume} {107}},\ \bibinfo {pages} {094035} (\bibinfo {year} {2023})},\ \Eprint {https://arxiv.org/abs/2303.13668} {arXiv:2303.13668 [hep-ph]} \BibitemShut {NoStop}%
\bibitem [{\citenamefont {Grocholski}\ \emph {et~al.}(2021)\citenamefont {Grocholski}, \citenamefont {Pire}, \citenamefont {Sznajder}, \citenamefont {Szymanowski},\ and\ \citenamefont {Wagner}}]{Grocholski:2021man}%
  \BibitemOpen
  \bibfield  {author} {\bibinfo {author} {\bibfnamefont {O.}~\bibnamefont {Grocholski}}, \bibinfo {author} {\bibfnamefont {B.}~\bibnamefont {Pire}}, \bibinfo {author} {\bibfnamefont {P.}~\bibnamefont {Sznajder}}, \bibinfo {author} {\bibfnamefont {L.}~\bibnamefont {Szymanowski}},\ and\ \bibinfo {author} {\bibfnamefont {J.}~\bibnamefont {Wagner}},\ }\bibfield  {title} {\bibinfo {title} {{Collinear factorization of diphoton photoproduction at next to leading order}},\ }\href {https://doi.org/10.1103/PhysRevD.104.114006} {\bibfield  {journal} {\bibinfo  {journal} {Phys. Rev. D}\ }\textbf {\bibinfo {volume} {104}},\ \bibinfo {pages} {114006} (\bibinfo {year} {2021})},\ \Eprint {https://arxiv.org/abs/2110.00048} {arXiv:2110.00048 [hep-ph]} \BibitemShut {NoStop}%
\bibitem [{\citenamefont {Grocholski}\ \emph {et~al.}(2022)\citenamefont {Grocholski}, \citenamefont {Pire}, \citenamefont {Sznajder}, \citenamefont {Szymanowski},\ and\ \citenamefont {Wagner}}]{Grocholski:2022rqj}%
  \BibitemOpen
  \bibfield  {author} {\bibinfo {author} {\bibfnamefont {O.}~\bibnamefont {Grocholski}}, \bibinfo {author} {\bibfnamefont {B.}~\bibnamefont {Pire}}, \bibinfo {author} {\bibfnamefont {P.}~\bibnamefont {Sznajder}}, \bibinfo {author} {\bibfnamefont {L.}~\bibnamefont {Szymanowski}},\ and\ \bibinfo {author} {\bibfnamefont {J.}~\bibnamefont {Wagner}},\ }\bibfield  {title} {\bibinfo {title} {{Phenomenology of diphoton photoproduction at next-to-leading order}},\ }\href {https://doi.org/10.1103/PhysRevD.105.094025} {\bibfield  {journal} {\bibinfo  {journal} {Phys. Rev. D}\ }\textbf {\bibinfo {volume} {105}},\ \bibinfo {pages} {094025} (\bibinfo {year} {2022})},\ \Eprint {https://arxiv.org/abs/2204.00396} {arXiv:2204.00396 [hep-ph]} \BibitemShut {NoStop}%
\bibitem [{\citenamefont {Duplan\v{c}i\'c}\ \emph {et~al.}(2018)\citenamefont {Duplan\v{c}i\'c}, \citenamefont {Passek-Kumeri\v{c}ki}, \citenamefont {Pire}, \citenamefont {Szymanowski},\ and\ \citenamefont {Wallon}}]{Duplancic:2018bum}%
  \BibitemOpen
  \bibfield  {author} {\bibinfo {author} {\bibfnamefont {G.}~\bibnamefont {Duplan\v{c}i\'c}}, \bibinfo {author} {\bibfnamefont {K.}~\bibnamefont {Passek-Kumeri\v{c}ki}}, \bibinfo {author} {\bibfnamefont {B.}~\bibnamefont {Pire}}, \bibinfo {author} {\bibfnamefont {L.}~\bibnamefont {Szymanowski}},\ and\ \bibinfo {author} {\bibfnamefont {S.}~\bibnamefont {Wallon}},\ }\bibfield  {title} {\bibinfo {title} {{Probing axial quark generalized parton distributions through exclusive photoproduction of a $\gamma\,\pi^\pm$ pair with a large invariant mass}},\ }\href {https://doi.org/10.1007/JHEP11(2018)179} {\bibfield  {journal} {\bibinfo  {journal} {J. High Energy Phys.}\ }\textbf {\bibinfo {volume} {11}},\ \bibinfo {pages} {179}},\ \Eprint {https://arxiv.org/abs/1809.08104} {arXiv:1809.08104 [hep-ph]} \BibitemShut {NoStop}%
\bibitem [{\citenamefont {Duplan\v{c}i\'c}\ \emph {et~al.}(2023{\natexlab{a}})\citenamefont {Duplan\v{c}i\'c}, \citenamefont {Nabeebaccus}, \citenamefont {Passek-Kumeri\v{c}ki}, \citenamefont {Pire}, \citenamefont {Szymanowski},\ and\ \citenamefont {Wallon}}]{Duplancic:2022ffo}%
  \BibitemOpen
  \bibfield  {author} {\bibinfo {author} {\bibfnamefont {G.}~\bibnamefont {Duplan\v{c}i\'c}}, \bibinfo {author} {\bibfnamefont {S.}~\bibnamefont {Nabeebaccus}}, \bibinfo {author} {\bibfnamefont {K.}~\bibnamefont {Passek-Kumeri\v{c}ki}}, \bibinfo {author} {\bibfnamefont {B.}~\bibnamefont {Pire}}, \bibinfo {author} {\bibfnamefont {L.}~\bibnamefont {Szymanowski}},\ and\ \bibinfo {author} {\bibfnamefont {S.}~\bibnamefont {Wallon}},\ }\bibfield  {title} {\bibinfo {title} {{Accessing chiral-even quark generalised parton distributions in the exclusive photoproduction of a \ensuremath{\gamma}\ensuremath{\pi}$^{±}$ pair with large invariant mass in both fixed-target and collider experiments}},\ }\href {https://doi.org/10.1007/JHEP03(2023)241} {\bibfield  {journal} {\bibinfo  {journal} {J. High Energy Phys.}\ }\textbf {\bibinfo {volume} {03}},\ \bibinfo {pages} {241}},\ \Eprint {https://arxiv.org/abs/2212.00655} {arXiv:2212.00655 [hep-ph]} \BibitemShut {NoStop}%
\bibitem [{\citenamefont {Duplan\v{c}i\'c}\ \emph {et~al.}(2023{\natexlab{b}})\citenamefont {Duplan\v{c}i\'c}, \citenamefont {Nabeebaccus}, \citenamefont {Passek-Kumeri\v{c}ki}, \citenamefont {Pire}, \citenamefont {Szymanowski},\ and\ \citenamefont {Wallon}}]{Duplancic:2023kwe}%
  \BibitemOpen
  \bibfield  {author} {\bibinfo {author} {\bibfnamefont {G.}~\bibnamefont {Duplan\v{c}i\'c}}, \bibinfo {author} {\bibfnamefont {S.}~\bibnamefont {Nabeebaccus}}, \bibinfo {author} {\bibfnamefont {K.}~\bibnamefont {Passek-Kumeri\v{c}ki}}, \bibinfo {author} {\bibfnamefont {B.}~\bibnamefont {Pire}}, \bibinfo {author} {\bibfnamefont {L.}~\bibnamefont {Szymanowski}},\ and\ \bibinfo {author} {\bibfnamefont {S.}~\bibnamefont {Wallon}},\ }\bibfield  {title} {\bibinfo {title} {{Probing chiral-even and chiral-odd leading twist quark generalized parton distributions through the exclusive photoproduction of a \ensuremath{\gamma}\ensuremath{\rho} pair}},\ }\href {https://doi.org/10.1103/PhysRevD.107.094023} {\bibfield  {journal} {\bibinfo  {journal} {Phys. Rev. D}\ }\textbf {\bibinfo {volume} {107}},\ \bibinfo {pages} {094023} (\bibinfo {year} {2023}{\natexlab{b}})},\ \Eprint {https://arxiv.org/abs/2302.12026} {arXiv:2302.12026 [hep-ph]} \BibitemShut {NoStop}%
\bibitem [{\citenamefont {Boussarie}\ \emph {et~al.}(2017)\citenamefont {Boussarie}, \citenamefont {Pire}, \citenamefont {Szymanowski},\ and\ \citenamefont {Wallon}}]{Boussarie:2016qop}%
  \BibitemOpen
  \bibfield  {author} {\bibinfo {author} {\bibfnamefont {R.}~\bibnamefont {Boussarie}}, \bibinfo {author} {\bibfnamefont {B.}~\bibnamefont {Pire}}, \bibinfo {author} {\bibfnamefont {L.}~\bibnamefont {Szymanowski}},\ and\ \bibinfo {author} {\bibfnamefont {S.}~\bibnamefont {Wallon}},\ }\bibfield  {title} {\bibinfo {title} {{Exclusive photoproduction of a $\gamma\,\rho$ pair with a large invariant mass}},\ }\href {https://doi.org/10.1007/JHEP02(2017)054} {\bibfield  {journal} {\bibinfo  {journal} {J. High Energy Phys.}\ }\textbf {\bibinfo {volume} {02}},\ \bibinfo {pages} {054}},\ \bibinfo {note} {[Erratum: J. High Energy Phys. 10, 029 (2018)]},\ \Eprint {https://arxiv.org/abs/1609.03830} {arXiv:1609.03830 [hep-ph]} \BibitemShut {NoStop}%
\bibitem [{\citenamefont {Qiu}\ and\ \citenamefont {Yu}(2024)}]{Qiu:2024mny}%
  \BibitemOpen
  \bibfield  {author} {\bibinfo {author} {\bibfnamefont {J.-W.}\ \bibnamefont {Qiu}}\ and\ \bibinfo {author} {\bibfnamefont {Z.}~\bibnamefont {Yu}},\ }\bibfield  {title} {\bibinfo {title} {{Extracting transition generalized parton distributions from hard exclusive pion-nucleon scattering}},\ }\href {https://doi.org/10.1103/PhysRevD.109.074023} {\bibfield  {journal} {\bibinfo  {journal} {Phys. Rev. D}\ }\textbf {\bibinfo {volume} {109}},\ \bibinfo {pages} {074023} (\bibinfo {year} {2024})},\ \Eprint {https://arxiv.org/abs/2401.13207} {arXiv:2401.13207 [hep-ph]} \BibitemShut {NoStop}%
\bibitem [{\citenamefont {Ji}(2014)}]{Ji:2014gla}%
  \BibitemOpen
  \bibfield  {author} {\bibinfo {author} {\bibfnamefont {X.}~\bibnamefont {Ji}},\ }\bibfield  {title} {\bibinfo {title} {{Parton Physics from Large-Momentum Effective Field Theory}},\ }\href {https://doi.org/10.1007/s11433-014-5492-3} {\bibfield  {journal} {\bibinfo  {journal} {Sci. China Phys. Mech. Astron.}\ }\textbf {\bibinfo {volume} {57}},\ \bibinfo {pages} {1407} (\bibinfo {year} {2014})},\ \Eprint {https://arxiv.org/abs/1404.6680} {arXiv:1404.6680 [hep-ph]} \BibitemShut {NoStop}%
\bibitem [{\citenamefont {Liu}\ and\ \citenamefont {Dong}(1994)}]{Liu:1993cv}%
  \BibitemOpen
  \bibfield  {author} {\bibinfo {author} {\bibfnamefont {K.-F.}\ \bibnamefont {Liu}}\ and\ \bibinfo {author} {\bibfnamefont {S.-J.}\ \bibnamefont {Dong}},\ }\bibfield  {title} {\bibinfo {title} {{Origin of difference between anti-d and anti-u partons in the nucleon}},\ }\href {https://doi.org/10.1103/PhysRevLett.72.1790} {\bibfield  {journal} {\bibinfo  {journal} {Phys. Rev. Lett.}\ }\textbf {\bibinfo {volume} {72}},\ \bibinfo {pages} {1790} (\bibinfo {year} {1994})},\ \Eprint {https://arxiv.org/abs/hep-ph/9306299} {arXiv:hep-ph/9306299 [hep-ph]} \BibitemShut {NoStop}%
\bibitem [{\citenamefont {Detmold}\ and\ \citenamefont {Lin}(2006)}]{Detmold:2005gg}%
  \BibitemOpen
  \bibfield  {author} {\bibinfo {author} {\bibfnamefont {W.}~\bibnamefont {Detmold}}\ and\ \bibinfo {author} {\bibfnamefont {C.~J.~D.}\ \bibnamefont {Lin}},\ }\bibfield  {title} {\bibinfo {title} {{Deep-inelastic scattering and the operator product expansion in lattice QCD}},\ }\href {https://doi.org/10.1103/PhysRevD.73.014501} {\bibfield  {journal} {\bibinfo  {journal} {Phys. Rev.}\ }\textbf {\bibinfo {volume} {D73}},\ \bibinfo {pages} {014501} (\bibinfo {year} {2006})},\ \Eprint {https://arxiv.org/abs/hep-lat/0507007} {arXiv:hep-lat/0507007 [hep-lat]} \BibitemShut {NoStop}%
\bibitem [{\citenamefont {Braun}\ and\ \citenamefont {Mueller}(2008)}]{Braun:2007wv}%
  \BibitemOpen
  \bibfield  {author} {\bibinfo {author} {\bibfnamefont {V.}~\bibnamefont {Braun}}\ and\ \bibinfo {author} {\bibfnamefont {D.}~\bibnamefont {Mueller}},\ }\bibfield  {title} {\bibinfo {title} {{Exclusive processes in position space and the pion distribution amplitude}},\ }\href {https://doi.org/10.1140/epjc/s10052-008-0608-4} {\bibfield  {journal} {\bibinfo  {journal} {Eur. Phys. J.}\ }\textbf {\bibinfo {volume} {C55}},\ \bibinfo {pages} {349} (\bibinfo {year} {2008})},\ \Eprint {https://arxiv.org/abs/0709.1348} {arXiv:0709.1348 [hep-ph]} \BibitemShut {NoStop}%
\bibitem [{\citenamefont {Chambers}\ \emph {et~al.}(2017)\citenamefont {Chambers} \emph {et~al.}}]{Chambers:2017dov}%
  \BibitemOpen
  \bibfield  {author} {\bibinfo {author} {\bibfnamefont {A.~J.}\ \bibnamefont {Chambers}} \emph {et~al.},\ }\bibfield  {title} {\bibinfo {title} {{Nucleon Structure Functions from Operator Product Expansion on the Lattice}},\ }\href {https://doi.org/10.1103/PhysRevLett.118.242001} {\bibfield  {journal} {\bibinfo  {journal} {Phys. Rev. Lett.}\ }\textbf {\bibinfo {volume} {118}},\ \bibinfo {pages} {242001} (\bibinfo {year} {2017})},\ \Eprint {https://arxiv.org/abs/1703.01153} {arXiv:1703.01153 [hep-lat]} \BibitemShut {NoStop}%
\bibitem [{\citenamefont {Ma}\ and\ \citenamefont {Qiu}(2018)}]{Ma:2017pxb}%
  \BibitemOpen
  \bibfield  {author} {\bibinfo {author} {\bibfnamefont {Y.-Q.}\ \bibnamefont {Ma}}\ and\ \bibinfo {author} {\bibfnamefont {J.-W.}\ \bibnamefont {Qiu}},\ }\bibfield  {title} {\bibinfo {title} {{Exploring Partonic Structure of Hadrons Using ab initio Lattice QCD Calculations}},\ }\href {https://doi.org/10.1103/PhysRevLett.120.022003} {\bibfield  {journal} {\bibinfo  {journal} {Phys. Rev. Lett.}\ }\textbf {\bibinfo {volume} {120}},\ \bibinfo {pages} {022003} (\bibinfo {year} {2018})},\ \Eprint {https://arxiv.org/abs/1709.03018} {arXiv:1709.03018 [hep-ph]} \BibitemShut {NoStop}%
\bibitem [{\citenamefont {Cichy}\ and\ \citenamefont {Constantinou}(2019)}]{Cichy:2018mum}%
  \BibitemOpen
  \bibfield  {author} {\bibinfo {author} {\bibfnamefont {K.}~\bibnamefont {Cichy}}\ and\ \bibinfo {author} {\bibfnamefont {M.}~\bibnamefont {Constantinou}},\ }\bibfield  {title} {\bibinfo {title} {{A guide to light-cone PDFs from Lattice QCD: an overview of approaches, techniques and results}},\ }\href {https://doi.org/10.1155/2019/3036904} {\bibfield  {journal} {\bibinfo  {journal} {Adv. High Energy Phys.}\ }\textbf {\bibinfo {volume} {2019}},\ \bibinfo {pages} {3036904} (\bibinfo {year} {2019})},\ \Eprint {https://arxiv.org/abs/1811.07248} {arXiv:1811.07248 [hep-lat]} \BibitemShut {NoStop}%
\bibitem [{\citenamefont {Radyushkin}(2020)}]{Radyushkin:2019mye}%
  \BibitemOpen
  \bibfield  {author} {\bibinfo {author} {\bibfnamefont {A.}~\bibnamefont {Radyushkin}},\ }\bibfield  {title} {\bibinfo {title} {{Theory and applications of parton pseudodistributions}},\ }\href {https://doi.org/10.1142/S0217751X20300021} {\bibfield  {journal} {\bibinfo  {journal} {Int. J. Mod. Phys. A}\ }\textbf {\bibinfo {volume} {35}},\ \bibinfo {pages} {2030002} (\bibinfo {year} {2020})},\ \Eprint {https://arxiv.org/abs/1912.04244} {arXiv:1912.04244 [hep-ph]} \BibitemShut {NoStop}%
\bibitem [{\citenamefont {Ji}\ \emph {et~al.}(2021)\citenamefont {Ji}, \citenamefont {Liu}, \citenamefont {Liu}, \citenamefont {Zhang},\ and\ \citenamefont {Zhao}}]{Ji:2020ect}%
  \BibitemOpen
  \bibfield  {author} {\bibinfo {author} {\bibfnamefont {X.}~\bibnamefont {Ji}}, \bibinfo {author} {\bibfnamefont {Y.-S.}\ \bibnamefont {Liu}}, \bibinfo {author} {\bibfnamefont {Y.}~\bibnamefont {Liu}}, \bibinfo {author} {\bibfnamefont {J.-H.}\ \bibnamefont {Zhang}},\ and\ \bibinfo {author} {\bibfnamefont {Y.}~\bibnamefont {Zhao}},\ }\bibfield  {title} {\bibinfo {title} {{Large-momentum effective theory}},\ }\href {https://doi.org/10.1103/RevModPhys.93.035005} {\bibfield  {journal} {\bibinfo  {journal} {Rev. Mod. Phys.}\ }\textbf {\bibinfo {volume} {93}},\ \bibinfo {pages} {035005} (\bibinfo {year} {2021})},\ \Eprint {https://arxiv.org/abs/2004.03543} {arXiv:2004.03543 [hep-ph]} \BibitemShut {NoStop}%
\bibitem [{\citenamefont {Constantinou}(2021)}]{Constantinou:2020pek}%
  \BibitemOpen
  \bibfield  {author} {\bibinfo {author} {\bibfnamefont {M.}~\bibnamefont {Constantinou}},\ }\bibfield  {title} {\bibinfo {title} {{The x-dependence of hadronic parton distributions: A review on the progress of lattice QCD}},\ }\href {https://doi.org/10.1140/epja/s10050-021-00353-7} {\bibfield  {journal} {\bibinfo  {journal} {Eur. Phys. J. A}\ }\textbf {\bibinfo {volume} {57}},\ \bibinfo {pages} {77} (\bibinfo {year} {2021})},\ \Eprint {https://arxiv.org/abs/2010.02445} {arXiv:2010.02445 [hep-lat]} \BibitemShut {NoStop}%
\bibitem [{\citenamefont {Cichy}(2022{\natexlab{a}})}]{Cichy:2021lih}%
  \BibitemOpen
  \bibfield  {author} {\bibinfo {author} {\bibfnamefont {K.}~\bibnamefont {Cichy}},\ }\bibfield  {title} {\bibinfo {title} {{Progress in $x$-dependent partonic distributions from lattice QCD}},\ }in\ \href {https://doi.org/10.22323/1.396.0017} {\emph {\bibinfo {booktitle} {Proceedings of The 38th International Symposium on Lattice Field Theory {\textemdash} PoS(LATTICE2021)}}},\ Vol.\ \bibinfo {volume} {396}\ (\bibinfo {year} {2022})\ p.\ \bibinfo {pages} {017}\BibitemShut {NoStop}%
\bibitem [{\citenamefont {Cichy}(2022{\natexlab{b}})}]{Cichy:2021ewm}%
  \BibitemOpen
  \bibfield  {author} {\bibinfo {author} {\bibfnamefont {K.}~\bibnamefont {Cichy}},\ }\bibfield  {title} {\bibinfo {title} {{Overview of lattice calculations of the x-dependence of PDFs, GPDs and TMDs}},\ }\href {https://doi.org/10.1051/epjconf/202225801005} {\bibfield  {journal} {\bibinfo  {journal} {EPJ Web Conf.}\ }\textbf {\bibinfo {volume} {258}},\ \bibinfo {pages} {01005} (\bibinfo {year} {2022}{\natexlab{b}})},\ \Eprint {https://arxiv.org/abs/2111.04552} {arXiv:2111.04552 [hep-lat]} \BibitemShut {NoStop}%
\bibitem [{\citenamefont {Chen}\ \emph {et~al.}(2020)\citenamefont {Chen}, \citenamefont {Lin},\ and\ \citenamefont {Zhang}}]{Chen:2019lcm}%
  \BibitemOpen
  \bibfield  {author} {\bibinfo {author} {\bibfnamefont {J.-W.}\ \bibnamefont {Chen}}, \bibinfo {author} {\bibfnamefont {H.-W.}\ \bibnamefont {Lin}},\ and\ \bibinfo {author} {\bibfnamefont {J.-H.}\ \bibnamefont {Zhang}},\ }\bibfield  {title} {\bibinfo {title} {{Pion generalized parton distribution from lattice QCD}},\ }\href {https://doi.org/10.1016/j.nuclphysb.2020.114940} {\bibfield  {journal} {\bibinfo  {journal} {Nucl. Phys. B}\ }\textbf {\bibinfo {volume} {952}},\ \bibinfo {pages} {114940} (\bibinfo {year} {2020})},\ \Eprint {https://arxiv.org/abs/1904.12376} {arXiv:1904.12376 [hep-lat]} \BibitemShut {NoStop}%
\bibitem [{\citenamefont {Alexandrou}\ \emph {et~al.}(2020)\citenamefont {Alexandrou}, \citenamefont {Cichy}, \citenamefont {Constantinou}, \citenamefont {Hadjiyiannakou}, \citenamefont {Jansen}, \citenamefont {Scapellato},\ and\ \citenamefont {Steffens}}]{Alexandrou:2020zbe}%
  \BibitemOpen
  \bibfield  {author} {\bibinfo {author} {\bibfnamefont {C.}~\bibnamefont {Alexandrou}}, \bibinfo {author} {\bibfnamefont {K.}~\bibnamefont {Cichy}}, \bibinfo {author} {\bibfnamefont {M.}~\bibnamefont {Constantinou}}, \bibinfo {author} {\bibfnamefont {K.}~\bibnamefont {Hadjiyiannakou}}, \bibinfo {author} {\bibfnamefont {K.}~\bibnamefont {Jansen}}, \bibinfo {author} {\bibfnamefont {A.}~\bibnamefont {Scapellato}},\ and\ \bibinfo {author} {\bibfnamefont {F.}~\bibnamefont {Steffens}},\ }\bibfield  {title} {\bibinfo {title} {{Unpolarized and helicity generalized parton distributions of the proton within lattice QCD}},\ }\href {https://doi.org/10.1103/PhysRevLett.125.262001} {\bibfield  {journal} {\bibinfo  {journal} {Phys. Rev. Lett.}\ }\textbf {\bibinfo {volume} {125}},\ \bibinfo {pages} {262001} (\bibinfo {year} {2020})},\ \Eprint {https://arxiv.org/abs/2008.10573} {arXiv:2008.10573 [hep-lat]} \BibitemShut {NoStop}%
\bibitem [{\citenamefont {Alexandrou}\ \emph {et~al.}(2022)\citenamefont {Alexandrou}, \citenamefont {Cichy}, \citenamefont {Constantinou}, \citenamefont {Hadjiyiannakou}, \citenamefont {Jansen}, \citenamefont {Scapellato},\ and\ \citenamefont {Steffens}}]{Alexandrou:2021bbo}%
  \BibitemOpen
  \bibfield  {author} {\bibinfo {author} {\bibfnamefont {C.}~\bibnamefont {Alexandrou}}, \bibinfo {author} {\bibfnamefont {K.}~\bibnamefont {Cichy}}, \bibinfo {author} {\bibfnamefont {M.}~\bibnamefont {Constantinou}}, \bibinfo {author} {\bibfnamefont {K.}~\bibnamefont {Hadjiyiannakou}}, \bibinfo {author} {\bibfnamefont {K.}~\bibnamefont {Jansen}}, \bibinfo {author} {\bibfnamefont {A.}~\bibnamefont {Scapellato}},\ and\ \bibinfo {author} {\bibfnamefont {F.}~\bibnamefont {Steffens}},\ }\bibfield  {title} {\bibinfo {title} {{Transversity GPDs of the proton from lattice QCD}},\ }\href {https://doi.org/10.1103/PhysRevD.105.034501} {\bibfield  {journal} {\bibinfo  {journal} {Phys. Rev. D}\ }\textbf {\bibinfo {volume} {105}},\ \bibinfo {pages} {034501} (\bibinfo {year} {2022})},\ \Eprint {https://arxiv.org/abs/2108.10789} {arXiv:2108.10789 [hep-lat]} \BibitemShut {NoStop}%
\bibitem [{\citenamefont {Bhattacharya}\ \emph {et~al.}(2022)\citenamefont {Bhattacharya}, \citenamefont {Cichy}, \citenamefont {Constantinou}, \citenamefont {Dodson}, \citenamefont {Gao}, \citenamefont {Metz}, \citenamefont {Mukherjee}, \citenamefont {Scapellato}, \citenamefont {Steffens},\ and\ \citenamefont {Zhao}}]{Bhattacharya:2022aob}%
  \BibitemOpen
  \bibfield  {author} {\bibinfo {author} {\bibfnamefont {S.}~\bibnamefont {Bhattacharya}}, \bibinfo {author} {\bibfnamefont {K.}~\bibnamefont {Cichy}}, \bibinfo {author} {\bibfnamefont {M.}~\bibnamefont {Constantinou}}, \bibinfo {author} {\bibfnamefont {J.}~\bibnamefont {Dodson}}, \bibinfo {author} {\bibfnamefont {X.}~\bibnamefont {Gao}}, \bibinfo {author} {\bibfnamefont {A.}~\bibnamefont {Metz}}, \bibinfo {author} {\bibfnamefont {S.}~\bibnamefont {Mukherjee}}, \bibinfo {author} {\bibfnamefont {A.}~\bibnamefont {Scapellato}}, \bibinfo {author} {\bibfnamefont {F.}~\bibnamefont {Steffens}},\ and\ \bibinfo {author} {\bibfnamefont {Y.}~\bibnamefont {Zhao}},\ }\bibfield  {title} {\bibinfo {title} {{Generalized parton distributions from lattice QCD with asymmetric momentum transfer: Unpolarized quarks}},\ }\href {https://doi.org/10.1103/PhysRevD.106.114512} {\bibfield  {journal} {\bibinfo  {journal} {Phys. Rev. D}\ }\textbf {\bibinfo {volume} {106}},\ \bibinfo {pages} {114512} (\bibinfo {year} {2022})},\ \Eprint
  {https://arxiv.org/abs/2209.05373} {arXiv:2209.05373 [hep-lat]} \BibitemShut {NoStop}%
\bibitem [{\citenamefont {Cichy}\ \emph {et~al.}(2023)\citenamefont {Cichy} \emph {et~al.}}]{Cichy:2023dgk}%
  \BibitemOpen
  \bibfield  {author} {\bibinfo {author} {\bibfnamefont {K.}~\bibnamefont {Cichy}} \emph {et~al.},\ }\bibfield  {title} {\bibinfo {title} {{Generalized Parton Distributions from Lattice QCD}},\ }\href {https://doi.org/10.5506/APhysPolBSupp.16.7-A6} {\bibfield  {journal} {\bibinfo  {journal} {Acta Phys. Polon. Supp.}\ }\textbf {\bibinfo {volume} {16}},\ \bibinfo {pages} {7} (\bibinfo {year} {2023})},\ \Eprint {https://arxiv.org/abs/2304.14970} {arXiv:2304.14970 [hep-lat]} \BibitemShut {NoStop}%
\bibitem [{\citenamefont {Bhattacharya}\ \emph {et~al.}(2023{\natexlab{a}})\citenamefont {Bhattacharya}, \citenamefont {Cichy}, \citenamefont {Constantinou}, \citenamefont {Gao}, \citenamefont {Metz}, \citenamefont {Miller}, \citenamefont {Mukherjee}, \citenamefont {Petreczky}, \citenamefont {Steffens},\ and\ \citenamefont {Zhao}}]{Bhattacharya:2023ays}%
  \BibitemOpen
  \bibfield  {author} {\bibinfo {author} {\bibfnamefont {S.}~\bibnamefont {Bhattacharya}}, \bibinfo {author} {\bibfnamefont {K.}~\bibnamefont {Cichy}}, \bibinfo {author} {\bibfnamefont {M.}~\bibnamefont {Constantinou}}, \bibinfo {author} {\bibfnamefont {X.}~\bibnamefont {Gao}}, \bibinfo {author} {\bibfnamefont {A.}~\bibnamefont {Metz}}, \bibinfo {author} {\bibfnamefont {J.}~\bibnamefont {Miller}}, \bibinfo {author} {\bibfnamefont {S.}~\bibnamefont {Mukherjee}}, \bibinfo {author} {\bibfnamefont {P.}~\bibnamefont {Petreczky}}, \bibinfo {author} {\bibfnamefont {F.}~\bibnamefont {Steffens}},\ and\ \bibinfo {author} {\bibfnamefont {Y.}~\bibnamefont {Zhao}},\ }\bibfield  {title} {\bibinfo {title} {{Moments of proton GPDs from the OPE of nonlocal quark bilinears up to NNLO}},\ }\href {https://doi.org/10.1103/PhysRevD.108.014507} {\bibfield  {journal} {\bibinfo  {journal} {Phys. Rev. D}\ }\textbf {\bibinfo {volume} {108}},\ \bibinfo {pages} {014507} (\bibinfo {year} {2023}{\natexlab{a}})},\ \Eprint
  {https://arxiv.org/abs/2305.11117} {arXiv:2305.11117 [hep-lat]} \BibitemShut {NoStop}%
\bibitem [{\citenamefont {Bhattacharya}\ \emph {et~al.}(2023{\natexlab{b}})\citenamefont {Bhattacharya}, \citenamefont {Cichy}, \citenamefont {Constantinou}, \citenamefont {Dodson}, \citenamefont {Metz}, \citenamefont {Scapellato},\ and\ \citenamefont {Steffens}}]{Bhattacharya:2023nmv}%
  \BibitemOpen
  \bibfield  {author} {\bibinfo {author} {\bibfnamefont {S.}~\bibnamefont {Bhattacharya}}, \bibinfo {author} {\bibfnamefont {K.}~\bibnamefont {Cichy}}, \bibinfo {author} {\bibfnamefont {M.}~\bibnamefont {Constantinou}}, \bibinfo {author} {\bibfnamefont {J.}~\bibnamefont {Dodson}}, \bibinfo {author} {\bibfnamefont {A.}~\bibnamefont {Metz}}, \bibinfo {author} {\bibfnamefont {A.}~\bibnamefont {Scapellato}},\ and\ \bibinfo {author} {\bibfnamefont {F.}~\bibnamefont {Steffens}},\ }\bibfield  {title} {\bibinfo {title} {{Chiral-even axial twist-3 GPDs of the proton from lattice QCD}},\ }\href {https://doi.org/10.1103/PhysRevD.108.054501} {\bibfield  {journal} {\bibinfo  {journal} {Phys. Rev. D}\ }\textbf {\bibinfo {volume} {108}},\ \bibinfo {pages} {054501} (\bibinfo {year} {2023}{\natexlab{b}})},\ \Eprint {https://arxiv.org/abs/2306.05533} {arXiv:2306.05533 [hep-lat]} \BibitemShut {NoStop}%
\bibitem [{\citenamefont {Bhattacharya}\ \emph {et~al.}(2024{\natexlab{a}})\citenamefont {Bhattacharya} \emph {et~al.}}]{Bhattacharya:2023jsc}%
  \BibitemOpen
  \bibfield  {author} {\bibinfo {author} {\bibfnamefont {S.}~\bibnamefont {Bhattacharya}} \emph {et~al.},\ }\bibfield  {title} {\bibinfo {title} {{Generalized parton distributions from lattice QCD with asymmetric momentum transfer: Axial-vector case}},\ }\href {https://doi.org/10.1103/PhysRevD.109.034508} {\bibfield  {journal} {\bibinfo  {journal} {Phys. Rev. D}\ }\textbf {\bibinfo {volume} {109}},\ \bibinfo {pages} {034508} (\bibinfo {year} {2024}{\natexlab{a}})},\ \Eprint {https://arxiv.org/abs/2310.13114} {arXiv:2310.13114 [hep-lat]} \BibitemShut {NoStop}%
\bibitem [{\citenamefont {Bhattacharya}\ \emph {et~al.}(2024{\natexlab{b}})\citenamefont {Bhattacharya}, \citenamefont {Cichy}, \citenamefont {Constantinou}, \citenamefont {Metz}, \citenamefont {Nurminen},\ and\ \citenamefont {Steffens}}]{Bhattacharya:2024qpp}%
  \BibitemOpen
  \bibfield  {author} {\bibinfo {author} {\bibfnamefont {S.}~\bibnamefont {Bhattacharya}}, \bibinfo {author} {\bibfnamefont {K.}~\bibnamefont {Cichy}}, \bibinfo {author} {\bibfnamefont {M.}~\bibnamefont {Constantinou}}, \bibinfo {author} {\bibfnamefont {A.}~\bibnamefont {Metz}}, \bibinfo {author} {\bibfnamefont {N.}~\bibnamefont {Nurminen}},\ and\ \bibinfo {author} {\bibfnamefont {F.}~\bibnamefont {Steffens}},\ }\bibfield  {title} {\bibinfo {title} {{Generalized parton distributions from the pseudodistribution approach on the lattice}},\ }\href {https://doi.org/10.1103/PhysRevD.110.054502} {\bibfield  {journal} {\bibinfo  {journal} {Phys. Rev. D}\ }\textbf {\bibinfo {volume} {110}},\ \bibinfo {pages} {054502} (\bibinfo {year} {2024}{\natexlab{b}})},\ \Eprint {https://arxiv.org/abs/2405.04414} {arXiv:2405.04414 [hep-lat]} \BibitemShut {NoStop}%
\bibitem [{\citenamefont {Dutrieux}\ \emph {et~al.}(2024)\citenamefont {Dutrieux}, \citenamefont {Edwards}, \citenamefont {Egerer}, \citenamefont {Karpie}, \citenamefont {Monahan}, \citenamefont {Orginos}, \citenamefont {Radyushkin}, \citenamefont {Richards}, \citenamefont {Romero},\ and\ \citenamefont {Zafeiropoulos}}]{HadStruc:2024rix}%
  \BibitemOpen
  \bibfield  {author} {\bibinfo {author} {\bibfnamefont {H.}~\bibnamefont {Dutrieux}}, \bibinfo {author} {\bibfnamefont {R.~G.}\ \bibnamefont {Edwards}}, \bibinfo {author} {\bibfnamefont {C.}~\bibnamefont {Egerer}}, \bibinfo {author} {\bibfnamefont {J.}~\bibnamefont {Karpie}}, \bibinfo {author} {\bibfnamefont {C.}~\bibnamefont {Monahan}}, \bibinfo {author} {\bibfnamefont {K.}~\bibnamefont {Orginos}}, \bibinfo {author} {\bibfnamefont {A.}~\bibnamefont {Radyushkin}}, \bibinfo {author} {\bibfnamefont {D.}~\bibnamefont {Richards}}, \bibinfo {author} {\bibfnamefont {E.}~\bibnamefont {Romero}},\ and\ \bibinfo {author} {\bibfnamefont {S.}~\bibnamefont {Zafeiropoulos}} (\bibinfo {collaboration} {HadStruc}),\ }\bibfield  {title} {\bibinfo {title} {{Towards unpolarized GPDs from pseudo-distributions}},\ }\href {https://doi.org/10.1007/JHEP08(2024)162} {\bibfield  {journal} {\bibinfo  {journal} {JHEP}\ }\textbf {\bibinfo {volume} {08}},\ \bibinfo {pages} {162}},\ \Eprint {https://arxiv.org/abs/2405.10304}
  {arXiv:2405.10304 [hep-lat]} \BibitemShut {NoStop}%
\bibitem [{\citenamefont {Ding}\ \emph {et~al.}(2024)\citenamefont {Ding}, \citenamefont {Gao}, \citenamefont {Mukherjee}, \citenamefont {Petreczky}, \citenamefont {Shi}, \citenamefont {Syritsyn},\ and\ \citenamefont {Zhao}}]{Ding:2024hkz}%
  \BibitemOpen
  \bibfield  {author} {\bibinfo {author} {\bibfnamefont {H.-T.}\ \bibnamefont {Ding}}, \bibinfo {author} {\bibfnamefont {X.}~\bibnamefont {Gao}}, \bibinfo {author} {\bibfnamefont {S.}~\bibnamefont {Mukherjee}}, \bibinfo {author} {\bibfnamefont {P.}~\bibnamefont {Petreczky}}, \bibinfo {author} {\bibfnamefont {Q.}~\bibnamefont {Shi}}, \bibinfo {author} {\bibfnamefont {S.}~\bibnamefont {Syritsyn}},\ and\ \bibinfo {author} {\bibfnamefont {Y.}~\bibnamefont {Zhao}},\ }\href@noop {} {\bibinfo {title} {{Three-dimensional Imaging of Pion using Lattice QCD: Generalized Parton Distributions}}} (\bibinfo {year} {2024}),\ \Eprint {https://arxiv.org/abs/2407.03516} {arXiv:2407.03516 [hep-lat]} \BibitemShut {NoStop}%
\bibitem [{\citenamefont {Diehl}\ \emph {et~al.}(2005)\citenamefont {Diehl}, \citenamefont {Feldmann}, \citenamefont {Jakob},\ and\ \citenamefont {Kroll}}]{Diehl:2004cx}%
  \BibitemOpen
  \bibfield  {author} {\bibinfo {author} {\bibfnamefont {M.}~\bibnamefont {Diehl}}, \bibinfo {author} {\bibfnamefont {T.}~\bibnamefont {Feldmann}}, \bibinfo {author} {\bibfnamefont {R.}~\bibnamefont {Jakob}},\ and\ \bibinfo {author} {\bibfnamefont {P.}~\bibnamefont {Kroll}},\ }\bibfield  {title} {\bibinfo {title} {{Generalized parton distributions from nucleon form-factor data}},\ }\href {https://doi.org/10.1140/epjc/s2004-02063-4} {\bibfield  {journal} {\bibinfo  {journal} {Eur. Phys. J. C}\ }\textbf {\bibinfo {volume} {39}},\ \bibinfo {pages} {1} (\bibinfo {year} {2005})},\ \Eprint {https://arxiv.org/abs/hep-ph/0408173} {arXiv:hep-ph/0408173} \BibitemShut {NoStop}%
\bibitem [{\citenamefont {Burkert}\ \emph {et~al.}(2023)\citenamefont {Burkert}, \citenamefont {Elouadrhiri}, \citenamefont {Girod}, \citenamefont {Lorc\'e}, \citenamefont {Schweitzer},\ and\ \citenamefont {Shanahan}}]{Burkert:2023wzr}%
  \BibitemOpen
  \bibfield  {author} {\bibinfo {author} {\bibfnamefont {V.~D.}\ \bibnamefont {Burkert}}, \bibinfo {author} {\bibfnamefont {L.}~\bibnamefont {Elouadrhiri}}, \bibinfo {author} {\bibfnamefont {F.~X.}\ \bibnamefont {Girod}}, \bibinfo {author} {\bibfnamefont {C.}~\bibnamefont {Lorc\'e}}, \bibinfo {author} {\bibfnamefont {P.}~\bibnamefont {Schweitzer}},\ and\ \bibinfo {author} {\bibfnamefont {P.~E.}\ \bibnamefont {Shanahan}},\ }\bibfield  {title} {\bibinfo {title} {{Colloquium: Gravitational form factors of the proton}},\ }\href {https://doi.org/10.1103/RevModPhys.95.041002} {\bibfield  {journal} {\bibinfo  {journal} {Rev. Mod. Phys.}\ }\textbf {\bibinfo {volume} {95}},\ \bibinfo {pages} {041002} (\bibinfo {year} {2023})},\ \Eprint {https://arxiv.org/abs/2303.08347} {arXiv:2303.08347 [hep-ph]} \BibitemShut {NoStop}%
\bibitem [{\citenamefont {Alexandrou}\ \emph {et~al.}(2018{\natexlab{a}})\citenamefont {Alexandrou} \emph {et~al.}}]{Alexandrou:2018egz}%
  \BibitemOpen
  \bibfield  {author} {\bibinfo {author} {\bibfnamefont {C.}~\bibnamefont {Alexandrou}} \emph {et~al.},\ }\bibfield  {title} {\bibinfo {title} {{Simulating twisted mass fermions at physical light, strange and charm quark masses}},\ }\href {https://doi.org/10.1103/PhysRevD.98.054518} {\bibfield  {journal} {\bibinfo  {journal} {Phys. Rev.}\ }\textbf {\bibinfo {volume} {D98}},\ \bibinfo {pages} {054518} (\bibinfo {year} {2018}{\natexlab{a}})},\ \Eprint {https://arxiv.org/abs/1807.00495} {arXiv:1807.00495 [hep-lat]} \BibitemShut {NoStop}%
\bibitem [{\citenamefont {Alexandrou}\ \emph {et~al.}(2021)\citenamefont {Alexandrou}, \citenamefont {Constantinou}, \citenamefont {Hadjiyiannakou}, \citenamefont {Jansen},\ and\ \citenamefont {Manigrasso}}]{Alexandrou:2021oih}%
  \BibitemOpen
  \bibfield  {author} {\bibinfo {author} {\bibfnamefont {C.}~\bibnamefont {Alexandrou}}, \bibinfo {author} {\bibfnamefont {M.}~\bibnamefont {Constantinou}}, \bibinfo {author} {\bibfnamefont {K.}~\bibnamefont {Hadjiyiannakou}}, \bibinfo {author} {\bibfnamefont {K.}~\bibnamefont {Jansen}},\ and\ \bibinfo {author} {\bibfnamefont {F.}~\bibnamefont {Manigrasso}},\ }\bibfield  {title} {\bibinfo {title} {{Flavor decomposition of the nucleon unpolarized, helicity, and transversity parton distribution functions from lattice QCD simulations}},\ }\href {https://doi.org/10.1103/PhysRevD.104.054503} {\bibfield  {journal} {\bibinfo  {journal} {Phys. Rev. D}\ }\textbf {\bibinfo {volume} {104}},\ \bibinfo {pages} {054503} (\bibinfo {year} {2021})},\ \Eprint {https://arxiv.org/abs/2106.16065} {arXiv:2106.16065 [hep-lat]} \BibitemShut {NoStop}%
\bibitem [{\citenamefont {Orginos}\ \emph {et~al.}(2017)\citenamefont {Orginos}, \citenamefont {Radyushkin}, \citenamefont {Karpie},\ and\ \citenamefont {Zafeiropoulos}}]{Orginos:2017kos}%
  \BibitemOpen
  \bibfield  {author} {\bibinfo {author} {\bibfnamefont {K.}~\bibnamefont {Orginos}}, \bibinfo {author} {\bibfnamefont {A.}~\bibnamefont {Radyushkin}}, \bibinfo {author} {\bibfnamefont {J.}~\bibnamefont {Karpie}},\ and\ \bibinfo {author} {\bibfnamefont {S.}~\bibnamefont {Zafeiropoulos}},\ }\bibfield  {title} {\bibinfo {title} {{Lattice QCD exploration of parton pseudo-distribution functions}},\ }\href {https://doi.org/10.1103/PhysRevD.96.094503} {\bibfield  {journal} {\bibinfo  {journal} {Phys. Rev.}\ }\textbf {\bibinfo {volume} {D96}},\ \bibinfo {pages} {094503} (\bibinfo {year} {2017})},\ \Eprint {https://arxiv.org/abs/1706.05373} {arXiv:1706.05373 [hep-ph]} \BibitemShut {NoStop}%
\bibitem [{\citenamefont {Radyushkin}(2018)}]{Radyushkin:2018cvn}%
  \BibitemOpen
  \bibfield  {author} {\bibinfo {author} {\bibfnamefont {A.}~\bibnamefont {Radyushkin}},\ }\bibfield  {title} {\bibinfo {title} {{One-loop evolution of parton pseudo-distribution functions on the lattice}},\ }\href {https://doi.org/10.1103/PhysRevD.98.014019} {\bibfield  {journal} {\bibinfo  {journal} {Phys. Rev.}\ }\textbf {\bibinfo {volume} {D98}},\ \bibinfo {pages} {014019} (\bibinfo {year} {2018})},\ \Eprint {https://arxiv.org/abs/1801.02427} {arXiv:1801.02427 [hep-ph]} \BibitemShut {NoStop}%
\bibitem [{\citenamefont {Zhang}\ \emph {et~al.}(2018)\citenamefont {Zhang}, \citenamefont {Chen},\ and\ \citenamefont {Monahan}}]{Zhang:2018ggy}%
  \BibitemOpen
  \bibfield  {author} {\bibinfo {author} {\bibfnamefont {J.-H.}\ \bibnamefont {Zhang}}, \bibinfo {author} {\bibfnamefont {J.-W.}\ \bibnamefont {Chen}},\ and\ \bibinfo {author} {\bibfnamefont {C.}~\bibnamefont {Monahan}},\ }\bibfield  {title} {\bibinfo {title} {{Parton distribution functions from reduced Ioffe-time distributions}},\ }\href {https://doi.org/10.1103/PhysRevD.97.074508} {\bibfield  {journal} {\bibinfo  {journal} {Phys. Rev. D}\ }\textbf {\bibinfo {volume} {97}},\ \bibinfo {pages} {074508} (\bibinfo {year} {2018})},\ \Eprint {https://arxiv.org/abs/1801.03023} {arXiv:1801.03023 [hep-ph]} \BibitemShut {NoStop}%
\bibitem [{\citenamefont {Izubuchi}\ \emph {et~al.}(2018)\citenamefont {Izubuchi}, \citenamefont {Ji}, \citenamefont {Jin}, \citenamefont {Stewart},\ and\ \citenamefont {Zhao}}]{Izubuchi:2018srq}%
  \BibitemOpen
  \bibfield  {author} {\bibinfo {author} {\bibfnamefont {T.}~\bibnamefont {Izubuchi}}, \bibinfo {author} {\bibfnamefont {X.}~\bibnamefont {Ji}}, \bibinfo {author} {\bibfnamefont {L.}~\bibnamefont {Jin}}, \bibinfo {author} {\bibfnamefont {I.~W.}\ \bibnamefont {Stewart}},\ and\ \bibinfo {author} {\bibfnamefont {Y.}~\bibnamefont {Zhao}},\ }\bibfield  {title} {\bibinfo {title} {{Factorization Theorem Relating Euclidean and Light-Cone Parton Distributions}},\ }\href {https://doi.org/10.1103/PhysRevD.98.056004} {\bibfield  {journal} {\bibinfo  {journal} {Phys. Rev.}\ }\textbf {\bibinfo {volume} {D98}},\ \bibinfo {pages} {056004} (\bibinfo {year} {2018})},\ \Eprint {https://arxiv.org/abs/1801.03917} {arXiv:1801.03917 [hep-ph]} \BibitemShut {NoStop}%
\bibitem [{\citenamefont {Radyushkin}(2019)}]{Radyushkin:2018nbf}%
  \BibitemOpen
  \bibfield  {author} {\bibinfo {author} {\bibfnamefont {A.}~\bibnamefont {Radyushkin}},\ }\bibfield  {title} {\bibinfo {title} {{Structure of parton quasi-distributions and their moments}},\ }\href {https://doi.org/10.1016/j.physletb.2018.11.047} {\bibfield  {journal} {\bibinfo  {journal} {Phys. Lett. B}\ }\textbf {\bibinfo {volume} {788}},\ \bibinfo {pages} {380} (\bibinfo {year} {2019})},\ \Eprint {https://arxiv.org/abs/1807.07509} {arXiv:1807.07509 [hep-ph]} \BibitemShut {NoStop}%
\bibitem [{\citenamefont {Li}\ \emph {et~al.}(2021)\citenamefont {Li}, \citenamefont {Ma},\ and\ \citenamefont {Qiu}}]{Li:2020xml}%
  \BibitemOpen
  \bibfield  {author} {\bibinfo {author} {\bibfnamefont {Z.-Y.}\ \bibnamefont {Li}}, \bibinfo {author} {\bibfnamefont {Y.-Q.}\ \bibnamefont {Ma}},\ and\ \bibinfo {author} {\bibfnamefont {J.-W.}\ \bibnamefont {Qiu}},\ }\bibfield  {title} {\bibinfo {title} {{Extraction of Next-to-Next-to-Leading-Order Parton Distribution Functions from Lattice QCD Calculations}},\ }\href {https://doi.org/10.1103/PhysRevLett.126.072001} {\bibfield  {journal} {\bibinfo  {journal} {Phys. Rev. Lett.}\ }\textbf {\bibinfo {volume} {126}},\ \bibinfo {pages} {072001} (\bibinfo {year} {2021})},\ \Eprint {https://arxiv.org/abs/2006.12370} {arXiv:2006.12370 [hep-ph]} \BibitemShut {NoStop}%
\bibitem [{\citenamefont {Bhat}\ \emph {et~al.}(2022)\citenamefont {Bhat}, \citenamefont {Chomicki}, \citenamefont {Cichy}, \citenamefont {Constantinou}, \citenamefont {Green},\ and\ \citenamefont {Scapellato}}]{Bhat:2022zrw}%
  \BibitemOpen
  \bibfield  {author} {\bibinfo {author} {\bibfnamefont {M.}~\bibnamefont {Bhat}}, \bibinfo {author} {\bibfnamefont {W.}~\bibnamefont {Chomicki}}, \bibinfo {author} {\bibfnamefont {K.}~\bibnamefont {Cichy}}, \bibinfo {author} {\bibfnamefont {M.}~\bibnamefont {Constantinou}}, \bibinfo {author} {\bibfnamefont {J.~R.}\ \bibnamefont {Green}},\ and\ \bibinfo {author} {\bibfnamefont {A.}~\bibnamefont {Scapellato}},\ }\bibfield  {title} {\bibinfo {title} {{Continuum limit of parton distribution functions from the pseudodistribution approach on the lattice}},\ }\href {https://doi.org/10.1103/PhysRevD.106.054504} {\bibfield  {journal} {\bibinfo  {journal} {Phys. Rev. D}\ }\textbf {\bibinfo {volume} {106}},\ \bibinfo {pages} {054504} (\bibinfo {year} {2022})},\ \Eprint {https://arxiv.org/abs/2205.07585} {arXiv:2205.07585 [hep-lat]} \BibitemShut {NoStop}%
\bibitem [{\citenamefont {Goloskokov}\ and\ \citenamefont {Kroll}(2007)}]{Goloskokov:2006hr}%
  \BibitemOpen
  \bibfield  {author} {\bibinfo {author} {\bibfnamefont {S.~V.}\ \bibnamefont {Goloskokov}}\ and\ \bibinfo {author} {\bibfnamefont {P.}~\bibnamefont {Kroll}},\ }\bibfield  {title} {\bibinfo {title} {{The Longitudinal cross-section of vector meson electroproduction}},\ }\href {https://doi.org/10.1140/epjc/s10052-007-0228-4} {\bibfield  {journal} {\bibinfo  {journal} {Eur. Phys. J. C}\ }\textbf {\bibinfo {volume} {50}},\ \bibinfo {pages} {829} (\bibinfo {year} {2007})},\ \Eprint {https://arxiv.org/abs/hep-ph/0611290} {arXiv:hep-ph/0611290} \BibitemShut {NoStop}%
\bibitem [{\citenamefont {Goloskokov}\ and\ \citenamefont {Kroll}(2009)}]{Goloskokov:2008ib}%
  \BibitemOpen
  \bibfield  {author} {\bibinfo {author} {\bibfnamefont {S.~V.}\ \bibnamefont {Goloskokov}}\ and\ \bibinfo {author} {\bibfnamefont {P.}~\bibnamefont {Kroll}},\ }\bibfield  {title} {\bibinfo {title} {{The Target asymmetry in hard vector-meson electroproduction and parton angular momenta}},\ }\href {https://doi.org/10.1140/epjc/s10052-008-0833-x} {\bibfield  {journal} {\bibinfo  {journal} {Eur. Phys. J. C}\ }\textbf {\bibinfo {volume} {59}},\ \bibinfo {pages} {809} (\bibinfo {year} {2009})},\ \Eprint {https://arxiv.org/abs/0809.4126} {arXiv:0809.4126 [hep-ph]} \BibitemShut {NoStop}%
\bibitem [{\citenamefont {Vanderhaeghen}\ \emph {et~al.}(1999)\citenamefont {Vanderhaeghen}, \citenamefont {Guichon},\ and\ \citenamefont {Guidal}}]{Vanderhaeghen:1999xj}%
  \BibitemOpen
  \bibfield  {author} {\bibinfo {author} {\bibfnamefont {M.}~\bibnamefont {Vanderhaeghen}}, \bibinfo {author} {\bibfnamefont {P.~A.~M.}\ \bibnamefont {Guichon}},\ and\ \bibinfo {author} {\bibfnamefont {M.}~\bibnamefont {Guidal}},\ }\bibfield  {title} {\bibinfo {title} {{Deeply virtual electroproduction of photons and mesons on the nucleon: Leading order amplitudes and power corrections}},\ }\href {https://doi.org/10.1103/PhysRevD.60.094017} {\bibfield  {journal} {\bibinfo  {journal} {Phys. Rev. D}\ }\textbf {\bibinfo {volume} {60}},\ \bibinfo {pages} {094017} (\bibinfo {year} {1999})},\ \Eprint {https://arxiv.org/abs/hep-ph/9905372} {arXiv:hep-ph/9905372} \BibitemShut {NoStop}%
\bibitem [{\citenamefont {Berthou}\ \emph {et~al.}(2018)\citenamefont {Berthou} \emph {et~al.}}]{Berthou:2015oaw}%
  \BibitemOpen
  \bibfield  {author} {\bibinfo {author} {\bibfnamefont {B.}~\bibnamefont {Berthou}} \emph {et~al.},\ }\bibfield  {title} {\bibinfo {title} {{PARTONS: PARtonic Tomography Of Nucleon Software}: {A computing framework for the phenomenology of Generalized Parton Distributions}},\ }\href {https://doi.org/10.1140/epjc/s10052-018-5948-0} {\bibfield  {journal} {\bibinfo  {journal} {Eur. Phys. J. C}\ }\textbf {\bibinfo {volume} {78}},\ \bibinfo {pages} {478} (\bibinfo {year} {2018})},\ \Eprint {https://arxiv.org/abs/1512.06174} {arXiv:1512.06174 [hep-ph]} \BibitemShut {NoStop}%
\bibitem [{\citenamefont {Moutarde}\ \emph {et~al.}(2018)\citenamefont {Moutarde}, \citenamefont {Sznajder},\ and\ \citenamefont {Wagner}}]{Moutarde:2018kwr}%
  \BibitemOpen
  \bibfield  {author} {\bibinfo {author} {\bibfnamefont {H.}~\bibnamefont {Moutarde}}, \bibinfo {author} {\bibfnamefont {P.}~\bibnamefont {Sznajder}},\ and\ \bibinfo {author} {\bibfnamefont {J.}~\bibnamefont {Wagner}},\ }\bibfield  {title} {\bibinfo {title} {{Border and skewness functions from a leading order fit to DVCS data}},\ }\href {https://doi.org/10.1140/epjc/s10052-018-6359-y} {\bibfield  {journal} {\bibinfo  {journal} {Eur. Phys. J. C}\ }\textbf {\bibinfo {volume} {78}},\ \bibinfo {pages} {890} (\bibinfo {year} {2018})},\ \Eprint {https://arxiv.org/abs/1807.07620} {arXiv:1807.07620 [hep-ph]} \BibitemShut {NoStop}%
\bibitem [{\citenamefont {Pumplin}\ \emph {et~al.}(2002)\citenamefont {Pumplin}, \citenamefont {Stump}, \citenamefont {Huston}, \citenamefont {Lai}, \citenamefont {Nadolsky},\ and\ \citenamefont {Tung}}]{Pumplin:2002vw}%
  \BibitemOpen
  \bibfield  {author} {\bibinfo {author} {\bibfnamefont {J.}~\bibnamefont {Pumplin}}, \bibinfo {author} {\bibfnamefont {D.~R.}\ \bibnamefont {Stump}}, \bibinfo {author} {\bibfnamefont {J.}~\bibnamefont {Huston}}, \bibinfo {author} {\bibfnamefont {H.~L.}\ \bibnamefont {Lai}}, \bibinfo {author} {\bibfnamefont {P.~M.}\ \bibnamefont {Nadolsky}},\ and\ \bibinfo {author} {\bibfnamefont {W.~K.}\ \bibnamefont {Tung}},\ }\bibfield  {title} {\bibinfo {title} {{New generation of parton distributions with uncertainties from global QCD analysis}},\ }\href {https://doi.org/10.1088/1126-6708/2002/07/012} {\bibfield  {journal} {\bibinfo  {journal} {JHEP}\ }\textbf {\bibinfo {volume} {07}},\ \bibinfo {pages} {012}},\ \Eprint {https://arxiv.org/abs/hep-ph/0201195} {arXiv:hep-ph/0201195} \BibitemShut {NoStop}%
\bibitem [{\citenamefont {Martin}\ \emph {et~al.}(2009)\citenamefont {Martin}, \citenamefont {Stirling}, \citenamefont {Thorne},\ and\ \citenamefont {Watt}}]{Martin:2009iq}%
  \BibitemOpen
  \bibfield  {author} {\bibinfo {author} {\bibfnamefont {A.~D.}\ \bibnamefont {Martin}}, \bibinfo {author} {\bibfnamefont {W.~J.}\ \bibnamefont {Stirling}}, \bibinfo {author} {\bibfnamefont {R.~S.}\ \bibnamefont {Thorne}},\ and\ \bibinfo {author} {\bibfnamefont {G.}~\bibnamefont {Watt}},\ }\bibfield  {title} {\bibinfo {title} {{Parton distributions for the LHC}},\ }\href {https://doi.org/10.1140/epjc/s10052-009-1072-5} {\bibfield  {journal} {\bibinfo  {journal} {Eur. Phys. J. C}\ }\textbf {\bibinfo {volume} {63}},\ \bibinfo {pages} {189} (\bibinfo {year} {2009})},\ \Eprint {https://arxiv.org/abs/0901.0002} {arXiv:0901.0002 [hep-ph]} \BibitemShut {NoStop}%
\bibitem [{\citenamefont {Buckley}\ \emph {et~al.}(2015)\citenamefont {Buckley}, \citenamefont {Ferrando}, \citenamefont {Lloyd}, \citenamefont {Nordstr\"om}, \citenamefont {Page}, \citenamefont {R\"ufenacht}, \citenamefont {Sch\"onherr},\ and\ \citenamefont {Watt}}]{Buckley:2014ana}%
  \BibitemOpen
  \bibfield  {author} {\bibinfo {author} {\bibfnamefont {A.}~\bibnamefont {Buckley}}, \bibinfo {author} {\bibfnamefont {J.}~\bibnamefont {Ferrando}}, \bibinfo {author} {\bibfnamefont {S.}~\bibnamefont {Lloyd}}, \bibinfo {author} {\bibfnamefont {K.}~\bibnamefont {Nordstr\"om}}, \bibinfo {author} {\bibfnamefont {B.}~\bibnamefont {Page}}, \bibinfo {author} {\bibfnamefont {M.}~\bibnamefont {R\"ufenacht}}, \bibinfo {author} {\bibfnamefont {M.}~\bibnamefont {Sch\"onherr}},\ and\ \bibinfo {author} {\bibfnamefont {G.}~\bibnamefont {Watt}},\ }\bibfield  {title} {\bibinfo {title} {{LHAPDF6: parton density access in the LHC precision era}},\ }\href {https://doi.org/10.1140/epjc/s10052-015-3318-8} {\bibfield  {journal} {\bibinfo  {journal} {Eur. Phys. J. C}\ }\textbf {\bibinfo {volume} {75}},\ \bibinfo {pages} {132} (\bibinfo {year} {2015})},\ \Eprint {https://arxiv.org/abs/1412.7420} {arXiv:1412.7420 [hep-ph]} \BibitemShut {NoStop}%
\bibitem [{\citenamefont {Ball}\ \emph {et~al.}(2015)\citenamefont {Ball} \emph {et~al.}}]{NNPDF:2014otw}%
  \BibitemOpen
  \bibfield  {author} {\bibinfo {author} {\bibfnamefont {R.~D.}\ \bibnamefont {Ball}} \emph {et~al.} (\bibinfo {collaboration} {NNPDF}),\ }\bibfield  {title} {\bibinfo {title} {{Parton distributions for the LHC Run II}},\ }\href {https://doi.org/10.1007/JHEP04(2015)040} {\bibfield  {journal} {\bibinfo  {journal} {JHEP}\ }\textbf {\bibinfo {volume} {04}},\ \bibinfo {pages} {040}},\ \Eprint {https://arxiv.org/abs/1410.8849} {arXiv:1410.8849 [hep-ph]} \BibitemShut {NoStop}%
\bibitem [{\citenamefont {Kumeri\v{c}ki}\ and\ \citenamefont {M\"uller}(2016)}]{Kumericki:2015lhb}%
  \BibitemOpen
  \bibfield  {author} {\bibinfo {author} {\bibfnamefont {K.}~\bibnamefont {Kumeri\v{c}ki}}\ and\ \bibinfo {author} {\bibfnamefont {D.}~\bibnamefont {M\"uller}},\ }\bibfield  {title} {\bibinfo {title} {{Description and interpretation of DVCS measurements}},\ }\href {https://doi.org/10.1051/epjconf/201611201012} {\bibfield  {journal} {\bibinfo  {journal} {EPJ Web Conf.}\ }\textbf {\bibinfo {volume} {112}},\ \bibinfo {pages} {01012} (\bibinfo {year} {2016})},\ \Eprint {https://arxiv.org/abs/1512.09014} {arXiv:1512.09014 [hep-ph]} \BibitemShut {NoStop}%
\bibitem [{\citenamefont {Polyakov}\ and\ \citenamefont {Shuvaev}(2002)}]{Polyakov:2002wz}%
  \BibitemOpen
  \bibfield  {author} {\bibinfo {author} {\bibfnamefont {M.~V.}\ \bibnamefont {Polyakov}}\ and\ \bibinfo {author} {\bibfnamefont {A.~G.}\ \bibnamefont {Shuvaev}},\ }\href@noop {} {\bibinfo {title} {{On 'dual' parametrizations of generalized parton distributions}}} (\bibinfo {year} {2002}),\ \Eprint {https://arxiv.org/abs/hep-ph/0207153} {arXiv:hep-ph/0207153} \BibitemShut {NoStop}%
\bibitem [{\citenamefont {M\"uller}\ \emph {et~al.}(2015)\citenamefont {M\"uller}, \citenamefont {Polyakov},\ and\ \citenamefont {Semenov-Tian-Shansky}}]{Muller:2014wxa}%
  \BibitemOpen
  \bibfield  {author} {\bibinfo {author} {\bibfnamefont {D.}~\bibnamefont {M\"uller}}, \bibinfo {author} {\bibfnamefont {M.~V.}\ \bibnamefont {Polyakov}},\ and\ \bibinfo {author} {\bibfnamefont {K.~M.}\ \bibnamefont {Semenov-Tian-Shansky}},\ }\bibfield  {title} {\bibinfo {title} {{Dual parametrization of generalized parton distributions in two equivalent representations}},\ }\href {https://doi.org/10.1007/JHEP03(2015)052} {\bibfield  {journal} {\bibinfo  {journal} {JHEP}\ }\textbf {\bibinfo {volume} {03}},\ \bibinfo {pages} {052}},\ \Eprint {https://arxiv.org/abs/1412.4165} {arXiv:1412.4165 [hep-ph]} \BibitemShut {NoStop}%
\bibitem [{\citenamefont {Dutrieux}\ \emph {et~al.}(2022)\citenamefont {Dutrieux}, \citenamefont {Dutrieux}, \citenamefont {Grocholski}, \citenamefont {Grocholski}, \citenamefont {Moutarde}, \citenamefont {Moutarde}, \citenamefont {Sznajder},\ and\ \citenamefont {Sznajder}}]{Dutrieux:2021wll}%
  \BibitemOpen
  \bibfield  {author} {\bibinfo {author} {\bibfnamefont {H.}~\bibnamefont {Dutrieux}}, \bibinfo {author} {\bibfnamefont {H.}~\bibnamefont {Dutrieux}}, \bibinfo {author} {\bibfnamefont {O.}~\bibnamefont {Grocholski}}, \bibinfo {author} {\bibfnamefont {O.}~\bibnamefont {Grocholski}}, \bibinfo {author} {\bibfnamefont {H.}~\bibnamefont {Moutarde}}, \bibinfo {author} {\bibfnamefont {H.}~\bibnamefont {Moutarde}}, \bibinfo {author} {\bibfnamefont {P.}~\bibnamefont {Sznajder}},\ and\ \bibinfo {author} {\bibfnamefont {P.}~\bibnamefont {Sznajder}},\ }\bibfield  {title} {\bibinfo {title} {{Artificial neural network modelling of generalised parton distributions}},\ }\href {https://doi.org/10.1140/epjc/s10052-022-10211-5} {\bibfield  {journal} {\bibinfo  {journal} {Eur. Phys. J. C}\ }\textbf {\bibinfo {volume} {82}},\ \bibinfo {pages} {252} (\bibinfo {year} {2022})},\ \bibinfo {note} {[Erratum: Eur. Phys. J. C 82, 389 (2022)]},\ \Eprint {https://arxiv.org/abs/2112.10528} {arXiv:2112.10528 [hep-ph]} \BibitemShut {NoStop}%
\bibitem [{\citenamefont {Alexandrou}\ \emph {et~al.}(2018{\natexlab{b}})\citenamefont {Alexandrou}, \citenamefont {Cichy}, \citenamefont {Constantinou}, \citenamefont {Jansen}, \citenamefont {Scapellato},\ and\ \citenamefont {Steffens}}]{Alexandrou:2018pbm}%
  \BibitemOpen
  \bibfield  {author} {\bibinfo {author} {\bibfnamefont {C.}~\bibnamefont {Alexandrou}}, \bibinfo {author} {\bibfnamefont {K.}~\bibnamefont {Cichy}}, \bibinfo {author} {\bibfnamefont {M.}~\bibnamefont {Constantinou}}, \bibinfo {author} {\bibfnamefont {K.}~\bibnamefont {Jansen}}, \bibinfo {author} {\bibfnamefont {A.}~\bibnamefont {Scapellato}},\ and\ \bibinfo {author} {\bibfnamefont {F.}~\bibnamefont {Steffens}},\ }\bibfield  {title} {\bibinfo {title} {{Light-Cone Parton Distribution Functions from Lattice QCD}},\ }\href {https://doi.org/10.1103/PhysRevLett.121.112001} {\bibfield  {journal} {\bibinfo  {journal} {Phys. Rev. Lett.}\ }\textbf {\bibinfo {volume} {121}},\ \bibinfo {pages} {112001} (\bibinfo {year} {2018}{\natexlab{b}})},\ \Eprint {https://arxiv.org/abs/1803.02685} {arXiv:1803.02685 [hep-lat]} \BibitemShut {NoStop}%
\bibitem [{\citenamefont {Diehl}\ and\ \citenamefont {Kroll}(2013)}]{Diehl:2013xca}%
  \BibitemOpen
  \bibfield  {author} {\bibinfo {author} {\bibfnamefont {M.}~\bibnamefont {Diehl}}\ and\ \bibinfo {author} {\bibfnamefont {P.}~\bibnamefont {Kroll}},\ }\bibfield  {title} {\bibinfo {title} {{Nucleon form factors, generalized parton distributions and quark angular momentum}},\ }\href {https://doi.org/10.1140/epjc/s10052-013-2397-7} {\bibfield  {journal} {\bibinfo  {journal} {Eur. Phys. J. C}\ }\textbf {\bibinfo {volume} {73}},\ \bibinfo {pages} {2397} (\bibinfo {year} {2013})},\ \Eprint {https://arxiv.org/abs/1302.4604} {arXiv:1302.4604 [hep-ph]} \BibitemShut {NoStop}%
\bibitem [{\citenamefont {Arrington}\ \emph {et~al.}(2007)\citenamefont {Arrington}, \citenamefont {Melnitchouk},\ and\ \citenamefont {Tjon}}]{Arrington:2007ux}%
  \BibitemOpen
  \bibfield  {author} {\bibinfo {author} {\bibfnamefont {J.}~\bibnamefont {Arrington}}, \bibinfo {author} {\bibfnamefont {W.}~\bibnamefont {Melnitchouk}},\ and\ \bibinfo {author} {\bibfnamefont {J.~A.}\ \bibnamefont {Tjon}},\ }\bibfield  {title} {\bibinfo {title} {{Global analysis of proton elastic form factor data with two-photon exchange corrections}},\ }\href {https://doi.org/10.1103/PhysRevC.76.035205} {\bibfield  {journal} {\bibinfo  {journal} {Phys. Rev. C}\ }\textbf {\bibinfo {volume} {76}},\ \bibinfo {pages} {035205} (\bibinfo {year} {2007})},\ \Eprint {https://arxiv.org/abs/0707.1861} {arXiv:0707.1861 [nucl-ex]} \BibitemShut {NoStop}%
\bibitem [{\citenamefont {Milbrath}\ \emph {et~al.}(1998)\citenamefont {Milbrath} \emph {et~al.}}]{Milbrath:1997de}%
  \BibitemOpen
  \bibfield  {author} {\bibinfo {author} {\bibfnamefont {B.~D.}\ \bibnamefont {Milbrath}} \emph {et~al.} (\bibinfo {collaboration} {Bates FPP}),\ }\bibfield  {title} {\bibinfo {title} {{A Comparison of polarization observables in electron scattering from the proton and deuteron}},\ }\href {https://doi.org/10.1103/PhysRevLett.80.452} {\bibfield  {journal} {\bibinfo  {journal} {Phys. Rev. Lett.}\ }\textbf {\bibinfo {volume} {80}},\ \bibinfo {pages} {452} (\bibinfo {year} {1998})},\ \bibinfo {note} {[Erratum: Phys. Rev. Lett. 82, 2221 (1999)]},\ \Eprint {https://arxiv.org/abs/nucl-ex/9712006} {arXiv:nucl-ex/9712006} \BibitemShut {NoStop}%
\bibitem [{\citenamefont {Pospischil}\ \emph {et~al.}(2001)\citenamefont {Pospischil} \emph {et~al.}}]{Pospischil:2001pp}%
  \BibitemOpen
  \bibfield  {author} {\bibinfo {author} {\bibfnamefont {T.}~\bibnamefont {Pospischil}} \emph {et~al.} (\bibinfo {collaboration} {A1}),\ }\bibfield  {title} {\bibinfo {title} {{Measurement of $G_{Ep}/G_{Mp}$ via polarization transfer at $Q^2 = 0.4~(\mathrm{GeV}/c)^2$}},\ }\href {https://doi.org/10.1007/s100500170046} {\bibfield  {journal} {\bibinfo  {journal} {Eur. Phys. J. A}\ }\textbf {\bibinfo {volume} {12}},\ \bibinfo {pages} {125} (\bibinfo {year} {2001})}\BibitemShut {NoStop}%
\bibitem [{\citenamefont {Gayou}\ \emph {et~al.}(2001)\citenamefont {Gayou} \emph {et~al.}}]{Gayou:2001qt}%
  \BibitemOpen
  \bibfield  {author} {\bibinfo {author} {\bibfnamefont {O.}~\bibnamefont {Gayou}} \emph {et~al.},\ }\bibfield  {title} {\bibinfo {title} {{Measurements of the elastic electromagnetic form-factor ratio $\mu_p G_{Ep} / G_{Mp}$ via polarization transfer}},\ }\href {https://doi.org/10.1103/PhysRevC.64.038202} {\bibfield  {journal} {\bibinfo  {journal} {Phys. Rev. C}\ }\textbf {\bibinfo {volume} {64}},\ \bibinfo {pages} {038202} (\bibinfo {year} {2001})}\BibitemShut {NoStop}%
\bibitem [{\citenamefont {Gayou}\ \emph {et~al.}(2002)\citenamefont {Gayou} \emph {et~al.}}]{Gayou:2001qd}%
  \BibitemOpen
  \bibfield  {author} {\bibinfo {author} {\bibfnamefont {O.}~\bibnamefont {Gayou}} \emph {et~al.} (\bibinfo {collaboration} {Jefferson Lab Hall A}),\ }\bibfield  {title} {\bibinfo {title} {{Measurement of $G_{Ep} / G_{Mp}$ in polarized-e p $\to$ e polarized-p to $Q^2 = 5.6~\mathrm{GeV}^2$}},\ }\href {https://doi.org/10.1103/PhysRevLett.88.092301} {\bibfield  {journal} {\bibinfo  {journal} {Phys. Rev. Lett.}\ }\textbf {\bibinfo {volume} {88}},\ \bibinfo {pages} {092301} (\bibinfo {year} {2002})},\ \Eprint {https://arxiv.org/abs/nucl-ex/0111010} {arXiv:nucl-ex/0111010} \BibitemShut {NoStop}%
\bibitem [{\citenamefont {Punjabi}\ \emph {et~al.}(2005)\citenamefont {Punjabi} \emph {et~al.}}]{Punjabi:2005wq}%
  \BibitemOpen
  \bibfield  {author} {\bibinfo {author} {\bibfnamefont {V.}~\bibnamefont {Punjabi}} \emph {et~al.},\ }\bibfield  {title} {\bibinfo {title} {{Proton elastic form-factor ratios to $Q^2 = 3.5~\mathrm{GeV}^2$ by polarization transfer}},\ }\href {https://doi.org/10.1103/PhysRevC.71.055202} {\bibfield  {journal} {\bibinfo  {journal} {Phys. Rev. C}\ }\textbf {\bibinfo {volume} {71}},\ \bibinfo {pages} {055202} (\bibinfo {year} {2005})},\ \bibinfo {note} {[Erratum: Phys. Rev. C 71, 069902 (2005)]},\ \Eprint {https://arxiv.org/abs/nucl-ex/0501018} {arXiv:nucl-ex/0501018} \BibitemShut {NoStop}%
\bibitem [{\citenamefont {MacLachlan}\ \emph {et~al.}(2006)\citenamefont {MacLachlan} \emph {et~al.}}]{MacLachlan:2006vw}%
  \BibitemOpen
  \bibfield  {author} {\bibinfo {author} {\bibfnamefont {G.}~\bibnamefont {MacLachlan}} \emph {et~al.},\ }\bibfield  {title} {\bibinfo {title} {{The ratio of proton electromagnetic form factors via recoil polarimetry at $Q^2 = 1.13~(\mathrm{GeV}/c)^2$}},\ }\href {https://doi.org/10.1016/j.nuclphysa.2005.09.012} {\bibfield  {journal} {\bibinfo  {journal} {Nucl. Phys. A}\ }\textbf {\bibinfo {volume} {764}},\ \bibinfo {pages} {261} (\bibinfo {year} {2006})}\BibitemShut {NoStop}%
\bibitem [{\citenamefont {Puckett}\ \emph {et~al.}(2010)\citenamefont {Puckett} \emph {et~al.}}]{Puckett:2010ac}%
  \BibitemOpen
  \bibfield  {author} {\bibinfo {author} {\bibfnamefont {A.~J.~R.}\ \bibnamefont {Puckett}} \emph {et~al.},\ }\bibfield  {title} {\bibinfo {title} {{Recoil Polarization Measurements of the Proton Electromagnetic Form Factor Ratio to $Q^2$ = 8.5 GeV$^2$}},\ }\href {https://doi.org/10.1103/PhysRevLett.104.242301} {\bibfield  {journal} {\bibinfo  {journal} {Phys. Rev. Lett.}\ }\textbf {\bibinfo {volume} {104}},\ \bibinfo {pages} {242301} (\bibinfo {year} {2010})},\ \Eprint {https://arxiv.org/abs/1005.3419} {arXiv:1005.3419 [nucl-ex]} \BibitemShut {NoStop}%
\bibitem [{\citenamefont {Paolone}\ \emph {et~al.}(2010)\citenamefont {Paolone} \emph {et~al.}}]{Paolone:2010qc}%
  \BibitemOpen
  \bibfield  {author} {\bibinfo {author} {\bibfnamefont {M.}~\bibnamefont {Paolone}} \emph {et~al.},\ }\bibfield  {title} {\bibinfo {title} {{Polarization Transfer in the $^{4}\mathrm{He}(\vec{e},e'\vec{p})^{3}\mathrm{H}$ Reaction at $Q^2$ = 0.8 and 1.3 (GeV/c)$^2$}},\ }\href {https://doi.org/10.1103/PhysRevLett.105.072001} {\bibfield  {journal} {\bibinfo  {journal} {Phys. Rev. Lett.}\ }\textbf {\bibinfo {volume} {105}},\ \bibinfo {pages} {072001} (\bibinfo {year} {2010})},\ \Eprint {https://arxiv.org/abs/1002.2188} {arXiv:1002.2188 [nucl-ex]} \BibitemShut {NoStop}%
\bibitem [{\citenamefont {Ron}\ \emph {et~al.}(2011)\citenamefont {Ron} \emph {et~al.}}]{Ron:2011rd}%
  \BibitemOpen
  \bibfield  {author} {\bibinfo {author} {\bibfnamefont {G.}~\bibnamefont {Ron}} \emph {et~al.} (\bibinfo {collaboration} {Jefferson Lab Hall A}),\ }\bibfield  {title} {\bibinfo {title} {{Low $Q^2$ measurements of the proton form factor ratio $\mu_p G_E / G_M$}},\ }\href {https://doi.org/10.1103/PhysRevC.84.055204} {\bibfield  {journal} {\bibinfo  {journal} {Phys. Rev. C}\ }\textbf {\bibinfo {volume} {84}},\ \bibinfo {pages} {055204} (\bibinfo {year} {2011})},\ \Eprint {https://arxiv.org/abs/1103.5784} {arXiv:1103.5784 [nucl-ex]} \BibitemShut {NoStop}%
\bibitem [{\citenamefont {Zhan}\ \emph {et~al.}(2011)\citenamefont {Zhan} \emph {et~al.}}]{Zhan:2011ji}%
  \BibitemOpen
  \bibfield  {author} {\bibinfo {author} {\bibfnamefont {X.}~\bibnamefont {Zhan}} \emph {et~al.},\ }\bibfield  {title} {\bibinfo {title} {{High-Precision Measurement of the Proton Elastic Form Factor Ratio $\mu_pG_E/G_M$ at low $Q^2$}},\ }\href {https://doi.org/10.1016/j.physletb.2011.10.002} {\bibfield  {journal} {\bibinfo  {journal} {Phys. Lett. B}\ }\textbf {\bibinfo {volume} {705}},\ \bibinfo {pages} {59} (\bibinfo {year} {2011})},\ \Eprint {https://arxiv.org/abs/1102.0318} {arXiv:1102.0318 [nucl-ex]} \BibitemShut {NoStop}%
\bibitem [{\citenamefont {Anklin}\ \emph {et~al.}(1998)\citenamefont {Anklin} \emph {et~al.}}]{Anklin:1998ae}%
  \BibitemOpen
  \bibfield  {author} {\bibinfo {author} {\bibfnamefont {H.}~\bibnamefont {Anklin}} \emph {et~al.},\ }\bibfield  {title} {\bibinfo {title} {{Precise measurements of the neutron magnetic form-factor}},\ }\href {https://doi.org/10.1016/S0370-2693(98)00442-0} {\bibfield  {journal} {\bibinfo  {journal} {Phys. Lett. B}\ }\textbf {\bibinfo {volume} {428}},\ \bibinfo {pages} {248} (\bibinfo {year} {1998})}\BibitemShut {NoStop}%
\bibitem [{\citenamefont {Kubon}\ \emph {et~al.}(2002)\citenamefont {Kubon} \emph {et~al.}}]{Kubon:2001rj}%
  \BibitemOpen
  \bibfield  {author} {\bibinfo {author} {\bibfnamefont {G.}~\bibnamefont {Kubon}} \emph {et~al.},\ }\bibfield  {title} {\bibinfo {title} {{Precise neutron magnetic form-factors}},\ }\href {https://doi.org/10.1016/S0370-2693(01)01386-7} {\bibfield  {journal} {\bibinfo  {journal} {Phys. Lett. B}\ }\textbf {\bibinfo {volume} {524}},\ \bibinfo {pages} {26} (\bibinfo {year} {2002})},\ \Eprint {https://arxiv.org/abs/nucl-ex/0107016} {arXiv:nucl-ex/0107016} \BibitemShut {NoStop}%
\bibitem [{\citenamefont {Anklin}\ \emph {et~al.}(1994)\citenamefont {Anklin} \emph {et~al.}}]{Anklin:1994ae}%
  \BibitemOpen
  \bibfield  {author} {\bibinfo {author} {\bibfnamefont {H.}~\bibnamefont {Anklin}} \emph {et~al.},\ }\bibfield  {title} {\bibinfo {title} {{Precision measurement of the neutron magnetic form-factor}},\ }\href {https://doi.org/10.1016/0370-2693(94)90538-X} {\bibfield  {journal} {\bibinfo  {journal} {Phys. Lett. B}\ }\textbf {\bibinfo {volume} {336}},\ \bibinfo {pages} {313} (\bibinfo {year} {1994})}\BibitemShut {NoStop}%
\bibitem [{\citenamefont {Lachniet}\ \emph {et~al.}(2009)\citenamefont {Lachniet} \emph {et~al.}}]{Lachniet:2008qf}%
  \BibitemOpen
  \bibfield  {author} {\bibinfo {author} {\bibfnamefont {J.}~\bibnamefont {Lachniet}} \emph {et~al.} (\bibinfo {collaboration} {CLAS}),\ }\bibfield  {title} {\bibinfo {title} {{A Precise Measurement of the Neutron Magnetic Form Factor $G^n_M$ in the Few-GeV$^2$ Region}},\ }\href {https://doi.org/10.1103/PhysRevLett.102.192001} {\bibfield  {journal} {\bibinfo  {journal} {Phys. Rev. Lett.}\ }\textbf {\bibinfo {volume} {102}},\ \bibinfo {pages} {192001} (\bibinfo {year} {2009})},\ \Eprint {https://arxiv.org/abs/0811.1716} {arXiv:0811.1716 [nucl-ex]} \BibitemShut {NoStop}%
\bibitem [{\citenamefont {Anderson}\ \emph {et~al.}(2007)\citenamefont {Anderson} \emph {et~al.}}]{Anderson:2006jp}%
  \BibitemOpen
  \bibfield  {author} {\bibinfo {author} {\bibfnamefont {B.}~\bibnamefont {Anderson}} \emph {et~al.} (\bibinfo {collaboration} {Jefferson Lab E95-001}),\ }\bibfield  {title} {\bibinfo {title} {{Extraction of the Neutron Magnetic Form Factor from Quasi-elastic $^{3}\vec{\mathrm{He}}(\vec{e},e')$ at Q$^2$ = 0.1 - 0.6 (GeV/c)$^2$}},\ }\href {https://doi.org/10.1103/PhysRevC.75.034003} {\bibfield  {journal} {\bibinfo  {journal} {Phys. Rev. C}\ }\textbf {\bibinfo {volume} {75}},\ \bibinfo {pages} {034003} (\bibinfo {year} {2007})},\ \Eprint {https://arxiv.org/abs/nucl-ex/0605006} {arXiv:nucl-ex/0605006} \BibitemShut {NoStop}%
\bibitem [{\citenamefont {Herberg}\ \emph {et~al.}(1999)\citenamefont {Herberg} \emph {et~al.}}]{Herberg:1999ud}%
  \BibitemOpen
  \bibfield  {author} {\bibinfo {author} {\bibfnamefont {C.}~\bibnamefont {Herberg}} \emph {et~al.},\ }\bibfield  {title} {\bibinfo {title} {{Determination of the neutron electric form-factor in the $\mathrm{D}(e,e' n)p$ reaction and the influence of nuclear binding}},\ }\href {https://doi.org/10.1007/s100500050268} {\bibfield  {journal} {\bibinfo  {journal} {Eur. Phys. J. A}\ }\textbf {\bibinfo {volume} {5}},\ \bibinfo {pages} {131} (\bibinfo {year} {1999})}\BibitemShut {NoStop}%
\bibitem [{\citenamefont {Glazier}\ \emph {et~al.}(2005)\citenamefont {Glazier} \emph {et~al.}}]{Glazier:2004ny}%
  \BibitemOpen
  \bibfield  {author} {\bibinfo {author} {\bibfnamefont {D.~I.}\ \bibnamefont {Glazier}} \emph {et~al.},\ }\bibfield  {title} {\bibinfo {title} {{Measurement of the electric form-factor of the neutron at $Q^2 = 0.3~(\mathrm{GeV}/c)^2$ to $0.8~(\mathrm{GeV}/c)^2$}},\ }\href {https://doi.org/10.1140/epja/i2004-10115-8} {\bibfield  {journal} {\bibinfo  {journal} {Eur. Phys. J. A}\ }\textbf {\bibinfo {volume} {24}},\ \bibinfo {pages} {101} (\bibinfo {year} {2005})},\ \Eprint {https://arxiv.org/abs/nucl-ex/0410026} {arXiv:nucl-ex/0410026} \BibitemShut {NoStop}%
\bibitem [{\citenamefont {Plaster}\ \emph {et~al.}(2006)\citenamefont {Plaster} \emph {et~al.}}]{Plaster:2005cx}%
  \BibitemOpen
  \bibfield  {author} {\bibinfo {author} {\bibfnamefont {B.}~\bibnamefont {Plaster}} \emph {et~al.} (\bibinfo {collaboration} {Jefferson Laboratory E93-038}),\ }\bibfield  {title} {\bibinfo {title} {{Measurements of the neutron electric to magnetic form factor ratio $\bm{G_{En}/G_{Mn}}$ via the $\bm{^{2}}$H$\bm{(}\vec{\bm{e}}\bm{,e'}\vec{\bm{n}}\bm{)^{1}}$H reaction to $\bm{Q^{2}=1.45}$ (GeV/$\bm{c}$)$\bm{^{2}}$}},\ }\href {https://doi.org/10.1103/PhysRevC.73.025205} {\bibfield  {journal} {\bibinfo  {journal} {Phys. Rev. C}\ }\textbf {\bibinfo {volume} {73}},\ \bibinfo {pages} {025205} (\bibinfo {year} {2006})},\ \Eprint {https://arxiv.org/abs/nucl-ex/0511025} {arXiv:nucl-ex/0511025} \BibitemShut {NoStop}%
\bibitem [{\citenamefont {Passchier}\ \emph {et~al.}(1999)\citenamefont {Passchier} \emph {et~al.}}]{Passchier:1999cj}%
  \BibitemOpen
  \bibfield  {author} {\bibinfo {author} {\bibfnamefont {I.}~\bibnamefont {Passchier}} \emph {et~al.},\ }\bibfield  {title} {\bibinfo {title} {{The Charge form-factor of the neutron from the reaction polarized H-2(polarized e, e-prime n) p}},\ }\href {https://doi.org/10.1103/PhysRevLett.82.4988} {\bibfield  {journal} {\bibinfo  {journal} {Phys. Rev. Lett.}\ }\textbf {\bibinfo {volume} {82}},\ \bibinfo {pages} {4988} (\bibinfo {year} {1999})},\ \Eprint {https://arxiv.org/abs/nucl-ex/9907012} {arXiv:nucl-ex/9907012} \BibitemShut {NoStop}%
\bibitem [{\citenamefont {Zhu}\ \emph {et~al.}(2001)\citenamefont {Zhu} \emph {et~al.}}]{Zhu:2001md}%
  \BibitemOpen
  \bibfield  {author} {\bibinfo {author} {\bibfnamefont {H.}~\bibnamefont {Zhu}} \emph {et~al.} (\bibinfo {collaboration} {E93026}),\ }\bibfield  {title} {\bibinfo {title} {{A Measurement of the electric form-factor of the neutron through polarized-d (polarized-e, e-prime n)p at $Q^2 = 0.5~(\mathrm{GeV}/c)^2$}},\ }\href {https://doi.org/10.1103/PhysRevLett.87.081801} {\bibfield  {journal} {\bibinfo  {journal} {Phys. Rev. Lett.}\ }\textbf {\bibinfo {volume} {87}},\ \bibinfo {pages} {081801} (\bibinfo {year} {2001})},\ \Eprint {https://arxiv.org/abs/nucl-ex/0105001} {arXiv:nucl-ex/0105001} \BibitemShut {NoStop}%
\bibitem [{\citenamefont {Warren}\ \emph {et~al.}(2004)\citenamefont {Warren} \emph {et~al.}}]{Warren:2003ma}%
  \BibitemOpen
  \bibfield  {author} {\bibinfo {author} {\bibfnamefont {G.}~\bibnamefont {Warren}} \emph {et~al.} (\bibinfo {collaboration} {Jefferson Lab E93-026}),\ }\bibfield  {title} {\bibinfo {title} {{Measurement of the electric form-factor of the neutron at $Q^2$ = 0.5 and 1.0 $\mathrm{GeV}^2/c^2$}},\ }\href {https://doi.org/10.1103/PhysRevLett.92.042301} {\bibfield  {journal} {\bibinfo  {journal} {Phys. Rev. Lett.}\ }\textbf {\bibinfo {volume} {92}},\ \bibinfo {pages} {042301} (\bibinfo {year} {2004})},\ \Eprint {https://arxiv.org/abs/nucl-ex/0308021} {arXiv:nucl-ex/0308021} \BibitemShut {NoStop}%
\bibitem [{\citenamefont {Geis}\ \emph {et~al.}(2008)\citenamefont {Geis} \emph {et~al.}}]{Geis:2008aa}%
  \BibitemOpen
  \bibfield  {author} {\bibinfo {author} {\bibfnamefont {E.}~\bibnamefont {Geis}} \emph {et~al.} (\bibinfo {collaboration} {BLAST}),\ }\bibfield  {title} {\bibinfo {title} {{The Charge Form Factor of the Neutron at Low Momentum Transfer from the H-2-polarized (e-polarized, e-prime n) p Reaction}},\ }\href {https://doi.org/10.1103/PhysRevLett.101.042501} {\bibfield  {journal} {\bibinfo  {journal} {Phys. Rev. Lett.}\ }\textbf {\bibinfo {volume} {101}},\ \bibinfo {pages} {042501} (\bibinfo {year} {2008})},\ \Eprint {https://arxiv.org/abs/0803.3827} {arXiv:0803.3827 [nucl-ex]} \BibitemShut {NoStop}%
\bibitem [{\citenamefont {Bermuth}\ \emph {et~al.}(2003)\citenamefont {Bermuth} \emph {et~al.}}]{Bermuth:2003qh}%
  \BibitemOpen
  \bibfield  {author} {\bibinfo {author} {\bibfnamefont {J.}~\bibnamefont {Bermuth}} \emph {et~al.},\ }\bibfield  {title} {\bibinfo {title} {{The Neutron charge form-factor and target analyzing powers from polarized-He-3 (polarized-e,e-prime n) scattering}},\ }\href {https://doi.org/10.1016/S0370-2693(03)00725-1} {\bibfield  {journal} {\bibinfo  {journal} {Phys. Lett. B}\ }\textbf {\bibinfo {volume} {564}},\ \bibinfo {pages} {199} (\bibinfo {year} {2003})},\ \Eprint {https://arxiv.org/abs/nucl-ex/0303015} {arXiv:nucl-ex/0303015} \BibitemShut {NoStop}%
\bibitem [{\citenamefont {Rohe}()}]{rohe_pc}%
  \BibitemOpen
  \bibfield  {author} {\bibinfo {author} {\bibfnamefont {D.}~\bibnamefont {Rohe}},\ }\href@noop {} {}\bibinfo {howpublished} {Private communication with M. Diehl and P. Kroll}\BibitemShut {NoStop}%
\bibitem [{\citenamefont {Riordan}\ \emph {et~al.}(2010)\citenamefont {Riordan} \emph {et~al.}}]{Riordan:2010id}%
  \BibitemOpen
  \bibfield  {author} {\bibinfo {author} {\bibfnamefont {S.}~\bibnamefont {Riordan}} \emph {et~al.},\ }\bibfield  {title} {\bibinfo {title} {{Measurements of the Electric Form Factor of the Neutron up to $Q^2=3.4~\mathrm{GeV}^2$ using the Reaction $^3\mathrm{He}^{\to}(e^{\to},e'n)pp$}},\ }\href {https://doi.org/10.1103/PhysRevLett.105.262302} {\bibfield  {journal} {\bibinfo  {journal} {Phys. Rev. Lett.}\ }\textbf {\bibinfo {volume} {105}},\ \bibinfo {pages} {262302} (\bibinfo {year} {2010})},\ \Eprint {https://arxiv.org/abs/1008.1738} {arXiv:1008.1738 [nucl-ex]} \BibitemShut {NoStop}%
\bibitem [{\citenamefont {Schiavilla}\ and\ \citenamefont {Sick}(2001)}]{Schiavilla:2001qe}%
  \BibitemOpen
  \bibfield  {author} {\bibinfo {author} {\bibfnamefont {R.}~\bibnamefont {Schiavilla}}\ and\ \bibinfo {author} {\bibfnamefont {I.}~\bibnamefont {Sick}},\ }\bibfield  {title} {\bibinfo {title} {{Neutron charge form-factor at large $q^2$}},\ }\href {https://doi.org/10.1103/PhysRevC.64.041002} {\bibfield  {journal} {\bibinfo  {journal} {Phys. Rev. C}\ }\textbf {\bibinfo {volume} {64}},\ \bibinfo {pages} {041002} (\bibinfo {year} {2001})},\ \Eprint {https://arxiv.org/abs/nucl-ex/0107004} {arXiv:nucl-ex/0107004} \BibitemShut {NoStop}%
\bibitem [{\citenamefont {Beringer}\ \emph {et~al.}(2012)\citenamefont {Beringer} \emph {et~al.}}]{Beringer:1900zz}%
  \BibitemOpen
  \bibfield  {author} {\bibinfo {author} {\bibfnamefont {J.}~\bibnamefont {Beringer}} \emph {et~al.} (\bibinfo {collaboration} {Particle Data Group}),\ }\bibfield  {title} {\bibinfo {title} {{Review of Particle Physics (RPP)}},\ }\href {https://doi.org/10.1103/PhysRevD.86.010001} {\bibfield  {journal} {\bibinfo  {journal} {Phys. Rev. D}\ }\textbf {\bibinfo {volume} {86}},\ \bibinfo {pages} {010001} (\bibinfo {year} {2012})}\BibitemShut {NoStop}%
\bibitem [{\citenamefont {James}\ and\ \citenamefont {Roos}(1975)}]{James:1975dr}%
  \BibitemOpen
  \bibfield  {author} {\bibinfo {author} {\bibfnamefont {F.}~\bibnamefont {James}}\ and\ \bibinfo {author} {\bibfnamefont {M.}~\bibnamefont {Roos}},\ }\bibfield  {title} {\bibinfo {title} {{Minuit: A System for Function Minimization and Analysis of the Parameter Errors and Correlations}},\ }\href {https://doi.org/10.1016/0010-4655(75)90039-9} {\bibfield  {journal} {\bibinfo  {journal} {Comput. Phys. Commun.}\ }\textbf {\bibinfo {volume} {10}},\ \bibinfo {pages} {343} (\bibinfo {year} {1975})}\BibitemShut {NoStop}%
\bibitem [{\citenamefont {Mitchell}(1998)}]{ga}%
  \BibitemOpen
  \bibfield  {author} {\bibinfo {author} {\bibfnamefont {M.}~\bibnamefont {Mitchell}},\ }\href@noop {} {\emph {\bibinfo {title} {An Introduction to Genetic Algorithms}}}\ (\bibinfo  {publisher} {MIT Press},\ \bibinfo {address} {Cambridge, MA, USA},\ \bibinfo {year} {1998})\BibitemShut {NoStop}%
\bibitem [{\citenamefont {Akhunzyanov}\ \emph {et~al.}(2019)\citenamefont {Akhunzyanov} \emph {et~al.}}]{COMPASS:2018pup}%
  \BibitemOpen
  \bibfield  {author} {\bibinfo {author} {\bibfnamefont {R.}~\bibnamefont {Akhunzyanov}} \emph {et~al.} (\bibinfo {collaboration} {COMPASS}),\ }\bibfield  {title} {\bibinfo {title} {{Transverse extension of partons in the proton probed in the sea-quark range by measuring the DVCS cross section}},\ }\href {https://doi.org/10.1016/j.physletb.2019.04.038} {\bibfield  {journal} {\bibinfo  {journal} {Phys. Lett. B}\ }\textbf {\bibinfo {volume} {793}},\ \bibinfo {pages} {188} (\bibinfo {year} {2019})},\ \bibinfo {note} {[Erratum: Phys. Lett. B 800, 135129 (2020)]},\ \Eprint {https://arxiv.org/abs/1802.02739} {arXiv:1802.02739 [hep-ex]} \BibitemShut {NoStop}%
\bibitem [{\citenamefont {Aschenauer}\ \emph {et~al.}(2013)\citenamefont {Aschenauer}, \citenamefont {Fazio}, \citenamefont {Kumericki},\ and\ \citenamefont {Mueller}}]{Aschenauer:2013hhw}%
  \BibitemOpen
  \bibfield  {author} {\bibinfo {author} {\bibfnamefont {E.-C.}\ \bibnamefont {Aschenauer}}, \bibinfo {author} {\bibfnamefont {S.}~\bibnamefont {Fazio}}, \bibinfo {author} {\bibfnamefont {K.}~\bibnamefont {Kumericki}},\ and\ \bibinfo {author} {\bibfnamefont {D.}~\bibnamefont {Mueller}},\ }\bibfield  {title} {\bibinfo {title} {{Deeply Virtual Compton Scattering at a Proposed High-Luminosity Electron-Ion Collider}},\ }\href {https://doi.org/10.1007/JHEP09(2013)093} {\bibfield  {journal} {\bibinfo  {journal} {JHEP}\ }\textbf {\bibinfo {volume} {09}},\ \bibinfo {pages} {093}},\ \Eprint {https://arxiv.org/abs/1304.0077} {arXiv:1304.0077 [hep-ph]} \BibitemShut {NoStop}%
\end{thebibliography}%
